\documentclass[12pt]{article}          

\usepackage{amsmath,amsthm}          
\usepackage{newtxtext,newtxmath}     
\usepackage{bbold}                    
\usepackage{bm}                        
\usepackage[T1]{fontenc}              
\usepackage[protrusion=true,expansion=false]{microtype}  

\usepackage[margin=1in]{geometry}      
\usepackage{setspace}
\usepackage{xcolor}                   
\usepackage{enumitem}                  
\usepackage{natbib}

\usepackage{booktabs}                 
\usepackage{array}                     
\usepackage{threeparttable}           
\usepackage{adjustbox}               
\usepackage{pdflscape}               
\usepackage{etoolbox}
\AtBeginEnvironment{table}{\singlespacing}

\makeatletter
\renewcommand{\TPTdoTablenotes}{%
  \par \prevdepth\z@ \hsize\textwidth \@parboxrestore
  \list{}{\topsep\z@skip \partopsep\z@skip \itemsep\z@ \parsep\z@
    \itemindent\z@ \TPTnoteSettings \let\makelabel\TPTnoteLabel}}
\makeatother

\usepackage{graphicx}                  
\usepackage{subcaption}
\usepackage[skip=4pt]{caption}        
\usepackage{float}                     
\AtBeginEnvironment{figure}{\singlespacing} 

\usepackage{endnotes}
\let\footnote=\endnote

\usepackage{footmisc}

\usepackage{xurl}                     
\usepackage{hyperref}                  
\hypersetup{colorlinks=true,
            citecolor=blue!55!black,
            linkcolor=blue!55!black,
            urlcolor=blue!55!black}  
            
\begin{document}

  \title{\bf \Large Taste or Strategy? Nationalistic Bias and Competitor Penalties in Figure Skating Judging}
  \author{Jialin Liu\\\small{Department of Economics} \\ \small{University of California, Davis}\\\small{\texttt{jalliu@ucdavis.edu} } \and Shu Shen\\\small{Department of Economics} \\\small{University of California, Davis}\\\small{\texttt{shushen@ucdavis.edu}} \thanks{This project started as a classroom demonstration example for using AI to collect data and help kickstart empirical research in an undergraduate Econometrics course in the University of California, Davis in the quarter of Spring 2026. The authors thank discussions from the students, and comments from colleagues in the UC Davis behavioral student group and the University of Victoria department of Economics. The authors used Claude (Anthropic) to help write and check the code that collects and processes the data and for editorial work. The authors take full responsibility for the contents of this paper.
  }}  
  \maketitle

\begin{abstract}

Many decisions rest on expert panels whose members have ties to some candidates, and favoritism is hard to tell from shared taste or better information. Using a rich dataset from figure skating, we find evidence of a strategic compatriot premium and a targeted penalty on the compatriot's close rivals. The premium falls along a randomly drawn start order and, before 2022/23, rises when close rivals have their own judges on the panel. The penalty grows with the compatriot's placement chances and fades once she is out of contention. A counterfactual analysis shows that these biases can shape close competitions.

\end{abstract}
\section{Introduction}

Many important decisions rest on the subjective judgments of panels of experts, who mark candidates competing for a limited number of opportunities, such as admissions, promotions, research grants (\citealp{li2017expertise}), election to academies (\citealp{fisman2018social}), sports competitions (\citealp{zitzewitz2006nationalism}), and the prizes, scholarships and positions awarded by top international ballet, piano and violin competitions (\citealp{ginsburgh2003expert}). Panel members often have ties to some of the candidates: a shared nationality or hometown, a candidate they once taught, or a common teacher. A favorable mark may then reflect favoritism, a shared taste \citep{zitzewitz2006nationalism}, or better information \citep{aigner1977statistical} if the evaluator is risk-averse. These are hard to tell apart because distortions in performance measures are rarely observed directly \citep{courty2008general}, and even when they are, few settings reveal how panels are formed, how each member evaluates each candidate, and which candidates are tied to which members.

In this paper, we utilize a rich dataset from figure skating competitions to separate the motives behind a judge's favorable mark that deviates from the rest of the judging panel. The sport is particularly suitable for the analysis as it reveals each judge's marks, the order of evaluation and the judges' national ties to the skaters. It also offers variation in what a judge knows and in what her mark can achieve. In some competitions the start order of the first segment is randomly drawn, so how much of the field a judge has seen when she marks her compatriot is random. The scores are announced after each performance, so a judge knows whether her compatriot is still in contention when she marks a rival. At major championships and in the individual events of the Olympics, the judging panels are randomly drawn from eligible federations, so whether a skater faces a compatriot judge is partially determined by chance. Our dataset covers the major international competitions in the ten seasons since the International Skating Union (ISU) ended anonymous judging in 2016, and we collected skater's ISU biography information in addition to the official competition markings. As far as we know, it is the most comprehensive dataset of individual judges' scores used to study any group judged sport. Many of the institutional features we use to tell favoritism apart from other unintentional factors have not been used before in the literature of subjective panel judging. 

In figure skating, the compatriot premium is well-documented. \cite{zitzewitz2006nationalism}, following the 2002 Salt Lake City scandal\footnote{In the pair skating competition at the 2002 Winter Olympic Games in Salt Lake City, French judge Marie-Reine Le Gougne admitted, and then later retracted the admission, that, under pressure from her federation, she had inflated the marks of the Russian pair in exchange for Russian support in the upcoming ice dance event.}, finds that judges score athletes from their own federations and those from other panel judges' federations higher. More recently, the judging of the ice dance at the 2026 Winter Olympic Games in Milan sparked heated online discussions\footnote{In the competition, Laurence Fournier Beaudry and Guillaume Cizeron of France won the rhythm dance by 0.46 points over Madison Chock and Evan Bates of the United States, and the free dance by 0.97 points. On the published per-judge marks of the free dance, the French judge had the American couple 7.71 points behind the French couple, while five of the other eight judges had the Americans ahead. In the rhythm dance, the French judge had them 5.74 points ahead.}. The literature (e.g., \citealp{zitzewitz2006nationalism,zitzewitz2014does,krumer2022nationalistic,ansorge1988international,heiniger2021judging,sandberg2018competing}, among many others) has long suggested nationalistic incentives behind the premium but could not disentangle from non-intentional explanations such as taste (\citealp{zitzewitz2006nationalism}) or familiarity (\citealp{aigner1977statistical}). Leveraging the institutional design of figure skating, we find that the premium is, at least in part, strategic. Even when the start order is purely random, a judge's premium for her compatriot declines as the competition goes on. Since neither taste or familiarity depend on a random start position, the premium of the last starter bounds how much these other factors can explain: at least a fifth of the average premium is something else. This declining pattern is consistent with strategic nationalistic scoring under uncertainty. When she marks an early-starting compatriot, a judge cannot tell how big a premium she needs to award to make it matter, whereas for a late-starting compatriot she knows better whether a premium is worthwhile and, if it is, how large it must be to change a placement. The compatriot premium also rises when the compatriot's close rivals, especially those who start later, have judges of their own federations on the panel. In the Beijing Olympic cycle, we find that the compatriot premium rises with the share of the compatriot's close rivals who have a judge of their own federation on the panel. The increase is larger when the other judges have a more biased reputation, which we estimate from the previous PyeongChang cycle and correct in the second step for its estimation error (\citealp{chen2026mostly}). 

Judges also mark down their compatriots' close rivals, and more importantly, in a targeted way. The finding is new to the literature, where evidence for the competitor penalty is thin and mixed\footnote{\citet{ansorge1988international} find that judges at the 1984 Olympic gymnastics competition scored the close rivals of their compatriots significantly lower. In contrast, \citet{sandberg2018competing} finds no significant penalty in dressage, and \citet{heiniger2021judging} find that the penalty in gymnastics is negligible in size and mostly statistically insignificant.}. The literature defines close rivals by placement outcomes. To avoid the reverse-causality problem, we instead define close rivals by the segment start order. Our rival definition is a strong proxy of the true rivalry in Segment 2 competitions, where the leaders after the first segment skate last. We find that a judge marks a rival who skates after her compatriot significantly lower, but only while the compatriot is still in contention, as is defined by the standings of the field when the rival skates. The competitor penalty also rises with the compatriot's chance of a good placement. In Segment 2, the penalty grows with the compatriot's start position, from about zero when the compatriot starts first to about half a percent of the scale when she starts last and competes for the podium. 

This paper also revisits the empirical finding of \citet{zitzewitz2006nationalism} that judges in figure skating give higher scores to other judges' compatriots. We call it the colleague-compatriot premium rather than vote trading since other reasons such as judges' courtesy to colleagues would produce it as well. We find that the evidence of colleague-compatriot premium in data is much weaker once we control for performance quality, such as the start position, the skater's prior placements at the World and World Junior Championships, the program's choreographer, etc. These controls are not needed in regressions that use within-performance judges score deviations for identification, but are crucial to cross-panel regressions for studying colleague-compatriot premium, home advantage, and the impact of panel composition on placement outcomes. We also find a compatriot premium on the technical panel. The premium is sizable, considering that the base values the technical panel sets enter every judge's score. The home ice advantage, in turn, is awarded by the guest judges rather than by the home federation's own judge.

The same cross-panel regressions discussed above are also subject to the non-convex average (negative weight) problem discussed in \citet{dechaisemartin2020two} and \citet{goldsmithpinkham2024contamination} for the use of multi-way fixed effects. The problem arises because the coefficient is identified from changes in panel composition across competitions, and at some competitions almost every performance has a compatriot judge on the panel. To our knowledge, this problem has not been discussed in the literature of subjective panel judging before, where cross-panel regression models are frequently used. In our analysis, the check of negative weight share and weight mass leads to better understandings of the model and points us to a subsample of competitions with randomly drawn judge panels, under which the problem nearly disappears.

With the above mentioned findings of strategic nationalistic scoring, we carry out a counterfactual analysis of the figure skating events at the 2026 Milan Winter Olympic Games. We show that the compatriot premium, the competitor penalty, and the colleague-compatriot premium, through the random draw of judge panels, can jointly shape close competitions. In the analysis we redraw admissible panels under ISU's draw rules for Winter Olympic Games and rescore each performance under each redraw. We find that the ice dance title changes hands in 45 to 57 percent of the redraws and the women's title 20 to 27 percent, depending on how the judges' honest disagreement is modeled. The ISU's trimmed mean offers little protection. We derive how much of one judge's bias a nine-judge trim removes, and trimming removes only a small part of the compatriot premium or the competitor penalty of a typical size. We therefore suggest keeping the federations of the leading skaters off the judging panel, when feasible, as the ISU already does for the technical panel. 

Last but not least, we would like to point out that, in this paper, we refrain from either selecting a set of biased judges with multiple testing procedures or ranking the judges, although we obtain judge-specific compatriot premium estimates as an intermediate step of our regression analysis. This is because a multiple testing procedure is more likely to single out judges who sit on more panels and come from larger federations, while the point estimates are unstable for judges that do not have a substantial presence in the data. Readers can reproduce the judge-specific estimates from our replication package, but we caution against reading too much into the numbers or using them as evidence against specific judges. To properly identify the most nationalistic judges in the data, proper statistical methods for ranking and selection such as \citet{armstrong2022robust, gu2023invidious,mogstad2024inference,andrews2024inference}, among others, need to be adopted.

The rest of the paper is organized as follows. Section~\ref{sec:insti} introduces the institutional background, points out the identification challenges in the literature, and describes the data. Section~\ref{sec:scorebias} presents our benchmark regressions, confirming a strong compatriot premium found in the literature and uncovering a new precisely estimated targeted competitor penalty. Section~\ref{sec:strategic} further explores strategic nationalistic scoring, showing that compatriot premium and competitor penalty are strategic, both varying along the start order, and the premium also responding to the panel composition. Section~\ref{sec:milan} examines how the judge panel composition can impact competition outcomes and carries out a counterfactual analysis of the 2026 Olympic events. Section~\ref{sec:conclusion} concludes and provides policy implications. The appendix gives further institutional background and reports robustness checks for the main results. Although our setting is figure skating, the findings carry lessons for other settings where economic decisions rest on subjective evaluations by expert panels. 

\section{Institutional Background and Data Summary} 
\label{sec:insti}

\subsection{Figure Skating Competitions and the Scoring Mechanism}
\label{sec:scoring}

Figure skating is a sport in which athletes or athlete pairs perform choreographed programs of jumps, spins, lifts, and other moves together with music. Scores of figure skating performances are fully determined by officials: a technical panel and a judge panel. Important figure skating competitions include, for the seniors, World, European, and Four Continents Championships, the Grand Prix (GP) series, and the GP Final, and, for the juniors, World Junior Championships, the Junior Grand Prix (JGP) series, and the JGP Final. Every four years, there are also the Winter Olympic Games, the Winter Youth Olympic Games, and the Asian Winter Games.\footnote{No other continent held continental winter games during our sample period.} All figure skating competitions span across multiple disciplines (Men, Women, Pair, Ice Dance)\footnote{Some Junior Grand Prix events hold no pairs competition. We do not study synchronized skating in this paper.} and sometimes also different tiers (Junior, Senior) and different types (Individual, Team). Regardless of discipline, tier, and type, each category of a figure skating competition consists of two segments, a short program (in ice dance, the rhythm dance, called the short dance before 2018/19) first, and 1-3 days later, a longer free skate (in ice dance, the free dance). 

The score of each segment has three parts: the Total Element Score (TES), the Program Component Score (PCS), and deductions. The Total Segment Score (TSS) is the TES plus the PCS minus the deductions. A skater's final ranking is determined by the sum of the two segment scores. Two groups of officials produce the TES. The technical panel identifies each element (e.g.\ jumps, spins, step sequences), assigns difficulty levels (Basic, then Level 1 through 4), and flags errors such as under-rotations, wrong edges, and falls. These calls determine the Base Value (BV) of each element. The technical panel consists of a Technical Controller and two Technical Specialists. During the performance the Technical Specialist identifies each element as it is executed and calls the level of difficulty when needed. The Technical Controller supervises and may correct a call. 

The judges mark how well each element was executed (on an integer scale from $-5$ to $+5$ post 2018/19 and from $-3$ to $+3$ before), and their marks are aggregated by a trimmed mean into the grade of execution (GOE). For each element, the single highest and lowest marks are dropped and the rest are converted into points and averaged. An element's score is its BV plus its GOE, and the TES is the sum over the program's elements. The technical panel's calls affect the GOE both directly, because GOE grades of jumps and step sequences are converted into points as a fixed share of the base value, and indirectly, through their influences on judges' execution markings.  

After the performance, each judge also marks the program components on a scale in steps of 0.25 up to 10.00. Since 2022/23 there are three components (five before): Composition, Presentation and Skating Skills. Each component's trimmed mean is multiplied by the segment's factor and the products are summed. Appendix \ref{app:protocol} gives out an example by working through the math with an official protocol of a performance.

Our dataset starts from the 2016/17 season because the ISU ended anonymous judging after the 2015/16 season. The protocols do not name the judges, but there is a separate Panel of Judges document that the ISU, or the organizing committee/federation for the events, publishes for every segment, which we use to identify the names of judges. In the appendix, we also explain in details how judges are linked to their federations.

\subsection{Judge Assignment and Performance Start Order}
\label{sec:jassign-sod}

In this section, we provide contexts for how judges are assigned to panels and how the start order of competitions is set. Judge assignment explains why some regression models used in the literature suffer from potential endogeneity and the non-convex average problem, and how choosing a specific subsample mitigates the problem. The start order is used to define competitors. 

Panel formation rules vary across competitions. In important competitions such as the individual events of the Olympics, the World Junior and World Championships, and the Europeans, the panel is drawn randomly. First, thirteen federations are drawn for each discipline, among eligible federations\footnote{A federation is eligible if it had a skater in the discipline at previous year's Championships or, for the Olympics, a skater who qualified for the Games at the previous World Championships.}, who nominate judges after being selected. Before the first segment, a nine judge panel is drawn out of the thirteen. The four not drawn to the first segment sit the second automatically. The other five seats for the second are then drawn from the nine who sat the first. Some other competitions like Four Continents and the Youth Olympic Games use a simplified random draw process which only draws federations but keep the same judges for both segments of a discipline. At the GP events, all judges are invited by the host. For the GP Final, the ISU invites the judges, with priority given to the federations with qualified skaters. At the JGP events, each federation with skaters entered may send 1-2 judges. When more judges are present than a discipline's nine seats, a panel draw takes place.

The start position is determined by different regimes. For the first segment, there are three regimes: open draw, which mostly happens in JGP events, reverse ranking based on the ISU World Standings, and seeded draw, where competitors are divided into groups by their standings with higher-ranked groups starting later and random draws within the group. In the second segment, skaters start in the reverse order of the first-segment result, either exactly or by a random draw within groups that are formed on the first-segment result. 

\subsection{Data Summary}
\label{sec:data}

The dataset covers the competitions listed in Section~\ref{sec:scoring} and the World Team Trophy from 2016/17 to 2025/26\footnote{The dataset does not cover the Challenger Series or smaller international competitions.}, except for 2020 Cup of China and the three 2020/21 GP events which are run with domestic panels because of COVID-19. We also exclude the team event of the 2020 Winter Youth Olympic Games, because the teams were formed with mixed skaters from different federations. All data entries are parsed from the official PDF protocols downloaded from the ISU results website or from the websites of the host federation or organizers.\footnote{The judges' marks are missing for four segments in the 2017 Asian Winter Games. These segments without judges' marks are not included in score regressions.} The regression sample has 172 competitions, 23,254 performances by 2,470 skaters, pairs, and couples from 78 federations, and 206,983 performance-judge pairs from 620 judges of 59 federations.

Table~\ref{tab:t21} summarizes the judges' grades and marks and their synthetic segment scores by era. ISU changed the marking scale twice during our ten seasons sample period. From 2018/19 GOE grade ranges changed from $\pm 3$ to $\pm 5$ grades and the second-half jump bonus are restricted to the last single jump of the short program and the last three of the free skating. From 2022/23 the five program components were reduced to three. Both changes took effect in the season after an Olympic Winter Games. Therefore, we divide the ten seasons into three eras: Era A consists of the last two seasons of the PyeongChang cycle; Era B is the Beijing cycle; Era C is the Milan cycle.

\begin{table}[htbp]
  \centering
  \caption{Data Summary By Discipline and Era}\label{tab:t21}
  \begin{threeparttable}
  \makebox[\textwidth]{\begin{adjustbox}{max width=\textwidth}\begin{tabular}{@{}l c c c r r@{}}
\toprule
& Judges per & & Minimum & & \\
Score type & performance & Range & increment & Mean & $N$ \\
\midrule
\multicolumn{6}{@{}l}{\textit{Era A. 2016/17--2017/18 (PyeongChang cycle)} (GOE $-3$ to $+3$, five components; 5,277 performances)} \\[2pt]
\quad GOE grade (per element) & 6--9 & $-3$ to $+3$ & 1 & 0.30 & 402,705 \\
\quad Program component mark & 6--9 & 0.25 to 10 & 0.25 & 6.37 & 235,030 \\
\quad Segment score (one per performance) & 6--9 & 3 to 223 & 0.01 & 78.61 & 5,277 \\
\addlinespace
\multicolumn{6}{@{}l}{\textit{Era B. 2018/19--2021/22 (Beijing cycle)} (GOE $-5$ to $+5$, five components; 7,424 performances)} \\[2pt]
\quad GOE grade (per element) & 5--9 & $-5$ to $+5$ & 1 & 0.72 & 557,277 \\
\quad Program component mark & 5--9 & 0.25 to 10 & 0.25 & 6.61 & 329,940 \\
\quad Segment score (one per performance) & 5--9 & 5 to 225 & 0.01 & 82.32 & 7,424 \\
\addlinespace
\multicolumn{6}{@{}l}{\textit{Era C. 2022/23--2025/26 (Milan cycle)} (GOE $-5$ to $+5$, three components; 10,553 performances)} \\[2pt]
\quad GOE grade (per element) & 5--9 & $-5$ to $+5$ & 1 & 0.47 & 793,515 \\
\quad Program component mark & 5--9 & 0.25 to 10 & 0.25 & 6.47 & 281,967 \\
\quad Segment score (one per performance) & 5--9 & 10 to 238 & 0.01 & 81.35 & 10,553 \\
\bottomrule
\end{tabular}
\end{adjustbox}}
    \begin{tablenotes}[flushleft]
      \footnotesize
      \item \textit{NOTE:} The data includes a total of 172 competitions from 2016/17 to 2025/26: the World Championships, the European Championships, the Four Continents Championships, the World Junior Championships; the Olympic Winter Games, the Winter Youth Olympic Games; the World Team Trophy, the Asian Winter Games, the Grand Prix, the Junior Grand Prix, and (Junior) Grand Prix Final. 
    \end{tablenotes}
  \end{threeparttable}
\end{table}

Table~\ref{tab:t22} shows that scores differ widely across disciplines and segments. The average performance score is 128.45 for Men FS and 50.96 for Women SP. To unify scores across disciplines, segments, and seasons, we express each judge's marks in percent of a reference value. Let $j$ index a judge and $p$ a performance. For the GOE, we define
\begin{equation*}
  \mathrm{goe\_pct}_{jp} \;=\; 100 \cdot
  \frac{\sum_{e} V_e(g_{jpe}) \;-\; \sum_{e} V_e(-G)}
       {\sum_{e} V_e(+G) \;-\; \sum_{e} V_e(-G)},
\end{equation*}
which places the GOE points judge $j$ awards on the range from the lowest to the highest she could award, given the BV set by the technical panel. In the formula, $g_{jpe}$ is the grade judge $j$ awarded to element $e$, $V_e(\cdot)$ converts a grade on that element into points under the rules of the season, and $\pm G$ are the highest/lowest grades the season allows. We measure the GOE rather than the TES because the latter is dominated by the BV, which makes up over 90\% of the TES on average and, being set by the technical panel, is the same for every judge on the panel. For the PCS, we define
\begin{equation*}
  \mathrm{pcs\_pct}_{jp} \;=\; 100 \cdot
  \frac{\sum_{k} c_{jpk}}{10 \cdot n_{\mathrm{comp}}},
\end{equation*}
which is the share of the full component score that judge $j$ awarded. Here $c_{jpk}$ is judge $j$'s mark on component $k$, and $n_{\mathrm{comp}}$ the number of components.

\begin{table}[htbp]
  \centering
  \caption{Data Summary By Discipline}\label{tab:t22}
  \begin{threeparttable}
  \makebox[\textwidth]{\begin{adjustbox}{max width=\textwidth}\begin{tabular}{@{}l rrr rrrr r@{}}
\toprule
& \multicolumn{3}{c}{Segment score (one per performance)} & \multicolumn{5}{c}{Synthetic judge-specific segment score} \\
\cmidrule(lr){2-4} \cmidrule(l){5-9}
& & & Standard & & & \multicolumn{2}{c}{Standard deviation} & \\
\cmidrule(l){7-8}
Discipline and segment & $N$ & Mean & deviation & $N$ & Mean & Overall & Within & Correlation \\
\midrule
Men SP & 3,603 & 66.58 & 16.93 & 32,252 & 66.56 & 17.10 & 2.85 & 0.987 \\
Men FS & 3,186 & 128.45 & 33.69 & 28,568 & 128.36 & 33.95 & 4.76 & 0.991 \\
Women SP & 4,297 & 50.96 & 13.75 & 38,475 & 50.98 & 13.87 & 2.30 & 0.988 \\
Women FS & 3,842 & 96.95 & 28.73 & 34,383 & 96.95 & 28.90 & 3.82 & 0.992 \\
Pairs SP & 1,669 & 57.78 & 11.57 & 14,941 & 57.74 & 11.78 & 2.50 & 0.980 \\
Pairs FS & 1,570 & 106.18 & 22.95 & 14,071 & 106.08 & 23.22 & 3.97 & 0.987 \\
Ice dance SD/RD & 2,690 & 59.81 & 13.67 & 23,458 & 59.95 & 13.89 & 2.80 & 0.982 \\
Ice dance FD & 2,397 & 91.72 & 20.75 & 20,835 & 92.04 & 20.96 & 3.65 & 0.986 \\
\midrule
All & 23,254 & 81.04 & 34.31 & 206,983 & 81.11 & 34.47 & 3.40 & 0.996 \\
\bottomrule
\end{tabular}
\end{adjustbox}}
    \begin{tablenotes}[flushleft]
      \footnotesize
  \item \textit{NOTE:} SP is the short program, the first segment of men, women, and pairs. FS is the free skate, the second segment of men, women, and pairs. SD and RD are the short dance and the rhythm dance, the first segment of ice dance through 2017/18 and from 2018/19. FD is the free dance, the second segment of ice dance. Within standard deviation is the standard deviation of scores across the judges of the same performance.
    \end{tablenotes}
  \end{threeparttable}
\end{table}

For the TSS, we use the judge's synthetic total, which is not an official score but the score a performance would have received had every judge on the panel given that judge's marks, measured against the scale of comparable performances:
\begin{equation*}
  \mathrm{tss\_pct}_{jp} \;=\; 100 \cdot \frac{s_{jp}}{M_{c(p)}},
\end{equation*}
where $s_{jp}$ is judge $j$'s synthetic segment score for performance $p$, and $M_{c(p)}$ is the highest score achieved in the performance's season $\times$ discipline $\times$ segment cell\footnote{The COVID-impacted 2020/21 season is pooled with 2019/20.}, by any skater other than the cell's top scorer so that one exceptional skater does not set the unit (e.g., Ilia Malinin's personal best in Men's free skate in 2025/26 is 11.9\% above the personal best of all other skaters). 

\begin{figure}[htbp]
  \centering
  \begin{subfigure}{\linewidth}
    \centering
    \includegraphics[width=0.9\linewidth,height=0.36\textheight,keepaspectratio]{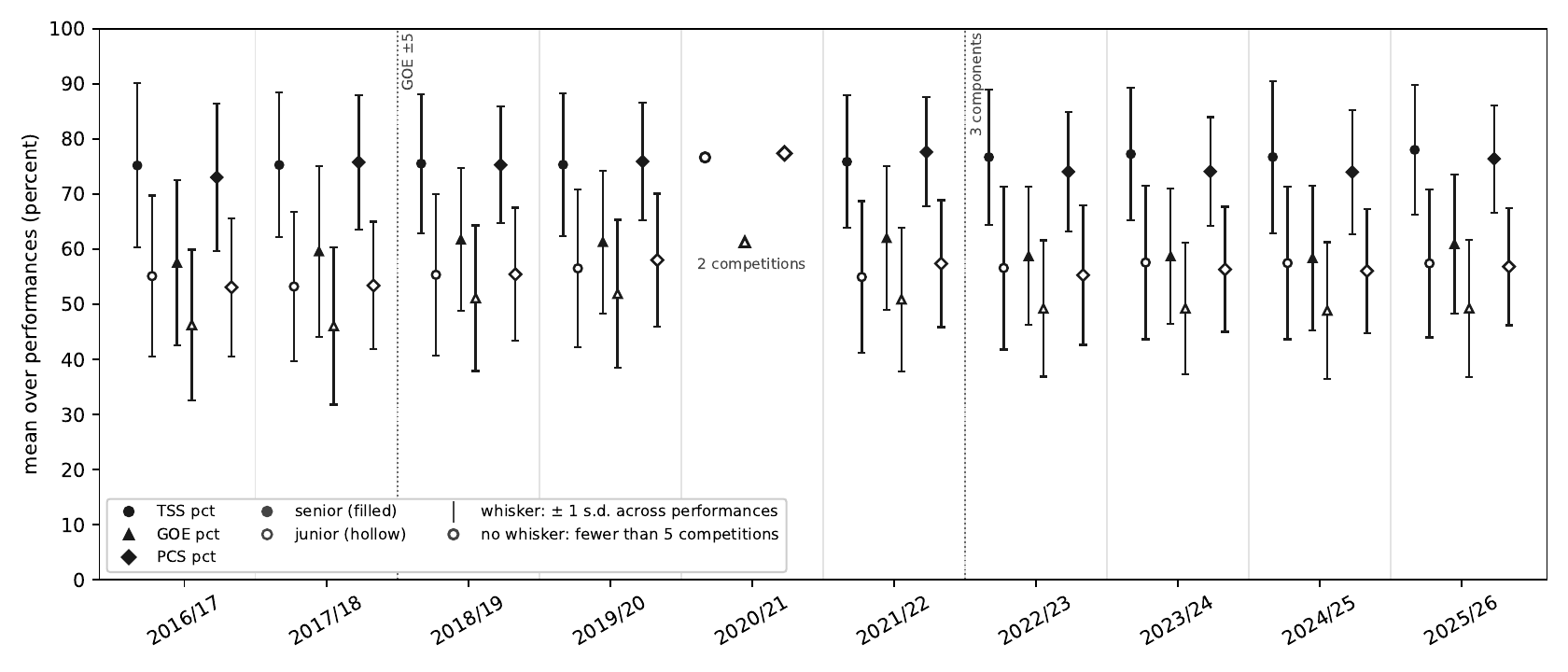}
    \caption{By season}
    \label{fig:f21a}
  \end{subfigure}

  \vspace{0.5em}

  \begin{subfigure}{\linewidth}
    \centering
    \includegraphics[width=0.9\linewidth,height=0.36\textheight,keepaspectratio]{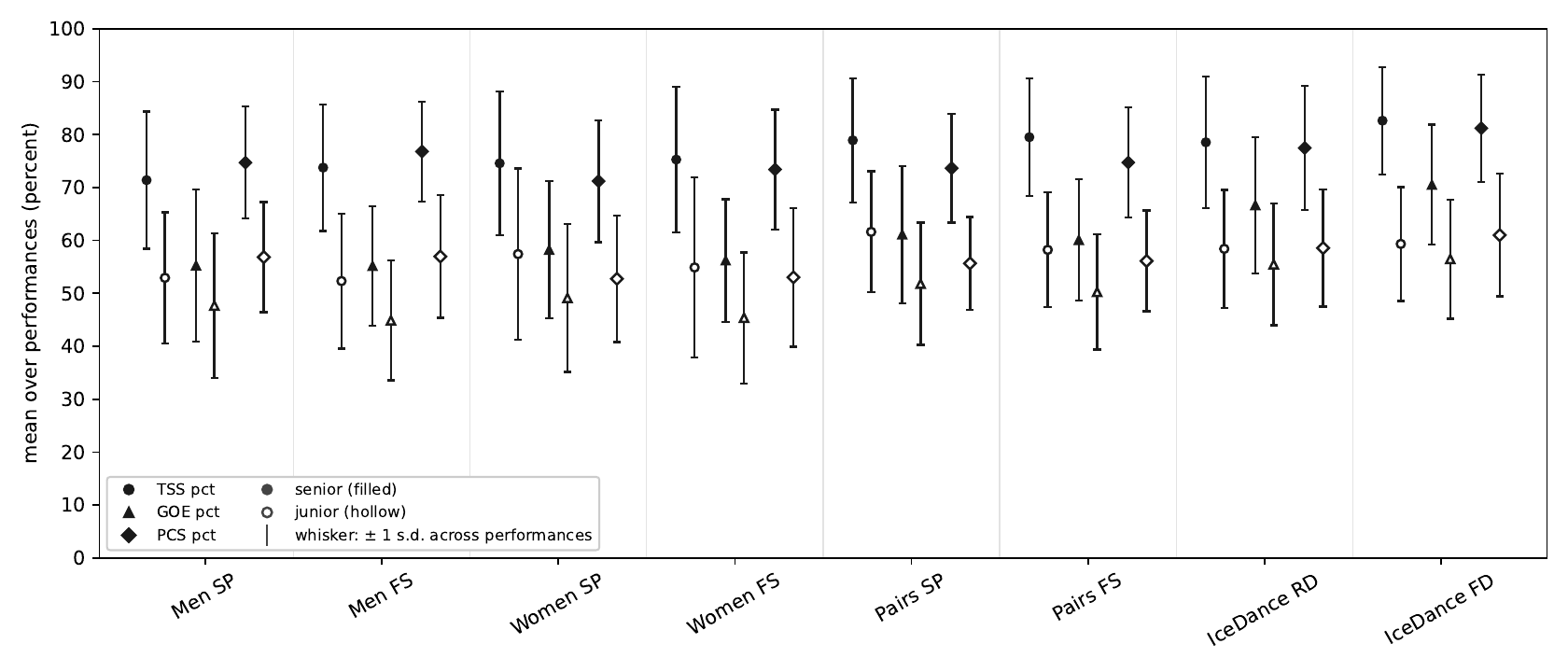}
    \caption{By discipline $\times$ program}
    \label{fig:f21b}
  \end{subfigure}

  \caption{Summary Of the Three Synthetic Score Percent Outcomes}
  \label{fig:f21}

  \begin{minipage}{0.95\linewidth}
    \footnotesize
    \textit{NOTE:} Panel (a): by season. The 2020/21 season has only two competitions and therefore drawn hollow. The dotted verticals mark the change of the GOE scale to $\pm$5 (2018/19) and of five program components to three (2022/23). Panel (b): by discipline $\times$ program. Ice Dance RD is the ice-dance short segment, the short dance through 2017/18 and the rhythm dance after.
  \end{minipage}
\end{figure}

Figure~\ref{fig:f21} shows that, after the normalization, the percent outcomes are comparable in scale across seasons, disciplines and segments. We do not remove the difference between the junior and senior tiers, which we view as a reflection of performance quality gap between the two age groups. For every performance, each judge on the panel has a GOE, a PCS, and a synthetic TSS pct outcome. The use of synthetic score outcomes allow us to control for performance fixed effects and explore judges' strategic behavior from within-performance score variations. 

\subsection{Existing Models and Their Identification Challenges}
\label{sec:challenge}

Let $i$ denote skaters (single skaters, pairs, and ice dance couples, all called skaters below), $j$ judges, $p$ performances, and $\mathrm{fed}(i)$ (resp.\ $\mathrm{fed}(j)$) the federation of skater $i$ (resp.\ judge $j$). Let $g(p)$ denote the segment of performance $p$ and $P(p)$ the panel of judges scoring the segment $g(p)$. Ideally, judge $j$'s mark on performance $p$ should only reflect her perception of the performance quality, and the perception is determined by the true quality of $p$ and unintended factors such as $j$'s judgment error and her overall leniency. In reality, many other factors come into play. Previous studies (\citealp{zitzewitz2006nationalism,sandberg2018competing,heiniger2021judging,krumer2022nationalistic}, among many others) have documented evidence of compatriot bias, vote trading, and home advantage using the following regression models and their variations:
\begin{align}
    Y_{ijp}&= \beta_1 \mathrm{C}_{ij} + \lambda_p+ \alpha_j+ \varepsilon_{ijp} \tag{1} \label{eq:baseline}\\
    Y_{ijp}&= \beta_1 \mathrm{C}_{ij} + \beta_H \mathrm{Home}_{p}+ \beta_O \mathrm{OP}_{ijp}+ \gamma_i+ \alpha_j+ \tau_{g(p)}+ \varepsilon_{ijp} \tag{1'} \label{eq:baseline-OP}.
\end{align}
In these equations, $\mathrm{C}_{ij}=\mathbf{1}\{\mathrm{fed}(j)=\mathrm{fed}(i)\}$ indicates whether skater $i$ and judge $j$ are from the same federation, $\mathrm{OP}_{ijp}= \mathbf{1}\{\exists\, j'\in P(p),\ j'\neq j:\ \mathrm{fed}(j')=\mathrm{fed}(i)\}$ indicates whether a judge other than $j$ on the panel is from skater $i$'s federation\footnote{No two judges from the same federations sit on the same panel. In other words, $\mathrm{C}_{ij}$ and $\mathrm{OP}_{ijp}$ cannot both be one for any $(i,j,p)$. }, $\mathrm{Home}_{p}$ indicates whether the skater is from the host country of the competition, $\lambda_p$ is the performance fixed effect (e.g., performance quality), $\alpha_j$ the judge fixed effect (e.g., leniency), $\gamma_i$ the skater fixed effect (e.g., ability or popularity that does not change over time), $\tau_{g(p)}$ the segment (competition $\times$ discipline $\times$ segment) fixed effect, and $\varepsilon_{ijp}$ the regression error. 

Model~\eqref{eq:baseline} is more robust than model~\eqref{eq:baseline-OP} because of the performance fixed effects. Even so, the compatriot premium $\beta_1$ estimated from~\eqref{eq:baseline} cannot reveal judge's intent. The premium can arise from a taste that judges share with their compatriot skaters, which \citet{zitzewitz2006nationalism} counts as compatriot unconscious bias. It can also arise from familiarity since judges typically know their compatriot skaters' ability more precisely, having watched them at domestic competitions. As \citet{aigner1977statistical} discuss, risk-averse evaluators may discount candidates whose quality they know less precisely, and therefore rate the candidates they have ties with higher on average even when their true quality is the same. In this paper, we argue that favoritism aimed at improving compatriot skaters' standings produces scoring strategies that vary with the information judges have when they mark, while shared taste or familiarity cannot produce these scoring patterns.

Model~\eqref{eq:baseline-OP} and its variants are often used in the literature to study the colleague-compatriot premium and home advantage. The equation replaces the performance fixed effect, i.e., the skater $\times$ segment fixed effect, with separate skater and segment fixed effects, because $\lambda_p$ would absorb the home ice indicator, and the other-panel compatriot indicator $\mathrm{OP}_{ijp}$ is perfectly collinear with $\mathrm{C}_{ij}$ and $\lambda_p$. As a result, the coefficients in~\eqref{eq:baseline-OP} may pick up the effects of uncontrolled performance-specific factors. For example, $\beta_\mathrm{O}$ is biased upward if federations choose where to send their judges when they can. As a skater develops, she may build a reputation among the judges and work with more capable choreographers on her programs. Her federation may also earn, through her results, the right to enter a judge at the major championships and the Olympics, and expecting her to place well, be more willing to send a judge to competitions where it has a say. The skater, segment, and judge fixed effects in~\eqref{eq:baseline-OP} cannot fully absorb such latent, time-varying ability, which in turn biases the estimates. To address this problem, we collect additional controls for performance quality from the skaters' season biography pages on the ISU website. We find that the colleague-compatriot premium shrinks substantially when these novel controls are included.

For inference, we cluster by competition and by athlete, using the two-way cluster-robust variance estimator of \citet{cameron2011robust}. Appendix~\ref{app:fedcluster} reports two robustness checks for the inference choice: clustering by competition and athlete federation\footnote{Competition and athlete federation clustering is not used as the default because in our estimation sample, the ten largest of the 78 federations account for 56 percent of the observations, while 52 federations have under 1 percent each. The effective number of federations, the inverse Herfindahl index of their shares, is 23 for all observations and 15 for the compatriot observations.} instead, and using the jackknife variation of two-way competition and athlete clustering following \citet{mackinnon2026jackknife}. Under both checks, the TSS estimates behind our main findings remain significant at the 5 percent level, except for the technical judge premium, which moves to the 10 percent level with the jackknife method.

Last but not least, we note that models~\eqref{eq:baseline} and~\eqref{eq:baseline-OP} impose a common premium, $\beta_{1}$, $\beta_{\mathrm{H}}$ and $\beta_{\mathrm{O}}$, for all observations. If the premium instead differs across observations, the OLS estimate is a weighted sum of the heterogeneous premia (\citealp{dechaisemartin2020two}) $\beta_{1,ijp}$, $\beta_{\mathrm{H},ijp}$ and  $\beta_{\mathrm{O},ijp}$. Then, by the Frisch-Waugh-Lovell theorem, $\hat\beta_1$, for instance, is an unbiased estimate of $\sum w_{ijp}\beta_{1,ijp}$ over the compatriot rows, with $w_{ijp}=\widetilde{\mathrm{C}}_{ijp}/\sum\widetilde{\mathrm{C}}_{ijp}$ where $\widetilde{\mathrm{C}}_{ijp}$ is the residual of regressing $\mathrm{C}_{ij}$ on other regressors including the performance and judge fixed effects. The weights can be negative when the predicted probability of $\mathrm{C}_{ij}=1$ exceeds one. Therefore, in all our analysis, we report both the share of observations with negative weights and the range of negative weight mass (i.e., $\sum_{w_{ijp}<0}\lvert w_{ijp}\rvert$). We find that the non-convex average problem arises with~\eqref{eq:baseline-OP} and other regression models that exploit cross-panel variations in the panel composition for identification. In Section~\ref{subsec:cross-panel}, we show how choosing a subsample of competitions with randomly drawn judge panels helps with the non-convex average problem in our empirical setting.

\subsection{Descriptive Analysis}
\label{sec:descriptive}

\begin{figure}[htbp]
  \centering
  \includegraphics[width=0.9\linewidth,height=0.72\textheight,keepaspectratio]{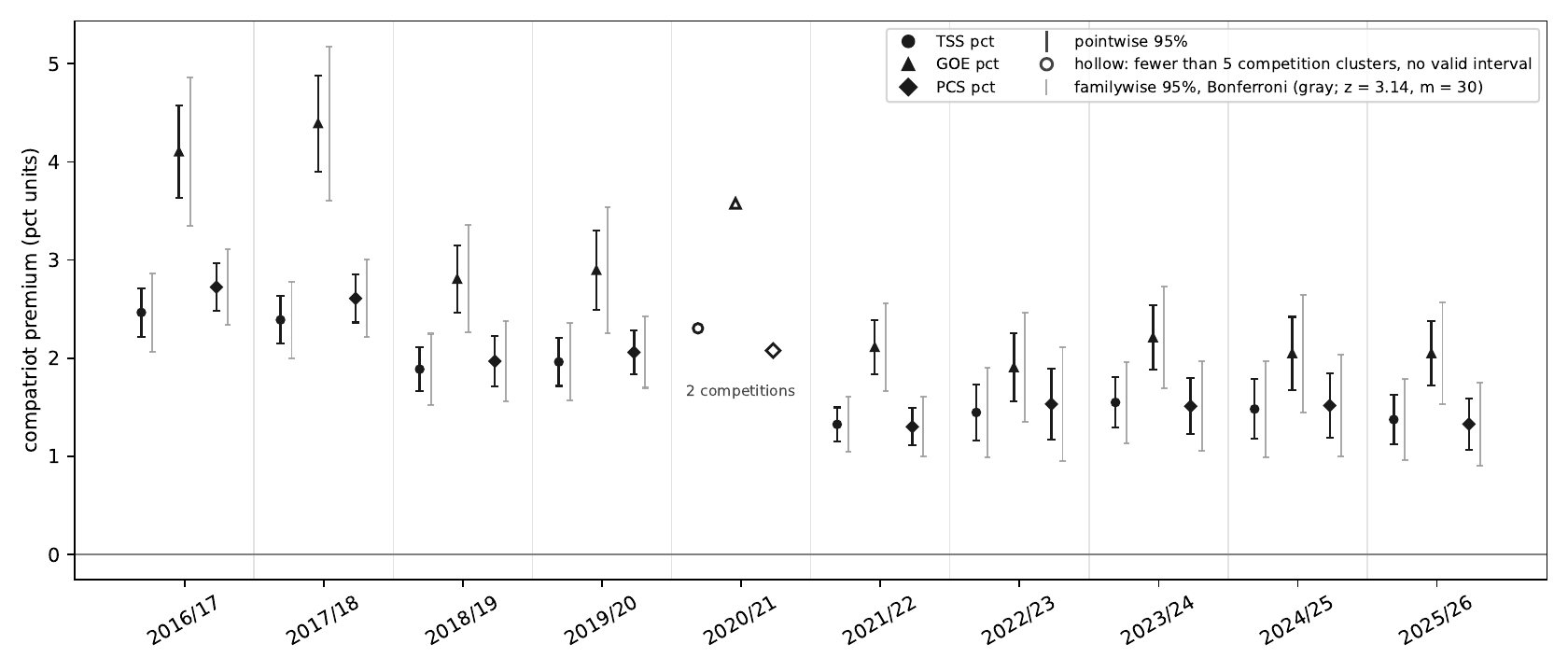}
  \caption{Compatriot Premium By Season}\label{fig:f22}
  \begin{minipage}{0.95\linewidth}
    \footnotesize
    \textit{NOTE:} Each estimate with its pointwise 95\% interval and the familywise 95\% interval. In our sample, the 2020/21 season has only two competitions and therefore no interval.
  \end{minipage}
\end{figure}

Before starting the main analysis, we use model~\eqref{eq:baseline} and its heterogeneity analysis extensions to describe the data. Figure~\ref{fig:f22} estimates the model with a separate compatriot coefficient for each season (Table~\ref{tab:tac1} in Appendix~\ref{app:hetero} reports the corresponding estimates). The premium is positive in every season and on every outcome, and statistically significant whenever a valid standard error can be computed.\footnote{For the pandemic season 2020/21, the subsample size is too small to compute valid two-way clustered standard error.} The seasons fall roughly into three periods: 2016/17 to 2017/18, 2018/19 to 2020/21, and 2021/22 to 2025/26. The premium is steady within each period and steps down in between: the TSS premium averages 2.43 percent in the first period, 1.92 in the second (COVID-impacted 2020/21 left out) and 1.44 in the third. The first drop follows the PyeongChang Olympics and coincides with several events that happened before the 2018/19 season: the ISU suspended two Chinese judges for national bias at the PyeongChang Olympics Games; the ISU Congress required the Officials Assessment Commission (OAC), which reviews every judge's marks after each ISU event, to report suspected national bias at every competition; and several rule changes  including the GOE scale change. The second drop came in the 2021/22 season. This coincides with the publication of an ISU Communication (2369) that added more details to the OAC reviews, requesting the commission to compare each judge's total score for a skater with totals for the two skaters placed immediately above and the two placed immediately below, and report all cases of possible nationalistic bias. Around the same time, the ISU suspended a judge for one year for national bias at the World Championships 2021. 

\begin{figure}[htbp]
  \centering
\includegraphics[width=0.9\linewidth,height=0.72\textheight,keepaspectratio]{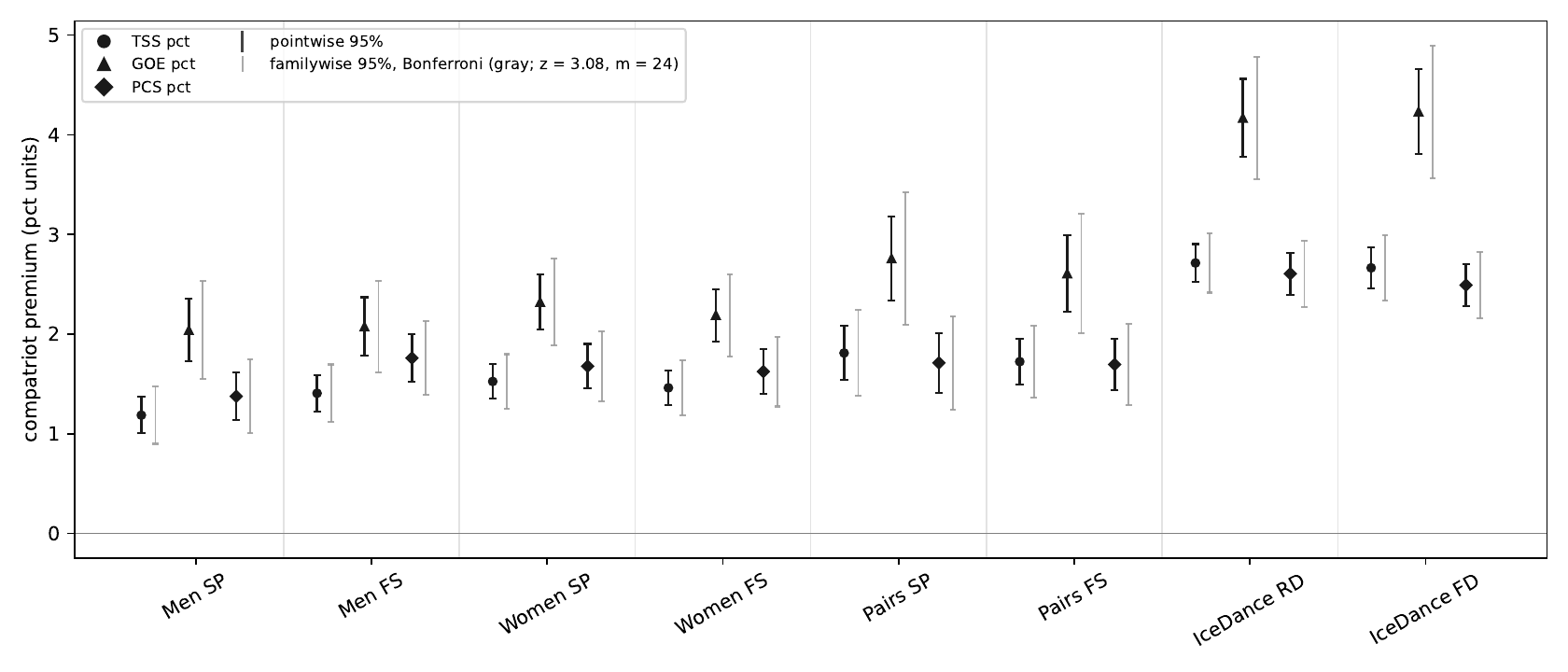}
  \caption{Compatriot Premium By Discipline $\times$ Segment}\label{fig:f23}
  \begin{minipage}{0.95\linewidth}
    \footnotesize
    \textit{NOTE:} Each estimate with its pointwise 95\% interval and the familywise 95\% interval.
  \end{minipage}
\end{figure}

Figure~\ref{fig:f23} plots the compatriot premium by discipline and segment, estimated with model~\eqref{eq:baseline} separately for each discipline $\times$ segment cell (Table~\ref{tab:tac2} in Appendix~\ref{app:hetero} reports the estimates). The discipline differences are large. Ice dance carries a TSS premium about twice the size of men's and women's, and the point estimates follow the ordering of men $<$ women $<$ pairs $<$ ice dance within each program on both TSS and GOE. On PCS, the three disciplines with jumps lie between 1.4-1.8 percent, while ice dance is above 2.5.

In Table~\ref{tab:tac3} and Table~\ref{tab:tac4} of the appendix, we also estimate the compatriot premium in each program component score, and the premium in GOE scores grouped by different types of elements in the singles competitions. Among the components, the premium is smallest on skating skills and largest on composition in every discipline and under both component systems, except for pairs in 2016/17–2021/22, where the premium is slightly larger on interpretation. Among the elements, step and choreographic sequences carry the largest premium in both programs of both singles disciplines rather than jumps and spins. The premium thus seems to be largest where the mark relies most on the judge's own subjective reading.

\section{Benchmark Models and Estimation Results}
\label{sec:scorebias}

\subsection{Compatriot Premium and Targeted Competitor Penalties}
\label{subsec:benchmark}

In this section, we estimate the compatriot premium together with the competitor penalty, a judge's tendency to mark down the rivals of her compatriot skaters. As discussed in the introduction, the evidence on a competitor penalty is mixed in the literature, with competitors defined by final placement results. We define competitors by the start order of a segment to avoid the two-way causality problem. The cost is that the start order identifies a compatriot's competitors with noise and biases the estimated penalty toward zero. Even so, we estimate the competitor penalty precisely and discover that the penalty is targeted. In the discussions below, we use competitor and rival interchangeably.

Let $S(p)$ denote the set of skaters performing in segment $g(p)$. Let $S_R(p)=\{sk(p') \in S(p): p'\neq p, |sn(p')-sn(p)|\leq 3\}$, where $sk(.)$ and $sn(.)$ denote the skater and the start number of a performance, respectively. A skater in set $S_R(p)$ is considered a close rival of $p$'s skater, as she comes from a different federation and starts within \textit{three} numbers of performance $p$. Consider the outcome equation:
 \begin{align}\label{eq:rival-perf}
   Y_{ijp} &= \beta_1 \mathrm{C}_{ij} + \beta_2 \mathrm{R}_{ijp} + \lambda_p + \alpha_j + \varepsilon_{ijp}. \tag{2}
 \end{align}
where $\mathrm{R}_{ijp}=\mathbf{1}\{\mathrm{C}_{ij}=0\}\cdot \mathbf{1}\{\exists i' \in S_R(p):\ \mathrm{fed}(i')=\mathrm{fed}(j)\}$ indicates if judge $j$ comes from the same federation as any of skater $i$'s close rivals. In the Appendix~\ref{app:altdef}, we use alternative definitions of $S_R(p)$ for robustness checks.

Columns (1)-(3) of Table~\ref{tab:t31} report estimates of model~\eqref{eq:rival-perf}, for each of the three percent-of-scale outcomes described in Section~\ref{sec:data}. On average, a judge marks a skater from her own federation 1.73 percent of the scale higher on TSS than the other judges on the same panel, 2.64 and 1.82 percent higher for GOE and PCS respectively. As discussed in Section~\ref{sec:challenge}, $\beta_1$ alone cannot separate nationalistic bias from other causes of the compatriot premium. We return to this question in Section~\ref{subsec:uncertainty}. Besides the compatriot premium, the table documents negative and statistically significant competitor penalties on all three outcomes: $-0.124$ on TSS, $-0.261$ on GOE, and $-0.080$ on PCS. 

Columns (4)-(6) of Table~\ref{tab:t31} repeat the analysis for TSS on three subsamples that differ in how the start order is determined. Column (4) uses the first segments whose start order is set by an open draw. Column (5) uses the other first segments, whose start order is set by a seeded draw. Column (6) uses all second segments, where the start order follows the result of the first segment therefore tracks who actually competes with whom. As is expected, the competitor penalty is statistically significant at the 5\% level only for the Segment 2 subsample. The slope coefficient is $-0.204$, about one eighth of the compatriot premium. 

Compared with the compatriot premium, the competitor penalty coefficient may seem small. However, the compatriot premium comes from a single judge, since a federation has at most one judge on the panel of a segment, whereas the competitor penalty comes from every rival judge on the panel. In our sample, a performance has on average 2.4 rival judges, and 47 percent of performances have three or more. Summing over the panel, the competitor penalty can add up to 36-48\% of the compatriot premium for a skater who faces three or four rival judges in Segment 2. And it only offers a lower bound, in absolute value, of the true competitor bias due to the attenuation bias.

Note that the rival penalty in Column (4) of Table~\ref{tab:t31} is marginally significant at the 10\% significance level under an open draw. We checked whether some competitions had been misclassified as open draws and found no sign of it. We therefore read this marginally significant estimate as either a chance rejection at the 10\% level or a small penalty that judges give to skaters who happen to perform close to their compatriots. Table~\ref{tab:t31} also reports the share of negative weights and the mass of negative weights for each regression estimates and finds no problems. The need to report such statistics is discussed in Section~\ref{sec:challenge}, following \citet{dechaisemartin2020two}. 

\begin{table}[htbp]
  \centering
  \caption{The Baseline Compatriot Premium and the Competitor Penalty}\label{tab:t31}
  \begin{threeparttable}
  \makebox[\textwidth]{\begin{adjustbox}{max width=\textwidth}\begin{tabular}{lrrrrrr}
\toprule
 & \multicolumn{3}{c}{\textbf{All rows}} & \multicolumn{3}{c}{\textbf{TSS pct, by how the start order is set}} \\
\cmidrule(lr){2-4}\cmidrule(lr){5-7}
\textbf{Variable} & \textbf{(1) TSS pct} & \textbf{(2) GOE pct} & \textbf{(3) PCS pct} & \textbf{(4) Segment 1, open draw} & \textbf{(5) Segment 1, seeded order} & \textbf{(6) Segment 2} \\
\midrule
Compatriot & +1.727*** & +2.644*** & +1.816*** & +1.611*** & +1.842*** & +1.700*** \\
 & (0.060) & (0.106) & (0.067) & (0.096) & (0.081) & (0.062) \\
Rival & -0.124*** & -0.261*** & -0.080*** & -0.086* & -0.032 & -0.204*** \\
 & (0.023) & (0.037) & (0.030) & (0.044) & (0.033) & (0.027) \\
\midrule
Performance FE & $\checkmark$ & $\checkmark$ & $\checkmark$ & $\checkmark$ & $\checkmark$ & $\checkmark$ \\
Judge FE & $\checkmark$ & $\checkmark$ & $\checkmark$ & $\checkmark$ & $\checkmark$ & $\checkmark$ \\
N & 206,983 & 206,983 & 206,983 & 46,208 & 62,918 & 97,857 \\
\# of Competitions (G) & 172 & 172 & 172 & 70 & 102 & 172 \\
Mass of negative weights & \multicolumn{3}{c}{0.000} & 0.000 & 0.000 & 0.000 \\
Share of negative weights & \multicolumn{3}{c}{0\%} & 0\% & 0\% & 0\% \\
\bottomrule
\end{tabular}
\end{adjustbox}}
    \begin{tablenotes}[flushleft]
      \footnotesize
      \item \textit{NOTE:} All ten seasons, 2016/17-2025/26. Segment 1, open draw: the first segments whose start order is an open lot (the JGP, the 2016/17 GP events and Final, and the 2025 Asian Winter Games). Segment 1, seeded order: the first segments whose order (or group order) follows the reverse of ISU World Standing. Segment 2: the free segments, whose order (or group order) follows the reverse of first segment's result. Two-way (competition + athlete) cluster-robust standard errors in parentheses, $t$ with $G-1$ degrees of freedom, $G$ the column's competitions; *** $p<0.01$, ** $p<0.05$, * $p<0.10$. Of the two alternative standard errors in Appendix~\ref{app:fedcluster}, clustering by competition and athlete federation changes no significance level, and the two-way cluster jackknife changes only that of Rival in column (3), from the 1 to the 5 percent level.
    \end{tablenotes}
  \end{threeparttable}
\end{table}

One may argue that, just like the compatriot premium, the competitor penalty may come from unintended factors. We therefore repeat the analysis of model~\ref{eq:rival-perf}, with the rivals split by whether they skate before or after the judges' compatriot and for those who skate after her, by whether the compatriot is still in contention. If the penalty came from a distaste for skating styles of other federations, or from judges marking neighboring performances harder, there would be no reason for the competitor penalty to vary with the start side or the compatriot’s standing after her performance.

Let $S^{+}_R(p)$ (resp.\ $S^{-}_R(p)$) denote the set of skaters starting within three numbers after (resp.\ before) performance $p$. We consider the following model of targeted competitor bias:
\begin{align}\label{eq:rival2-perf}
   Y_{ijp} =  \beta_1 \mathrm{C}_{ij} & + \beta_2^+ \mathrm{R}^+_{ijp}  + \beta_2^{-,in} \mathrm{R}^{-,in}_{ijp} + \beta_2^{-,out} \mathrm{R}^{-,out}_{ijp}  + \lambda_p + \alpha_j + \varepsilon_{ijp}, \tag{3}
\end{align}
where $R^{+}_{ijp}=\mathbf{1}\{C_{ij}=0\}\cdot\mathbf{1}\{\exists\, s \in S^+_R(p):\ \mathrm{fed}(s)=\mathrm{fed}(j)\}$ indicates if a skater of judge $j$'s federation starts within three numbers after performance $p$, $R^{-}_{ijp}=\mathbf{1}\{C_{ij}=0\}\cdot\mathbf{1}\{\exists\, s \in S^-_R(p):\ \mathrm{fed}(s)=\mathrm{fed}(j)\}$ indicates if a skater of judge $j$'s federation starts within three numbers before performance $p$, $R^{-,in}_{ijp}= R^{-}_{ijp}\cdot \mathbf{1}\{\exists\, s \in S^-_R(p):\ \mathrm{fed}(s)=\mathrm{fed}(j), \mathrm{In}_p(s)=1\}$ indicates whether such an earlier compatriot $s$ is in contention when performance $p$ starts, with $\mathrm{In}_p(s)=1$ if skater $s$ is still in contention when performance $p$ starts, that is, if her segment score ranks among the top three of the performances skated so far and, in a second segment, she also placed in the top eight after the first segment, and $R^{-,out}_{ijp}= R^{-}_{ijp}-R^{-,in}_{ijp}$ indicates that judge $j$'s earlier compatriots within three start numbers of performance $p$ are all already out of contention. 

\begin{table}[htbp]
  \centering
  \caption{How Rival Penalty Changes with Start Order and Compatriot's Realized Performance Quality}\label{tab:t31b}
  \begin{threeparttable}
  \makebox[\textwidth]{\begin{adjustbox}{max width=\textwidth}\begin{tabular}{lrrrrrr}
\toprule
 & \multicolumn{3}{c}{\textbf{All rows}} & \multicolumn{3}{c}{\textbf{TSS pct, by how the start order is set}} \\
\cmidrule(lr){2-4}\cmidrule(lr){5-7}
\textbf{Variable} & \textbf{(1) TSS pct} & \textbf{(2) GOE pct} & \textbf{(3) PCS pct} & \textbf{(4) Segment 1, open draw} & \textbf{(5) Segment 1, seeded order} & \textbf{(6) Segment 2} \\
\midrule
Compatriot & +1.726*** & +2.647*** & +1.814*** & +1.614*** & +1.847*** & +1.692*** \\
 & (0.060) & (0.106) & (0.067) & (0.096) & (0.081) & (0.062) \\
Rival starting earlier & -0.158*** & -0.260*** & -0.142*** & -0.082* & -0.047 & -0.262*** \\
 & (0.022) & (0.037) & (0.031) & (0.047) & (0.037) & (0.029) \\
Rival starting later, in contention & -0.101*** & -0.253*** & -0.032 & -0.093 & +0.023 & -0.264*** \\
 & (0.035) & (0.056) & (0.046) & (0.069) & (0.051) & (0.048) \\
Rival starting later, out of contention & -0.004 & -0.095** & +0.033 & -0.021 & -0.021 & +0.006 \\
 & (0.028) & (0.045) & (0.039) & (0.065) & (0.050) & (0.033) \\
\midrule
Performance FE & $\checkmark$ & $\checkmark$ & $\checkmark$ & $\checkmark$ & $\checkmark$ & $\checkmark$ \\
Judge FE & $\checkmark$ & $\checkmark$ & $\checkmark$ & $\checkmark$ & $\checkmark$ & $\checkmark$ \\
N & 206,983 & 206,983 & 206,983 & 46,208 & 62,918 & 97,857 \\
\# of Competitions (G) & 172 & 172 & 172 & 70 & 102 & 172 \\
Mass of negative weights & \multicolumn{3}{c}{0.000} & 0.000 & 0.000 & 0.000 \\
Share of negative weights & \multicolumn{3}{c}{0\%} & 0\% & 0\% & 0\% \\
\bottomrule
\end{tabular}
\end{adjustbox}}
    \begin{tablenotes}[flushleft]
      \footnotesize
      \item \textit{NOTE:} Columns (1)-(3) are on the whole sample. Columns (4)-(6) are on three subsamples partitioned by the start order mechanism; same definitions as Table~\ref{tab:t31}. Two-way (competition + athlete) cluster-robust standard errors in parentheses, $t$ with $G-1$ degrees of freedom, $G$ the column's competitions; *** $p<0.01$, ** $p<0.05$, * $p<0.10$. Of the two alternative standard errors in Appendix~\ref{app:fedcluster}, clustering by competition and athlete federation moves Rival starting later, in contention in column (1) from the 1 to the 5 percent level, Rival starting later, out of contention in column (2) to not significant, and Rival starting earlier in column (4) from the 10 to the 5 percent level. The two-way cluster jackknife changes no significance level.
    \end{tablenotes}
  \end{threeparttable}
\end{table}

Table~\ref{tab:t31b} reports the estimates for all three score outcomes on all observations and then for TSS only on the three start-order subsamples. Columns (1)-(3) show that judges mark down their compatriots' close rivals, unless the marked skater performs after the judge's compatriot, and the compatriot is already out of contention. The exception is a small GOE penalty on the rivals who start after a compatriot who is already out of contention ($-0.095$), and no significant PCS penalty on the rivals who start after a compatriot who is still in contention.\footnote{Under two-way clustering by competition and by the athlete's federation (note of Table~\ref{tab:t31b}), the small GOE penalty on $\mathrm{R}^{-,out}$ is no longer significant.} For rivals who start before the compatriot skater, although the judge does not know yet how her compatriot will perform, every point taken from the marked skater lowers the bar that her compatriot has to clear, which can help boost her confidence and ease her nervousness. When the compatriot has already skated and is out of contention, the penalty on TSS ($\beta_2^{-,out}$) is both small and statistically insignificant. Similar to findings in Table~\ref{tab:t31}, columns (4) and (5) of Table~\ref{tab:t31b} do not have any coefficients reporting statistical significance at the 5\% level. 

\subsection{Colleague-Compatriot Premium, Home Ice Advantage, and Technical Judge Premium} 
\label{subsec:cross-panel}

In this section, we study whether judges give higher marks to the compatriot skaters of their colleagues on the same panel. We call this the colleague-compatriot premium rather than vote trading (\citealp{zitzewitz2006nationalism}) because a positive premium can also have other behavioral explanations such as courtesy toward a colleague, since every judge's marks in our sample period are published together with their names. This section also studies the home ice advantage and the compatriot bias of the technical panel. 

For reasons discussed in Section~\ref{sec:challenge}, we replace the performance fixed effects in~\eqref{eq:rival-perf} with skater and segment fixed effects. Further, we include a vector of performance quality controls to reduce the potential endogeneity caused by dropping the performance fixed effects. Specifically, we consider the following cross-panel design:
\begin{align}
Y_{i j p} =& \beta_1 \mathrm{C}_{ij} +\beta_2 \mathrm{R}_{ijp} +\beta_{O} \mathrm{OP}_{ijp}+\beta_H \mathrm{Home}_{p}+\beta_{HC}\mathrm{C}_{ij} \times (\mathrm{Home}_{p}-\bar{\mathrm{Home}}_p)  \notag \\
& + \beta_C \mathrm{T^C}_{ip}+\beta_S \mathrm{T^S}_{ip} + \tilde Z_p'\gamma_1+ \gamma_i+ \alpha_j+ \tau_{g(p)}+ \varepsilon_{ijp} \tag{2'} \label{eq:crosspanel-rival-OP}
\end{align}
where $T^{C}_{ip}$ (resp.\ $T^{S}_{ip}$) indicates whether the skater and the technical controller (resp.\ one of two technical specialists) come from the same federation, and $\tilde Z_p=(Z_p', \mathrm{Start}_p)'$ where $Z_p$ is a vector of performance quality controls constructed using information from skater's bio page posted every season on the ISU website. The vector includes: prior placements at the Junior World and World Championships, whether first season in junior/senior events, and indicators of choreographers and of chosen music types. 

\begin{table}[htbp]
  \centering
  \caption{Colleague-Compatriot Premium, Home Ice Advantage, and Technical Judge Premium}\label{tab:t32}
  \begin{threeparttable}
  \makebox[\textwidth]{\begin{adjustbox}{max width=\textwidth}\begin{tabular}{lrrrrrrrrr}
\toprule
 & \multicolumn{3}{c}{\textbf{Without performance quality controls}} & \multicolumn{3}{c}{\textbf{With performance quality controls}} & \multicolumn{3}{c}{\textbf{Championships and Olympics}} \\
\cmidrule(lr){2-4}\cmidrule(lr){5-7}\cmidrule(lr){8-10}
\textbf{Variable} & \textbf{(1) TSS pct} & \textbf{(2) GOE pct} & \textbf{(3) PCS pct} & \textbf{(4) TSS pct} & \textbf{(5) GOE pct} & \textbf{(6) PCS pct} & \textbf{(7) TSS pct} & \textbf{(8) GOE pct} & \textbf{(9) PCS pct} \\
\midrule
Compatriot & +2.012*** & +2.776*** & +2.279*** & +1.860*** & +2.709*** & +2.082*** & +2.207*** & +3.260*** & +2.245*** \\
 & (0.140) & (0.203) & (0.105) & (0.130) & (0.197) & (0.094) & (0.236) & (0.400) & (0.181) \\
Rival & -0.084*** & -0.229*** & +0.033 & -0.124*** & -0.247*** & -0.024 & -0.094 & -0.281*** & -0.023 \\
 & (0.031) & (0.046) & (0.034) & (0.030) & (0.046) & (0.032) & (0.060) & (0.074) & (0.054) \\
Other-panel compatriot & +0.270** & +0.122 & +0.420*** & +0.135 & +0.063 & +0.246*** & +0.271 & +0.381 & +0.247* \\
 & (0.126) & (0.175) & (0.084) & (0.118) & (0.171) & (0.072) & (0.208) & (0.317) & (0.121) \\
Other-segment compatriot &  &  &  &  &  &  & +0.362 & +0.601 & +0.272 \\
 &  &  &  &  &  &  & (0.299) & (0.401) & (0.164) \\
Home ice & +0.634*** & +1.063*** & +0.473*** & +0.740*** & +1.140*** & +0.582*** & +1.201*** & +1.925*** & +0.888*** \\
 & (0.185) & (0.240) & (0.144) & (0.171) & (0.237) & (0.125) & (0.344) & (0.585) & (0.212) \\
Compatriot $\times$ home ice & -0.428*** & -0.539*** & -0.563*** & -0.412*** & -0.532*** & -0.534*** & +0.084 & +0.182 & +0.298 \\
 & (0.111) & (0.175) & (0.129) & (0.110) & (0.176) & (0.129) & (0.349) & (0.655) & (0.315) \\
Compatriot Technical Controller & +0.400** & +0.197 & -0.058 & +0.370** & +0.180 & -0.113 & +0.856** & +0.681 & +0.306 \\
 & (0.185) & (0.273) & (0.155) & (0.160) & (0.260) & (0.121) & (0.351) & (0.514) & (0.185) \\
Compatriot Technical Specialist & +0.283 & +0.041 & -0.020 & +0.292 & +0.035 & -0.001 & +0.250 & -0.196 & +0.060 \\
 & (0.196) & (0.230) & (0.133) & (0.180) & (0.221) & (0.116) & (0.310) & (0.451) & (0.200) \\
\midrule
$\tilde{Z}_{p}$ &  &  &  & $\checkmark$ & $\checkmark$ & $\checkmark$ & $\checkmark$ & $\checkmark$ & $\checkmark$ \\
Skater FE & $\checkmark$ & $\checkmark$ & $\checkmark$ & $\checkmark$ & $\checkmark$ & $\checkmark$ & $\checkmark$ & $\checkmark$ & $\checkmark$ \\
Segment FE & $\checkmark$ & $\checkmark$ & $\checkmark$ & $\checkmark$ & $\checkmark$ & $\checkmark$ & $\checkmark$ & $\checkmark$ & $\checkmark$ \\
Judge FE & $\checkmark$ & $\checkmark$ & $\checkmark$ & $\checkmark$ & $\checkmark$ & $\checkmark$ & $\checkmark$ & $\checkmark$ & $\checkmark$ \\
N & 206,983 & 206,983 & 206,983 & 206,983 & 206,983 & 206,983 & 55,276 & 55,276 & 55,276 \\
\# of Competitions (G) & 172 & 172 & 172 & 172 & 172 & 172 & 30 & 30 & 30 \\
Mass of negative weights, Other-panel/segment compatriot & \multicolumn{3}{c}{0.047} & \multicolumn{3}{c}{0.050} & \multicolumn{3}{c}{0.000--0.006} \\
Mass of negative weights, all other coefficients & \multicolumn{3}{c}{0.000--0.015} & \multicolumn{3}{c}{0.000--0.015} & \multicolumn{3}{c}{0.000--0.008} \\
Share of negative weights, Other-panel/segment compatriot & \multicolumn{3}{c}{18\%} & \multicolumn{3}{c}{18\%} & \multicolumn{3}{c}{1\%--4\%} \\
Share of negative weights, all other coefficients & \multicolumn{3}{c}{0\%--5\%} & \multicolumn{3}{c}{0\%--5\%} & \multicolumn{3}{c}{0\%--10\%} \\
\bottomrule
\end{tabular}
\end{adjustbox}}
    \begin{tablenotes}[flushleft]
      \footnotesize
      \item \textit{NOTE:} Columns (1)-(6) All rows, 2016/17-2025/26. Columns (7)-(9), World (Junior) Championships, European Championships, and Olympic Games. Two-way (competition + athlete) cluster-robust standard errors in parentheses, $t$ with $G-1$ degrees of freedom, $G$ the column's competitions; *** $p<0.01$, ** $p<0.05$, * $p<0.10$. Of the two alternative standard errors in Appendix~\ref{app:fedcluster}, clustering by competition and athlete federation moves Rival from the 1 to the 10 percent level in column (1) and makes it significant at the 10 percent level in column (7), and moves Other-panel compatriot from the 5 to the 10 percent level in column (1) and from the 10 to the 5 percent level in column (9). The two-way cluster jackknife moves Rival in column (1) from the 1 to the 5 percent level, and Compatriot Technical Controller in columns (1), (4) and (7) from the 5 to the 10 percent level.
    \end{tablenotes}
  \end{threeparttable}
\end{table}

Table~\ref{tab:t32} reports the main slope coefficients of model~\eqref{eq:crosspanel-rival-OP}. Comparing Columns (1)-(3) with Columns (4)-(6) confirms the importance of using performance quality controls in this cross-panel model. With them, the evidence for colleague-compatriot premium becomes much weaker: the TSS estimate of $\beta_O$ falls by about half and is no longer significant. Only the PCS estimate remains significant, at about 60 percent of its size once the controls are added. Skaters competing at home also receive higher marks, which may be influenced by crowd support from the audience, or the skater's familiarity with the venue. Interestingly, the table also reports a negative and statistically significant slope coefficient of the compatriot and home ice interaction term, implying that the guest judges award more home ice advantage than home federation's own judge. At the Championships and Olympics, where only 4.9 percent of the compatriot rows are on home ice, $\beta_{HC}$ is small and statistically insignificant.

The regressions also find compatriot bias among technical judges. A compatriot technical controller raises the TSS by 0.37 percent of the scale with the controls and by 0.86 at the Championships and Olympics, which is about a fifth and two fifths of a compatriot judge's premium. The technical specialist's coefficient is not significant in any column, and neither indicator is significant for the GOE or the PCS, which is consistent with the technical judges working mainly through setting the base values of the elements. Although the $\beta_C$ and $\beta_S$ estimates look small in the table, their effects on the official score are sizable. The base values the technical panel sets enter every judge's score, unlike the compatriot premium and the competitor penalty, which enter the official score through the panel's trimmed mean.

Notice that regressions in columns (1)-(6) report that 18 percent of the observations used to identify $\beta_{\mathrm{O}}$ enter its estimate with a negative weight, and the mass of negative weights is around 5 percent. As is discussed in Section~\ref{sec:challenge}, these negative weights arise because the coefficient is identified from changes in panel composition across competitions, and in some competitions almost every performance by a skater from a large federation has a compatriot judge on the panel. In columns (7)–(9), we repeat the analysis on a subsample of Championships and Olympic Games. At these competitions, federations are randomly drawn to form a pool of judges for each discipline, and the judges who sit each segment are drawn from that pool, which mitigates the problem of negative weights. As we see from the table, the share of negative weights behind $\beta_{\mathrm{O}}$ falls from 18 to 4 percent (the technical judges' shares rise to 10 percent, but their negative weights sum to less than 0.01), and the mass falls from 5 to 0.6 percent. With the new subsample, empirical findings from the full sample still holds. The other-panel compatriot premium is only significant for the PCS. Columns (7)–(9) also include an other-segment compatriot indicator, which is from the random panel draw rules of these championship and Olympics events, as judges not sitting on the current panel will still sit on the other segment's panel. Its coefficients are of the same size as the other-panel compatriot's but not significant due to lack of statistical precision. 

Table~\ref{tab:tac5-1} and~\ref{tab:tac5-2} in the appendix report the slope estimates of the performance controls used in columns (4)-(9) of Table~\ref{tab:t32}. Because the regressions include skater fixed effects, these coefficients compare a skater with herself. A skater placed at an earlier World Championships scores 2.8 percent of the scale more on the TSS, and one who has placed at an earlier World Junior Championships scores 1.3 percent more. In the first season in junior events, a skater scores 1.8 percent less, while the first season in senior events makes no significant difference. Some choreographer indicators are also producing large and significant coefficients, although they need not to be causal, as a skater may work with a better-known choreographer in seasons when she is in a better form.

\section{Strategic Nationalistic Scoring}
\label{sec:strategic}

\subsection{Start Order and the Nationalistic Bias}
\label{subsec:uncertainty}

Section~\ref{subsec:benchmark} showed that judges target their competitor penalties in a way that is hard to explain by unintentional factors such as taste, cultural differences, or simply judges being harsher on skaters who perform close to their compatriots. In this section, we ask the same question for compatriot premium: do judges target it too? Does a judge adjust the premium to how well her compatriot has skated, or to the compatriot's start position? If so, can this be explained by unintentional factors, or is it instead in line with nationalistic incentives? 

Extending model~\eqref{eq:rival2-perf}, Table~\ref{tab:t41} reports estimates of the following regression model:
\begin{align}\label{eq:start-open}
       Y_{ijp} =& \beta_1 \mathrm{C}_{ij} + \beta_2^+ \mathrm{R}^+_{ijp}  + \beta_2^{-,in} \mathrm{R}^{-,in}_{ijp} + \beta_2^{-,out} \mathrm{R}^{-,out}_{ijp}  + \delta_1 \mathrm{C}_{ij}\times (\mathrm{Start}_p-\mathrm{\bar{Start}}) \nonumber\\
       &+ \delta_2^+ \mathrm{R}^+_{ijp}\times (\mathrm{CStart}_{pj}-\mathrm{\bar{CStart}}^+)+\delta_2^{-,in} \mathrm{R}^{-,in}_{ijp}\times (\mathrm{CStart}_{pj}-\mathrm{\bar{CStart}}^{-,in})\nonumber \\
       &+\delta_2^{-,out}\mathrm{R}^{-,out}_{ijp}\times (\mathrm{CStart}_{pj}-\mathrm{\bar{CStart}}^{-,out}) +  \mathrm{C}_{ij}\times H_p' \gamma_1 + \mathrm{R}^+_{ijp}\times H_p' \gamma_2^+ \nonumber \\
       &  + \mathrm{R}_{ijp}^{-,in}\times H_p' \gamma_2^{-,in}+ \mathrm{R}_{ijp}^{-,out}\times H_p' \gamma_2^{-,out} + \lambda_p+\alpha_j + \varepsilon_{ijp}, \tag{4}
\end{align}
where $\mathrm{Start}_p$ is the start position of performance $p$ within its segment expressed as a fraction of the field, $\mathrm{CStart}_{pj}$ is the average start position, in the same units, of judge $j$'s compatriots who are the close rivals of performance $p$, $\mathrm{\bar{Start}}$ and $\mathrm{\bar{CStart}}^+$ (resp.\ $\mathrm{\bar{CStart}}^{-,in}/\mathrm{\bar{CStart}}^{-,out}$) are their means over the compatriot rows and over the early-starting rival rows (resp.\ the late-starting rival rows with the compatriot in/out of contention) used for normalization, and $H_p=(X_p', G_j',Z_p')'$, with $X_p$ the vector of discipline $\times$ segment $\times$ junior/senior $\times$ top 20 skater federations $\times$ season dummies, $G_j$ the top 20 judge federation dummies, and $Z_p$ the performance quality controls defined in Section~\ref{subsec:cross-panel}. The reason for controlling for $H_p$ is to avoid the interaction slopes $\delta_1$, $\delta_2^{+}$, $\delta_2^{-,in}$, and $\delta_2^{-,out}$ picking up judge marking variations from discipline, segment, tier, skater/judge federation or performance quality. All $H_p$ controls are demeaned as well.

\begin{table}[htbp]
  \centering
  \caption{Start Order, the Size of the Compatriot Premium and the Competitor Penalty}\label{tab:t41}
  \begin{threeparttable}
  \makebox[\textwidth]{\begin{adjustbox}{max width=\textwidth}\begin{tabular}{lrrrrrr}
\toprule
 & \multicolumn{3}{c}{\textbf{All rows}} & \multicolumn{3}{c}{\textbf{TSS pct, by how the start order is set}} \\
\cmidrule(lr){2-4}\cmidrule(lr){5-7}
\textbf{Variable} & \textbf{(1) TSS pct} & \textbf{(2) GOE pct} & \textbf{(3) PCS pct} & \textbf{(4) Segment 1, open draw} & \textbf{(5) Segment 1, seeded order} & \textbf{(6) Segment 2} \\
\midrule
Compatriot & +1.741*** & +2.665*** & +1.823*** & +1.641*** & +1.863*** & +1.710*** \\
 & (0.062) & (0.110) & (0.068) & (0.099) & (0.083) & (0.065) \\
Rival starting earlier & -0.133*** & -0.234*** & -0.112*** & -0.044 & -0.027 & -0.232*** \\
 & (0.023) & (0.038) & (0.032) & (0.046) & (0.038) & (0.029) \\
Rival starting later, in contention & -0.088** & -0.236*** & -0.033 & -0.047 & +0.022 & -0.246*** \\
 & (0.034) & (0.057) & (0.045) & (0.069) & (0.051) & (0.049) \\
Rival starting later, out of contention & -0.005 & -0.098** & +0.035 & -0.018 & -0.026 & +0.006 \\
 & (0.030) & (0.048) & (0.041) & (0.069) & (0.053) & (0.034) \\
Compatriot $\times$ start position & -0.711*** & -0.454*** & -1.162*** & -0.697*** & -0.536*** & -0.702*** \\
 & (0.094) & (0.167) & (0.121) & (0.192) & (0.185) & (0.129) \\
Rival starting earlier $\times$ compatriots' start position & -0.346*** & -0.525*** & -0.379*** & -0.176 & -0.238 & -0.468*** \\
 & (0.094) & (0.146) & (0.127) & (0.149) & (0.168) & (0.130) \\
Rival starting later, in contention $\times$ compatriots' start position & -0.553*** & -0.494** & -0.757*** & -0.434 & -0.419** & -0.734*** \\
 & (0.127) & (0.219) & (0.157) & (0.269) & (0.205) & (0.214) \\
Rival starting later, out of contention $\times$ compatriots' start position & -0.215* & -0.148 & -0.324* & +0.176 & -0.642* & -0.358** \\
 & (0.125) & (0.203) & (0.176) & (0.240) & (0.340) & (0.154) \\
\midrule
$\mathrm{C}_{jp}$, $\mathrm{R}^{+}_{jp}$, $\mathrm{R}^{-,in}_{jp}$ and $\mathrm{R}^{-,out}_{jp}$ $\cdot$ $H_{p}$ & $\checkmark$ & $\checkmark$ & $\checkmark$ & $\checkmark$ & $\checkmark$ & $\checkmark$ \\
Performance FE & $\checkmark$ & $\checkmark$ & $\checkmark$ & $\checkmark$ & $\checkmark$ & $\checkmark$ \\
Judge FE & $\checkmark$ & $\checkmark$ & $\checkmark$ & $\checkmark$ & $\checkmark$ & $\checkmark$ \\
N & 206,983 & 206,983 & 206,983 & 46,208 & 62,918 & 97,857 \\
\# of Competitions (G) & 172 & 172 & 172 & 70 & 102 & 172 \\
Mass of negative weights, main regressors & \multicolumn{3}{c}{0.000} & 0.000 & 0.000 & 0.000 \\
Mass of negative weights, Compatriot $\times$ start position & \multicolumn{3}{c}{0.036} & 0.010 & 0.063 & 0.045 \\
Mass of negative weights, the three Rival $\times$ compatriots' start position & \multicolumn{3}{c}{0.033--0.052} & 0.012--0.034 & 0.050--0.066 & 0.032--0.053 \\
Share of negative weights, main regressors & \multicolumn{3}{c}{0\%} & 0\% & 0\% & 0\% \\
Share of negative weights, Compatriot $\times$ start position & \multicolumn{3}{c}{18\%} & 10\% & 21\% & 21\% \\
Share of negative weights, the three Rival $\times$ compatriots' start position & \multicolumn{3}{c}{20\%--22\%} & 12\%--18\% & 22\%--24\% & 19\%--22\% \\
\bottomrule
\end{tabular}
\end{adjustbox}}
    \begin{tablenotes}[flushleft]
      \footnotesize
      \item \textit{NOTE:} $H_p=(X_p', G_j', Z_p')'$, where $X_{p}$ is the vector of discipline $\times$ segment defined by start order type (segment 1 open draw, segment 1 seed order, segment 2) $\times$ junior/senior $\times$ top 20 federations by the \# of performances (the other pooled) $\times$ season (with 2020/21 merged with 2019/20) dummies, $G_j$ is the vector of dummies for the judge’s federation among the same top 20 federations (the other pooled), and $Z_p$ is the same vector of performance quality controls used in Table~\ref{tab:t32}. The compatriots' start position is the average start position of the judge's compatriots within three start numbers of the marked skater. Each interaction is centered at its mean over the rows of its indicator. Two-way (competition + athlete) cluster-robust standard errors in parentheses, $t$ with $G-1$ degrees of freedom, $G$ the column's competitions; *** $p<0.01$, ** $p<0.05$, * $p<0.10$. Appendix~\ref{app:fedcluster} reports alternative standard errors clustered by competition and skater federation and with the two-way cluster jackknife.
    \end{tablenotes}
  \end{threeparttable}
\end{table}

Table~\ref{tab:t41} reports the regression results. First, we look at column (4), which uses the subsample of open-draw first segments. There, $\mathrm{Start}_p$ is random and the rivalry definition carries no information about competitiveness. It is not surprising that none of the three rival terms nor their interactions is statistically significant. The compatriot premium, however, adjusts with the random start order. The slope estimate of $\delta_1$ is negative and statistically significant. The compatriot premium is 1.64 on average, but the difference between the first starter and the last starter is as large as 0.70.

If the compatriot premium is pure taste, we would not expect it to depend on a randomly assigned start position. If, instead, the premium is at least partly aimed at improving the compatriot's placement, it may change with the start position, since the earlier the compatriot skates, the less the judge knows about the rest of the field, and the more uncertainty she faces in determining how large a premium would improve her compatriot's standing. If the compatriot skates late, most of that uncertainty has been resolved, and the judge can decide with high precision whether it is still worthwhile to mark her compatriot higher, and if the answer is yes, how much she should inflate the compatriot's score. Column (4) therefore bounds the role of taste. If the taste and unfamiliarity discount do not depend on the start position, their part of the compatriot premium is at most the premium of the last starter. Reading from column (4), the compatriot premium is 1.29 percent of the scale on TSS for the last starter, 1.99 percent of the scale for the first starter, and 1.64 on average. At least about a fifth of the average compatriot premium, therefore, needs another explanation.

The other columns of the table show that $\delta_1$ is stable regardless of how start order is determined. The interaction term of competitor penalty and compatriot start position, however, behaves differently, and in an informative way. In the open draw, where the start order says nothing about the standings, no rival slope is significant. In the second segments, where the order is the reverse of the first-segment result and the compatriot's start position hence correlates with the chance of a good placement, each rival term is marked down more the later the judge's compatriot starts. Moving the compatriot from the first to the last start position lowers her close rival's TSS by 0.47 percent of the scale if the rival skates before her, and by 0.73 percent of the scale if the rival skates after her while she is still in contention, both significant at 1 percent level. The seeded first segments, whose order follows the World Standings, show the same pattern but with smaller slopes. Tables~\ref{tab:tac6} and~\ref{tab:tac7} in the appendix repeat columns (4)-(6) of Table~\ref{tab:t41} under alternative definitions of being in contention and of a close rival. The empirical findings are qualitatively the same.

\begin{figure}[htbp]
  \centering
\includegraphics[width=0.9\linewidth,height=0.72\textheight,keepaspectratio]{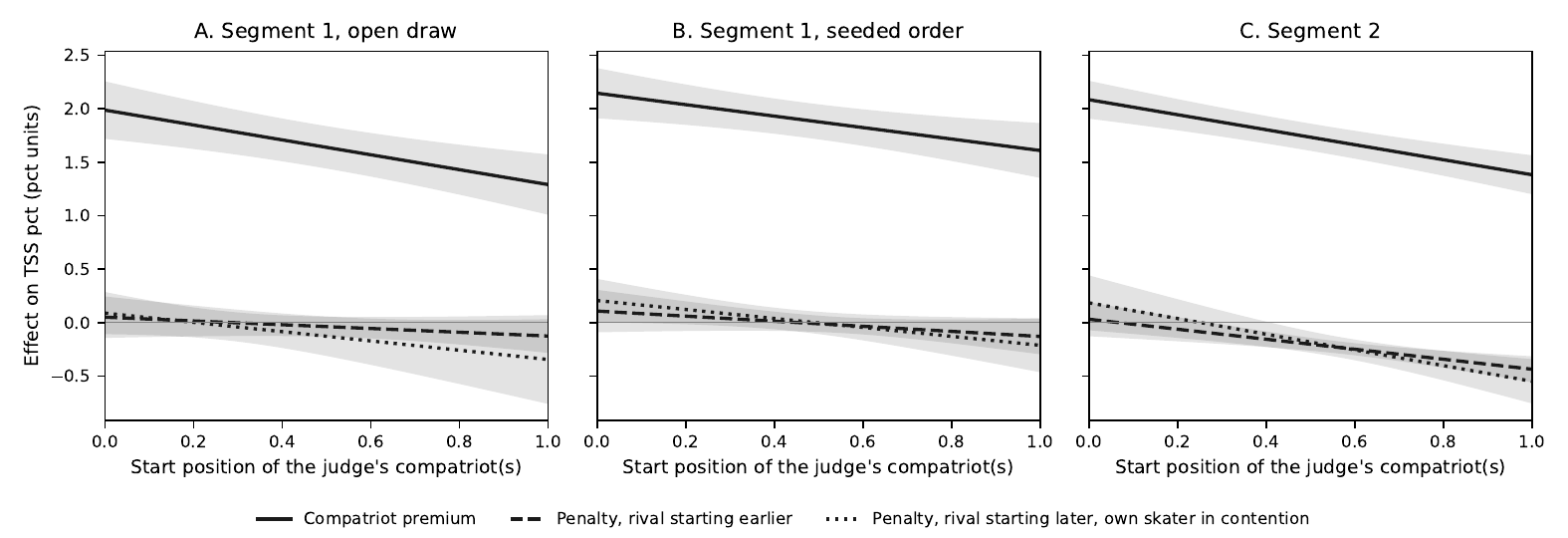}
  \caption{Compatriot Premium and Competitor Penalty Along the Compatriot's Start Position}\label{fig:f41}
  \begin{minipage}{0.95\linewidth}
    \footnotesize
    \textit{NOTE:} Lines are point estimates of the compatriot premium and the competitor penalty evaluated at different values of the start position. Bands are their associated pointwise 95\% confidence intervals.
  \end{minipage}
\end{figure}

Figure~\ref{fig:f41} plots how the compatriot premium and the competitor penalty on TSS respond to the start position of the judge's compatriot skater. The compatriot premium declines with the start position in much the same way across the three figures separately representing the open draw Segment 1 subsample, the seeded order Segment 1 subsample, and the Segment 2 subsample. In the two first segment subsamples, the competitor penalties are small with confidence bands including zero at almost every start position. For Segment 2, however, both the penalty on a rival who starts before the compatriot and the penalty on a rival who starts after a compatriot still in contention grow, in absolute value, with the compatriot's start position. The penalty on a rival who starts after a compatriot already out of contention is close to zero and is not drawn. 

\begin{figure}[htbp]
  \centering
\includegraphics[width=0.9\linewidth,height=0.72\textheight,keepaspectratio]{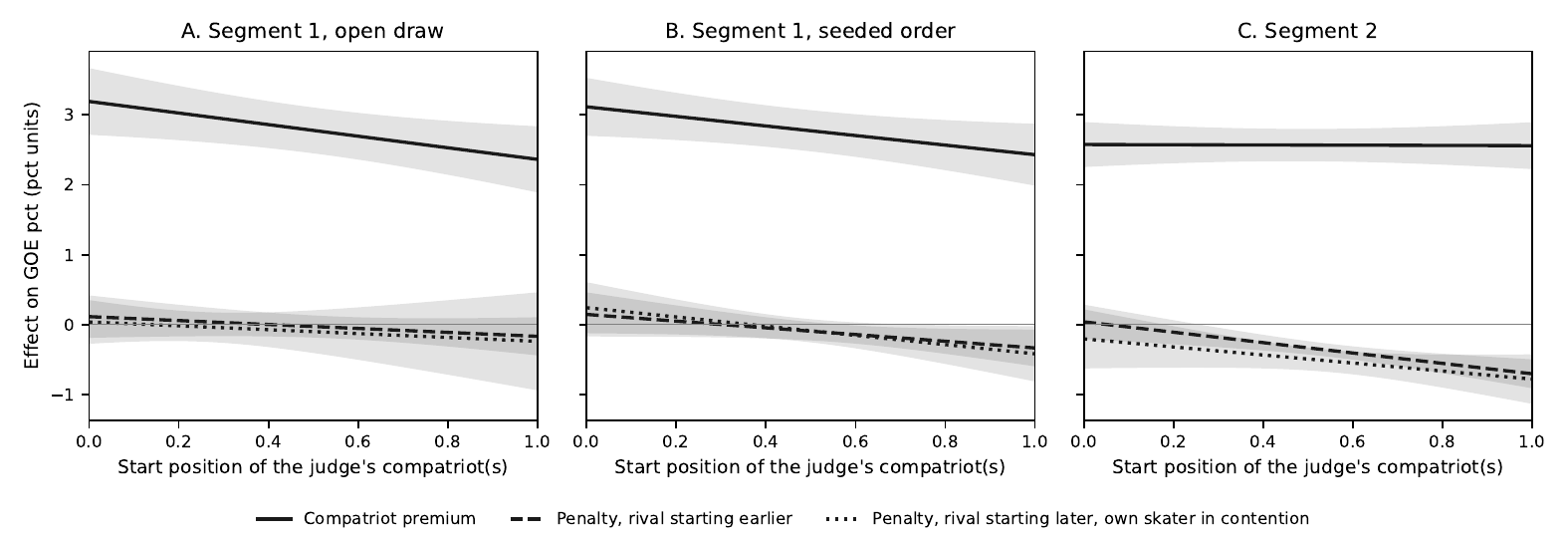}
\includegraphics[width=0.9\linewidth,height=0.72\textheight,keepaspectratio]{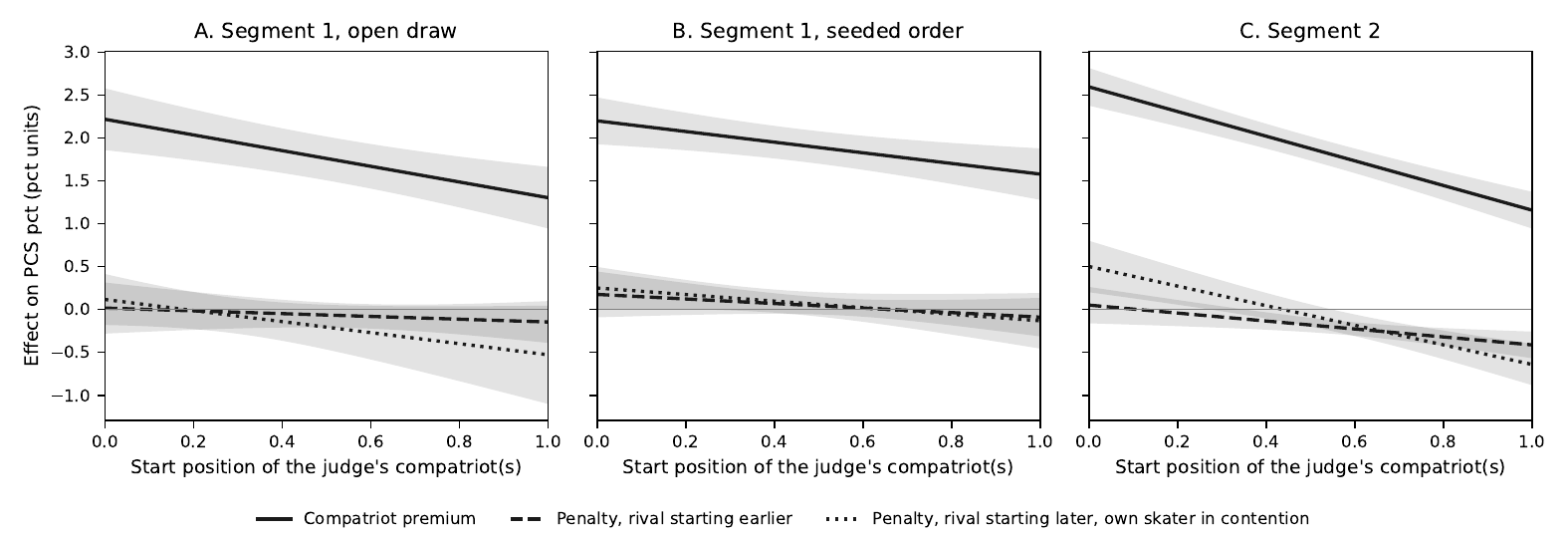}
  \caption{Compatriot Premium and Competitor Penalty On GOE and PCS}\label{fig:f42}
  \begin{minipage}{0.95\linewidth}
    \footnotesize
    \textit{NOTE:} Lines are point estimates of the compatriot premium and the competitor penalty evaluated at different values of the start position. Bands are their associated pointwise 95\% confidence intervals.
  \end{minipage}
\end{figure}

Figure~\ref{fig:f42} repeats the analysis of Figure~\ref{fig:f41} with the compatriot premium and the competitor penalties measured on GOE and PCS. In the first segments, the patterns we see in Figure~\ref{fig:f42} on TSS carries over to both components. The compatriot premium falls with the start position even when it is randomly drawn, while the competitor penalties are small with confidence bands including zero at every start position in the open draw. In the seeded order, the GOE penalties on both rival types grow with the start position, and their bands exclude zero when the compatriot starts in the last two fifths of the field (a rival starting earlier) or in its last quarter (a rival starting
later, while the compatriot is in contention). In Segment 2, the two components of TSS show different data patterns. On PCS, the compatriot premium falls steeply, from 2.60 to 1.16 percent of the scale, and the penalty on a rival who starts after a compatriot still in contention grows the most,
to 0.64 percent of the scale on average when the compatriot starts last. On GOE, the premium is flat with respect to the start position, while the rival penalties grow with the compatriot's start position, which in Segment 2 tracks her chance of having a good placement.

Why the GOE premium does not fall in Segment 2 is not obvious. One possibility is that it reflects performance quality that rises with the start order that our control vector $H_p$ fails to capture. If the GOE premium rises with the quality that the Segment 2 start order carries, that rise can offset the decline the premium would otherwise show along the start order. This explanation does not extend to the competitor penalties, however. On PCS, judges' scores vary less within the panel for the late starters (the standard deviation of a performance's PCS marks falls by 16 percent from the first to the last fifth of Segment 2, and by 15 percent among the judges with no stake in it), yet the penalty on a rival who starts after a compatriot still in contention grows the most there. 

The share and mass of negative weights for the compatriot indicator and the three rival indicators are calculated as discussed in Section~\ref{sec:challenge}. Those for the centered interaction terms are calculated following the non-binary case discussed in the appendix of \citet{dechaisemartin2020two}. Both measures are essentially zero for the indicators and larger for the interactions. Given the size and the precision of the slope estimates, we do not expect negative weight mass of at most 7\% to change the empirical findings of Table~\ref{tab:t32} qualitatively. 

We conclude this subsection by noting that, in the literature on group judged competitions, \citet{ginsburgh2003expert} find that the order of appearance influences the scores at a top international music competition, and even the musicians' later careers. Our paper, as far as we know, is the first that links start order with the compatriot premium and the competitor penalty. 

\subsection{Escalation in Response to Colleagues' Reputations}
\label{sec:escalation}

In this subsection, we bring together the strategic compatriot premium and targeted competitor penalty documented so far to study whether judges escalate their nationalistic bias in response to panel composition and colleagues' reputation. We ask, whether the marks a judge gives her compatriot and her compatriot's rivals depend on who else sits on the panel. Does the premium rise when more of her compatriot's close rivals have a compatriot judge on the panel? Does it rise further when those judges are known to be biased? Likewise, does a judge mark a rival down harder when the rival has a compatriot judge on the panel? Or does she spare the rival to avoid retaliation from the rival's compatriot judge? Escalation behavior can arise from judges' own preferences over the results of their compatriots, from federation incentives, or from both. 

To answer these questions, we use a two-step estimation procedure inspired by the teacher value-added literature (\citealp{chetty2014measuring,chetty2014measuringII,mulhern2023beyond}), which uses teacher fixed effect estimates from an earlier sample to study the impact of teacher quality on students' longer-term outcomes in a later sample. In the same spirit, we split our ten seasons of data into three eras, divided by the Olympic cycles. In Step 1, we use the data of an earlier era as the training  sample to estimate judge-specific compatriot premiums. In Step 2, we use these judge-specific estimates to approximate judges' reputation on judging fairness and ask whether the compatriot premium and/or competitor penalty a judge gives in a later era depend on how biased the compatriot judges of the rivals are known to be. 

The first step in the two-step estimation procedure uses the following model: \begin{align}\label{eq:step1}
  Y_{ijp} = \beta_{1j}\,\mathrm{C}_{ij} + \lambda_p + \alpha_j + \varepsilon_{ijp}, \tag{5}
\end{align}
with the estimates of $\beta_{1j}$ denoted by $\widehat{\beta}^{\,\mathcal{T}}_{1j}$. Recall that $S^{+}_R(p)$ (resp.\ $S^{-}_R(p)$) denotes the set of skaters starting within three positions after (resp.\ before) performance $p$. Let $\mathrm{W}^{+}_{p}=\frac{1}{|S^{+}_R(p)|}\sum_{s \in S^{+}_R(p)} \mathbf{1}\{\mathrm{fed}(s)\neq \mathrm{fed}(sk(p))\}\sum_{j'\in P(p)} \mathrm{C}_{sj'}\,\widehat{\beta}^{\,\mathcal{T}}_{1j'}$ be the average bias of skater $i$'s later rivals' compatriot judges, and $\mathrm{W}^{-,in}_{p}$ (resp.\ $\mathrm{W}^{-,out}_{p}$) be similarly defined average bias of the compatriot judges of skater $i$'s earlier rivals in (resp.\ out of) contention. Let $\mathrm{P}_p=\sum_{j'\in P(p)} \mathrm{C}_{sk(p)j'}\widehat{\beta}^{\,\mathcal{T}}_{1j'}$ be the trained premium of the compatriot judge of performance $p$. $\mathrm{P}_p=0$ if performance $p$ does not have a compatriot judge on the panel or the performance has a compatriot judge who is known to be unbiased. If a judge has no estimate from the training window, or is not measured there due to insufficient compatriot markings,\footnote{A judge has an estimate if she marks both a compatriot and a non-compatriot skater in the training window, and is measured if she marked compatriots in at least four competitions, marked four different compatriot skaters and thirty non-compatriot performances.} she is assigned the mean trained premium of the measured judges as her reputation. 

The second step in the two-step estimation procedure uses the following outcome equation:
\begin{align}\label{eq:step2} 
    Y_{ijp} = & \beta_1 \mathrm{C}_{ij} + \beta^{+}_2 \mathrm{R}^{+}_{ijp} + \beta^{-,in}_2 \mathrm{R}^{-,in}_{ijp}+ \beta^{-,out}_2 \mathrm{R}^{-,out}_{ijp} \notag \\
    &+ \beta_3^{-,in} \mathrm{C}_{ij}\times (\mathrm{W}^{-,in}_{p}-\bar{\mathrm{W}}^{-,in})+ \beta_3^{-,out} \mathrm{C}_{ij}\times (\mathrm{W}^{-,out}_{p}-\bar{\mathrm{W}}^{-,out}) + \beta_3^{+}\, \mathrm{C}_{ij}\times (\mathrm{W}^{+}_{p}-\bar{\mathrm{W}}^{+}) \notag \\
    &+\beta_4^{+} \mathrm{R}^{+}_{ijp}\times (\mathrm{P}_p - \bar{\mathrm{P}}^+) + \beta_4^{-,in} \mathrm{R}^{-,in}_{ijp} \times (\mathrm{P}_p - \bar{\mathrm{P}}^{-,in})+ \beta_4^{-,out} \mathrm{R}^{-,out}_{ijp}\times (\mathrm{P}_p - \bar{\mathrm{P}}^{-,out}) \notag \\
    &+ \mathrm{C}_{ij}\times \tilde H_p'\gamma_1 + \mathrm{R}^+_{ij}\times \tilde H_p'\gamma_2 + \mathrm{R}^{-,in}_{ij}\times \tilde H_p'\gamma_3+ \mathrm{R}^{-,out}_{ij}\times \tilde H_p'\gamma_4 + \lambda_p + \alpha_j + \varepsilon_{ijp}, \tag{6}
\end{align}
where $\tilde H_p=(H_p', \mathrm{Start}_p)'$ with $H_p$ defined as before. In the model, slope coefficients $\beta_3^{+}$ and $\beta_3^{-,in/out}$ measure whether a judge escalates the premium she awards her compatriot when her compatriot's close rivals' compatriot judges have a reputation for being more biased, or when more of her compatriot's close rivals have compatriot judges on the panel. Slope coefficients $\beta_4^{+}$ and $\beta_4^{-,in/out}$ measure whether a judge penalizes a rival more (a negative $\beta_4$) or less (a positive $\beta_4$) if the rival has a more biased compatriot judge on the panel. Interaction controls are included so that the key escalation coefficients do not pick up other sources of variations in the premium or penalty, such as skaters from larger federations starting later and at the same time having neighboring skaters also from larger federations and have compatriot judges sitting on the panel. 

Because $\mathrm{W}_p$ and $\mathrm{P}_p$ are built from the first-step estimates, the second-step coefficients are corrected\footnote{The corrected coefficient estimator is $(Z'Z - \Lambda)^{-1} Z'y$, where $Z$ collects the regressors and $\Lambda$ is the second moment of the first-step sampling error in the generated columns, computed from the two-way cluster-robust first-step covariance matrix.} for the first-step estimation error following~\citet{chen2026mostly}. For inference, our setup is slightly more general, as our first-step errors map into the second-step generated regressors through a linear design matrix with overlapping coverage: each judge's first-step estimate enters the generated regressors of many performances. As a result, our asymptotic variance formula has an extra first-step variance term, which the setting of~\citet{chen2026mostly} does not need, and allows for two-way clustering. Again, we cluster by competition and by skater~\citep{cameron2011robust} as the default and report in the appendix clustering by competition and by skater federation and the two-way cluster jackknife of~\citet{mackinnon2026jackknife} as robustness checks. No empirical findings change qualitatively with the inference method choice.

\begin{figure}[htbp]
  \centering
  \includegraphics[width=0.9\linewidth,height=0.72\textheight,keepaspectratio]{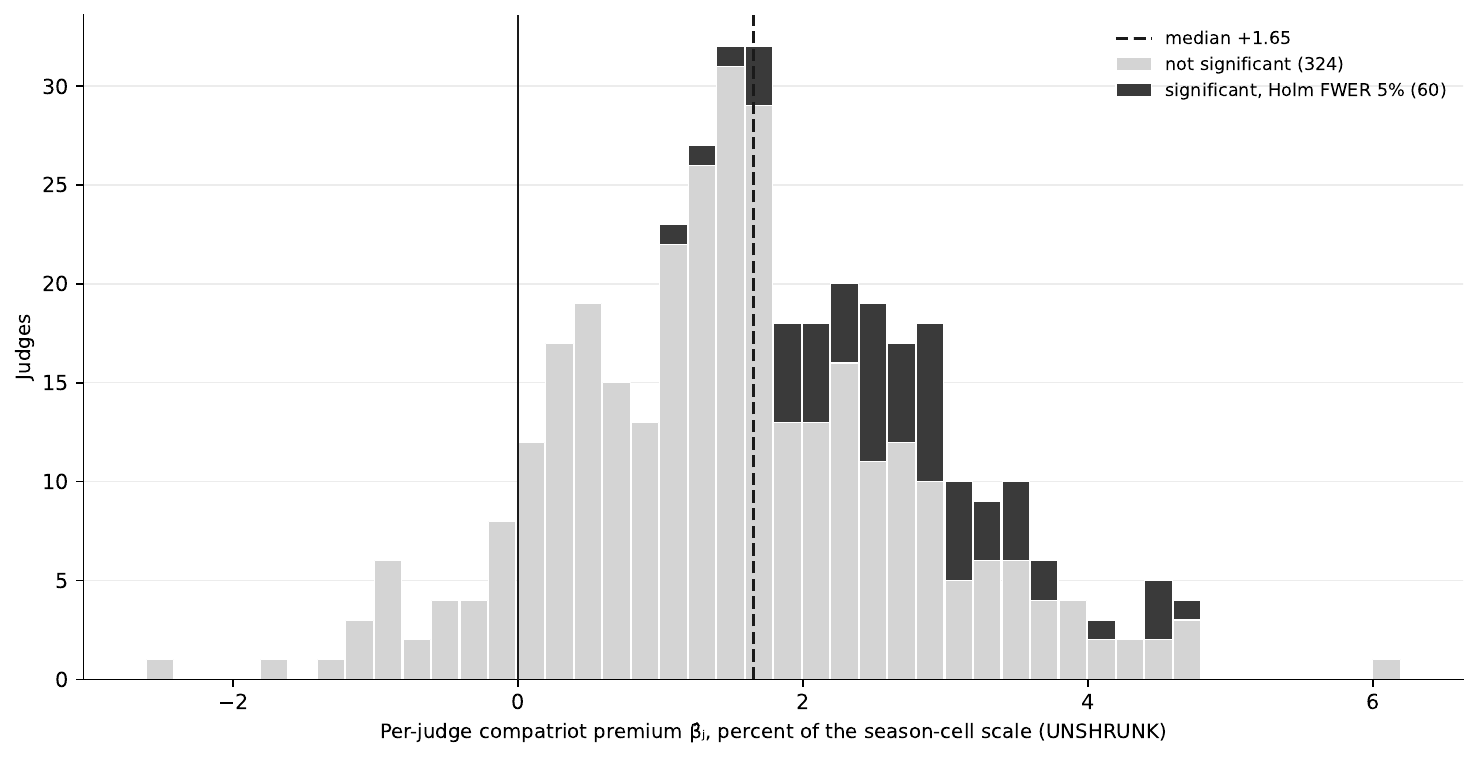}
  \caption{Judge-specific Compatriot Premium on TSS pct}\label{fig:f4}
  \begin{minipage}{0.95\linewidth}
    \footnotesize
    \textit{NOTE:} A judge's compatriot premium is considered measured and drawn on the graph only if the judges scored at least 4 competitions with at least 4 compatriot athletes and 30 non-compatriot performances over the ten seasons. 
  \end{minipage}
\end{figure}

Before discussing the regression results, we plot the distribution of judge-specific compatriot premium to show that judge-specific compatriot premium estimates are not pure noise. Figure~\ref{fig:f4} is estimated with all ten seasons of data\footnote{A judge with an ISU disciplinary record enters only through her seasons up to the incident.}, and a judge's compatriot premium estimate is considered measured if she has scored in at least four compatriot competitions, for four compatriot athletes and thirty non-compatriot performances. The figure shows that the median premium is 1.65 percent of the TSS season-cell scale. The 10th percentile is 0.13 and the 90th percentile is as high as 3.30. Multiple testing with Holm's step down identifies 60 judges with a statistically significant positive premium, while no judge has a statistically significant negative premium. Since these are raw estimates, the histogram is the distribution of the true premia with estimation noise, and its spread is larger than that of the true premia. 

\begin{table}[htbp]
  \centering
  \caption{Escalation of the Compatriot Premium and the Competitor Penalty: Era A $\rightarrow$ Era B}\label{tab:t42}
  \begin{threeparttable}
  \makebox[\textwidth]{\begin{adjustbox}{max width=\textwidth}\begin{tabular}{lrrrrrr}
\toprule
 & \multicolumn{3}{c}{\textbf{Segment 2}} & \multicolumn{3}{c}{\textbf{TSS pct}} \\
\cmidrule(lr){2-4}\cmidrule(lr){5-7}
\textbf{Variable} & \textbf{(1) TSS pct} & \textbf{(2) GOE pct} & \textbf{(3) PCS pct} & \textbf{(4) Segment 1, open draw} & \textbf{(5) Segment 1, seeded order} & \textbf{(6) All rows} \\
\midrule
Compatriot & +1.747*** & +2.644*** & +1.776*** & +1.530*** & +1.784*** & +1.716*** \\
 & (0.090) & (0.144) & (0.098) & (0.120) & (0.093) & (0.076) \\
Rival starting earlier & -0.220*** & -0.356*** & -0.200*** & +0.038 & -0.025 & -0.111** \\
 & (0.058) & (0.100) & (0.072) & (0.080) & (0.062) & (0.042) \\
Rival starting later, in contention & -0.125 & -0.273** & -0.100 & -0.052 & -0.043 & -0.075 \\
 & (0.079) & (0.127) & (0.095) & (0.163) & (0.074) & (0.055) \\
Rival starting later, out of contention & -0.034 & -0.134 & -0.016 & -0.037 & +0.072 & -0.007 \\
 & (0.051) & (0.089) & (0.071) & (0.118) & (0.093) & (0.041) \\
Compatriot $\times$ $\mathrm{W}_{\mathrm{earlier}}$, in contention & +0.052 & +0.153 & +0.009 & -0.116 & +0.000 & +0.002 \\
 & (0.072) & (0.132) & (0.079) & (0.141) & (0.079) & (0.056) \\
Compatriot $\times$ $\mathrm{W}_{\mathrm{earlier}}$, out of contention & +0.045 & -0.039 & +0.129 & -0.376* & -0.092 & -0.047 \\
 & (0.065) & (0.099) & (0.089) & (0.184) & (0.111) & (0.059) \\
Compatriot $\times$ $\mathrm{W}_{\mathrm{later}}$ & +0.192*** & +0.262** & +0.193** & -0.020 & +0.036 & +0.114** \\
 & (0.064) & (0.105) & (0.082) & (0.118) & (0.078) & (0.050) \\
Rival starting earlier $\times$ $\mathrm{P}_{p}$ & +0.059 & +0.085 & +0.063 & -0.011 & -0.001 & +0.020 \\
 & (0.039) & (0.062) & (0.050) & (0.072) & (0.047) & (0.025) \\
Rival starting later, in contention $\times$ $\mathrm{P}_{p}$ & +0.056 & +0.073 & +0.077 & -0.039 & +0.141* & +0.113** \\
 & (0.085) & (0.147) & (0.099) & (0.107) & (0.082) & (0.053) \\
Rival starting later, out of contention $\times$ $\mathrm{P}_{p}$ & +0.035 & +0.051 & +0.032 & +0.014 & -0.169 & -0.022 \\
 & (0.049) & (0.082) & (0.061) & (0.116) & (0.109) & (0.037) \\
\midrule
$\mathrm{C}_{jp}$, $\mathrm{R}^{+}_{jp}$, $\mathrm{R}^{-,in}_{jp}$ and $\mathrm{R}^{-,out}_{jp}$ $\cdot$ $\tilde{H}_{p}$ & $\checkmark$ & $\checkmark$ & $\checkmark$ & $\checkmark$ & $\checkmark$ & $\checkmark$ \\
Performance FE & $\checkmark$ & $\checkmark$ & $\checkmark$ & $\checkmark$ & $\checkmark$ & $\checkmark$ \\
Judge FE & $\checkmark$ & $\checkmark$ & $\checkmark$ & $\checkmark$ & $\checkmark$ & $\checkmark$ \\
N & 31,312 & 31,312 & 31,312 & 13,814 & 20,862 & 65,988 \\
\# of Competitions (G) & 57 & 57 & 57 & 21 & 36 & 57 \\
Mass of negative weights, $\mathrm{C}_{jp}$, $\mathrm{R}^{+}_{jp}$, $\mathrm{R}^{-,in}_{jp}$, $\mathrm{R}^{-,out}_{jp}$ & \multicolumn{3}{c}{0.000} & 0.000 & 0.000 & 0.000 \\
Mass of negative weights, $\mathrm{C}_{jp} \times \mathrm{W}_{\mathrm{later}}$ & \multicolumn{3}{c}{0.034} & 0.039 & 0.031 & 0.032 \\
Mass of negative weights, all other interactions & \multicolumn{3}{c}{0.027--0.075} & 0.029--0.065 & 0.028--0.075 & 0.029--0.046 \\
Share of negative weights, $\mathrm{C}_{jp}$, $\mathrm{R}^{+}_{jp}$, $\mathrm{R}^{-,in}_{jp}$, $\mathrm{R}^{-,out}_{jp}$ & \multicolumn{3}{c}{0\%} & 0\% & 0\% & 0\% \\
Share of negative weights, $\mathrm{C}_{jp} \times \mathrm{W}_{\mathrm{later}}$ & \multicolumn{3}{c}{20\%} & 19\% & 18\% & 19\% \\
Share of negative weights, all other interactions & \multicolumn{3}{c}{16\%--27\%} & 15\%--23\% & 17\%--22\% & 17\%--23\% \\
\bottomrule
\end{tabular}
\end{adjustbox}}
    \begin{tablenotes}[flushleft]
      \scriptsize
      \item \textit{NOTE:} Era A: 2016/17-2017/18. Era B: 2018/19-2021/22. Each interaction is centered at its mean over the rows of its indicator. Two-way (competition + athlete) cluster-robust standard errors in parentheses, with a term for the estimation error of the first-step trained premia added to the variance. $t$ with $G-1$ degrees of freedom, $G$ the column's competitions; *** $p<0.01$, ** $p<0.05$, * $p<0.10$. Appendix~\ref{app:fedcluster} reports alternative standard errors clustered by competition and skater federation and with the two-way cluster jackknife.
    \end{tablenotes}
  \end{threeparttable}
\end{table}

Table~\ref{tab:t42} reports the estimates of equation~\eqref{eq:step2}, with the premia trained on all Era A data and escalation response evaluated on the Segment 2 subsample of Era B. The compatriot premium rises with the trained premia of the later rivals. The coefficient is significant at the 1 percent level on TSS and at the 5 percent level on both GOE and PCS. The earlier rivals' compatriot judges do not move the premium, whether those rivals are still in contention or not. In the first segments, no $\beta_3$ is significant even at the 5\% significance level. The competitor penalty does not respond to the rival's own compatriot judge. The three $\beta_4$ terms are not jointly significant. One estimate, for later rivals still in contention on the full sample, is significant at the 5\% level. This can be from randomness in testing, or if one really wants to interpret it, a judge may spare a rival whose compatriot judge is known to be more biased.

\begin{table}[htbp]
  \centering
  \caption{Escalation of the Compatriot Premium and the Competitor Penalty: Era B $\rightarrow$ Era C}\label{tab:t42b}
  \begin{threeparttable}
  \makebox[\textwidth]{\begin{adjustbox}{max width=\textwidth}\begin{tabular}{lrrrrrr}
\toprule
 & \multicolumn{3}{c}{\textbf{Segment 2}} & \multicolumn{3}{c}{\textbf{TSS pct}} \\
\cmidrule(lr){2-4}\cmidrule(lr){5-7}
\textbf{Variable} & \textbf{(1) TSS pct} & \textbf{(2) GOE pct} & \textbf{(3) PCS pct} & \textbf{(4) Segment 1, open draw} & \textbf{(5) Segment 1, seeded order} & \textbf{(6) All rows} \\
\midrule
Compatriot & +1.392*** & +1.802*** & +1.513*** & +1.149*** & +1.605*** & +1.417*** \\
 & (0.086) & (0.109) & (0.102) & (0.141) & (0.114) & (0.084) \\
Rival starting earlier & -0.227*** & -0.340*** & -0.204*** & -0.125* & -0.006 & -0.141*** \\
 & (0.039) & (0.053) & (0.058) & (0.065) & (0.048) & (0.031) \\
Rival starting later, in contention & -0.186** & -0.409*** & -0.087 & +0.063 & +0.069 & -0.025 \\
 & (0.071) & (0.105) & (0.094) & (0.120) & (0.073) & (0.051) \\
Rival starting later, out of contention & +0.058 & -0.043 & +0.141* & -0.098 & -0.014 & +0.008 \\
 & (0.052) & (0.073) & (0.073) & (0.100) & (0.074) & (0.047) \\
Compatriot $\times$ $\mathrm{W}_{\mathrm{earlier}}$, in contention & +0.006 & +0.079 & -0.038 & +0.150 & +0.065 & +0.065 \\
 & (0.098) & (0.162) & (0.114) & (0.243) & (0.108) & (0.060) \\
Compatriot $\times$ $\mathrm{W}_{\mathrm{earlier}}$, out of contention & +0.060 & +0.026 & +0.119 & -0.155 & -0.206 & -0.023 \\
 & (0.068) & (0.105) & (0.105) & (0.232) & (0.185) & (0.060) \\
Compatriot $\times$ $\mathrm{W}_{\mathrm{later}}$ & -0.004 & +0.042 & -0.039 & +0.252 & +0.097 & +0.053 \\
 & (0.056) & (0.094) & (0.065) & (0.169) & (0.105) & (0.058) \\
Rival starting earlier $\times$ $\mathrm{P}_{p}$ & -0.027 & -0.036 & -0.025 & -0.092 & -0.179*** & -0.084** \\
 & (0.044) & (0.058) & (0.063) & (0.080) & (0.059) & (0.032) \\
Rival starting later, in contention $\times$ $\mathrm{P}_{p}$ & -0.016 & -0.119 & +0.062 & +0.022 & -0.064 & -0.040 \\
 & (0.092) & (0.119) & (0.124) & (0.129) & (0.060) & (0.054) \\
Rival starting later, out of contention $\times$ $\mathrm{P}_{p}$ & -0.108** & -0.110 & -0.146** & +0.220* & +0.040 & +0.004 \\
 & (0.051) & (0.070) & (0.072) & (0.121) & (0.082) & (0.042) \\
\midrule
$\mathrm{C}_{jp}$, $\mathrm{R}^{+}_{jp}$, $\mathrm{R}^{-,in}_{jp}$ and $\mathrm{R}^{-,out}_{jp}$ $\cdot$ $\tilde{H}_{p}$ & $\checkmark$ & $\checkmark$ & $\checkmark$ & $\checkmark$ & $\checkmark$ & $\checkmark$ \\
Performance FE & $\checkmark$ & $\checkmark$ & $\checkmark$ & $\checkmark$ & $\checkmark$ & $\checkmark$ \\
Judge FE & $\checkmark$ & $\checkmark$ & $\checkmark$ & $\checkmark$ & $\checkmark$ & $\checkmark$ \\
N & 44,679 & 44,679 & 44,679 & 20,290 & 29,020 & 93,989 \\
\# of Competitions (G) & 76 & 76 & 76 & 28 & 48 & 76 \\
Mass of negative weights, $\mathrm{C}_{jp}$, $\mathrm{R}^{+}_{jp}$, $\mathrm{R}^{-,in}_{jp}$, $\mathrm{R}^{-,out}_{jp}$ & \multicolumn{3}{c}{0.000} & 0.000 & 0.000 & 0.000 \\
Mass of negative weights, $\mathrm{C}_{jp} \times \mathrm{W}_{\mathrm{later}}$ & \multicolumn{3}{c}{0.023} & 0.038 & 0.027 & 0.024 \\
Mass of negative weights, all other interactions & \multicolumn{3}{c}{0.025--0.071} & 0.047--0.067 & 0.024--0.049 & 0.032--0.044 \\
Share of negative weights, $\mathrm{C}_{jp}$, $\mathrm{R}^{+}_{jp}$, $\mathrm{R}^{-,in}_{jp}$, $\mathrm{R}^{-,out}_{jp}$ & \multicolumn{3}{c}{0\%} & 0\% & 0\% & 0\% \\
Share of negative weights, $\mathrm{C}_{jp} \times \mathrm{W}_{\mathrm{later}}$ & \multicolumn{3}{c}{17\%} & 21\% & 15\% & 16\% \\
Share of negative weights, all other interactions & \multicolumn{3}{c}{14\%--27\%} & 14\%--19\% & 16\%--20\% & 14\%--22\% \\
\bottomrule
\end{tabular}
\end{adjustbox}}
    \begin{tablenotes}[flushleft]
      \scriptsize
      \item \textit{NOTE:} Era B: 2018/19-2021/22. Era C: 2022/23-2025/26. Each interaction is centered at its mean over the rows of its indicator. Two-way (competition + athlete) cluster-robust standard errors in parentheses, with a term for the estimation error of the first-step trained premia added to the variance. $t$ with $G-1$ degrees of freedom, $G$ the column's competitions; *** $p<0.01$, ** $p<0.05$, * $p<0.10$. Appendix~\ref{app:fedcluster} reports alternative standard errors clustered by competition and skater federation and with the two-way cluster jackknife.
    \end{tablenotes}
  \end{threeparttable}
\end{table}

Table~\ref{tab:t42b} repeats the analysis of Table~\ref{tab:t42} but with the premia trained on Era B and escalation evaluated on Era C. We see that none of the $\beta_3$ estimates is significant at even the 10\% significance level. Some $\beta_4$ estimates for earlier starting rivals with protected judges are negative and significant at the 5\%, suggesting that for Segment 1 with seeded draws in Era C, judges mark an earlier rival down harder when the rival's compatriot judge has a larger trained premium. The change from Era B to Era C is in line with the general decline in the compatriot premium shown in Figure~\ref{fig:f22} of Section~\ref{sec:descriptive}. The average premium in Segment 2 is 1.392 in Era C (Table~\ref{tab:t42b}) against 1.747 in Era B (Table~\ref{tab:t42}), and the measured judges' trained premia average 2.52 when trained in Era A but drop to 1.79 when trained in Era B.

\begin{table}[htbp]
  \centering
  \caption{Table~\ref{tab:t42} Robustness Check: First-step Judge-specific Compatriot Premium All Set to One}\label{tab:t42c}
  \begin{threeparttable}
  \makebox[\textwidth]{\begin{adjustbox}{max width=\textwidth}\begin{tabular}{lrrrrrr}
\toprule
 & \multicolumn{3}{c}{\textbf{Segment 2}} & \multicolumn{3}{c}{\textbf{TSS pct}} \\
\cmidrule(lr){2-4}\cmidrule(lr){5-7}
\textbf{Variable} & \textbf{(1) TSS pct} & \textbf{(2) GOE pct} & \textbf{(3) PCS pct} & \textbf{(4) Segment 1, open draw} & \textbf{(5) Segment 1, seeded order} & \textbf{(6) All rows} \\
\midrule
Compatriot & +1.747*** & +2.641*** & +1.780*** & +1.531*** & +1.786*** & +1.716*** \\
 & (0.090) & (0.144) & (0.098) & (0.120) & (0.093) & (0.076) \\
Rival starting earlier & -0.221*** & -0.358*** & -0.201*** & +0.036 & -0.026 & -0.112*** \\
 & (0.059) & (0.100) & (0.072) & (0.080) & (0.063) & (0.042) \\
Rival starting later, in contention & -0.126 & -0.274** & -0.101 & -0.053 & -0.040 & -0.075 \\
 & (0.079) & (0.127) & (0.094) & (0.162) & (0.074) & (0.055) \\
Rival starting later, out of contention & -0.034 & -0.135 & -0.016 & -0.039 & +0.074 & -0.007 \\
 & (0.051) & (0.089) & (0.071) & (0.118) & (0.092) & (0.041) \\
Compatriot $\times$ share earlier, in contention & +0.147 & +0.433 & +0.039 & -0.369 & +0.035 & -0.002 \\
 & (0.186) & (0.329) & (0.223) & (0.338) & (0.209) & (0.144) \\
Compatriot $\times$ share earlier, out of contention & +0.129 & -0.114 & +0.359 & -1.129** & -0.172 & -0.136 \\
 & (0.166) & (0.254) & (0.223) & (0.470) & (0.338) & (0.158) \\
Compatriot $\times$ share later & +0.423*** & +0.525** & +0.467** & -0.139 & +0.100 & +0.238* \\
 & (0.151) & (0.253) & (0.195) & (0.329) & (0.179) & (0.121) \\
Rival starting earlier $\times$ compatriot judge & +0.182* & +0.189 & +0.243* & -0.013 & -0.033 & +0.056 \\
 & (0.096) & (0.150) & (0.136) & (0.189) & (0.132) & (0.066) \\
Rival starting later, in contention $\times$ compatriot judge & +0.206 & +0.105 & +0.370 & +0.029 & +0.535*** & +0.367*** \\
 & (0.214) & (0.392) & (0.250) & (0.253) & (0.192) & (0.136) \\
Rival starting later, out of contention $\times$ compatriot judge & +0.075 & +0.108 & +0.074 & -0.005 & -0.411 & -0.073 \\
 & (0.137) & (0.216) & (0.174) & (0.260) & (0.281) & (0.099) \\
\midrule
$\mathrm{C}_{jp}$, $\mathrm{R}^{+}_{jp}$, $\mathrm{R}^{-,in}_{jp}$ and $\mathrm{R}^{-,out}_{jp}$ $\cdot$ $\tilde{H}_{p}$ & $\checkmark$ & $\checkmark$ & $\checkmark$ & $\checkmark$ & $\checkmark$ & $\checkmark$ \\
Performance FE & $\checkmark$ & $\checkmark$ & $\checkmark$ & $\checkmark$ & $\checkmark$ & $\checkmark$ \\
Judge FE & $\checkmark$ & $\checkmark$ & $\checkmark$ & $\checkmark$ & $\checkmark$ & $\checkmark$ \\
N & 31,312 & 31,312 & 31,312 & 13,814 & 20,862 & 65,988 \\
\# of Competitions (G) & 57 & 57 & 57 & 21 & 36 & 57 \\
Mass of negative weights, $\mathrm{C}_{jp}$, $\mathrm{R}^{+}_{jp}$, $\mathrm{R}^{-,in}_{jp}$, $\mathrm{R}^{-,out}_{jp}$ & \multicolumn{3}{c}{0.000} & 0.000 & 0.000 & 0.000 \\
Mass of negative weights, $\mathrm{C}_{jp} \times$ share later & \multicolumn{3}{c}{0.032} & 0.042 & 0.029 & 0.031 \\
Mass of negative weights, all other interactions & \multicolumn{3}{c}{0.020--0.087} & 0.026--0.071 & 0.032--0.075 & 0.021--0.050 \\
Share of negative weights, $\mathrm{C}_{jp}$, $\mathrm{R}^{+}_{jp}$, $\mathrm{R}^{-,in}_{jp}$, $\mathrm{R}^{-,out}_{jp}$ & \multicolumn{3}{c}{0\%} & 0\% & 0\% & 0\% \\
Share of negative weights, $\mathrm{C}_{jp} \times$ share later & \multicolumn{3}{c}{20\%} & 20\% & 21\% & 19\% \\
Share of negative weights, all other interactions & \multicolumn{3}{c}{14\%--28\%} & 15\%--24\% & 17\%--22\% & 16\%--25\% \\
\bottomrule
\end{tabular}
\end{adjustbox}}
    \begin{tablenotes}[flushleft]
      \footnotesize
      \item \textit{NOTE:} Era A: 2016/17-2017/18. Era B: 2018/19-2021/22. Each interaction is centered at its mean over the rows of its indicator. Two-way (competition + athlete) cluster-robust standard errors in parentheses, $t$ with $G-1$ degrees of freedom, $G$ the column's competitions; *** $p<0.01$, ** $p<0.05$, * $p<0.10$. Appendix~\ref{app:fedcluster} reports alternative standard errors clustered by competition and skater federation and with the two-way cluster jackknife.
    \end{tablenotes}
  \end{threeparttable}
\end{table}

Notice that if we set every $\widehat{\beta}^{\,\mathcal{T}}_{1j}$ to 1 in the second step regression model in~\eqref{eq:step2}, $W^{+}_p$ and $W^{-,in/out}_p$ reduce to the shares of later and earlier rivals protected by a compatriot judge on the panel and $P_p$ reduces to an indicator for whether performance $p$ has a compatriot judge on the panel. We estimate this special case in Table~\ref{tab:t42c}. We find that the pattern of Table~\ref{tab:t42} holds even if we simply consider the proportion of rivals protected by their own compatriot judges. In Segment 2, the compatriot premium rises by 0.42 percent of the scale on TSS when all of the later rivals are protected by a compatriot judge. 

Tables~\ref{tab:tac10} and~\ref{tab:tac11} in the Appendix repeat Tables~\ref{tab:t42} and~\ref{tab:t42b} for senior and junior events, Segment 2. In the Era A $\to$ Era B analysis, the escalation of the compatriot premium comes mostly from the junior events. There, the premium rises with the trained premia of the later rivals' compatriot judges on TSS and PCS, and also with those of the earlier rivals still in contention on TSS and GOE. Since the scores of these earlier rivals are known when the compatriot performs, junior-event judges seem to respond not only to a threat that is yet to come but also to one that has already been realized. This is not revealed in Table~\ref{tab:t42}, as the analysis there pools the senior and junior events together. Table~\ref{tab:tac10} also shows that the competitor penalty responds to the rival's compatriot judge only in the senior events, and only for the later rivals of a compatriot already out of contention. A senior-event judge marks such a rival higher relative to the rest of the panel, the larger the trained premium of the rival's compatriot judge. This data pattern is in line with the colleague-compatriot premium of Section~\ref{subsec:cross-panel}, which the literature reads as vote trading~\citep{zitzewitz2006nationalism}. But since the three $\beta_4$ terms are not jointly significant in Table~\ref{tab:tac10} (p-value$=0.17-0.22$ across the three outcomes), we read the result with cautious. In the Era B $\to$ Era C analysis, no escalation behavior is found significant at the 5 percent level in either tier, which is in line with our readings of Table~\ref{tab:t42b}.

Table~\ref{tab:tac12} nest Table~\ref{tab:tac10} and Table~\ref{tab:t42c} for senior and junior events, asking whether the escalation responds to how many of the rivals are protected by a compatriot judge or to how biased these judges are known to be. So the regressions include both $\mathrm{C}\times \mathrm{W}$ and its share counterpart in Table~\ref{tab:t42c}, and both $\mathrm{R}\times \mathrm{P}$ and its compatriot judge presence counterpart. Because the two sets of terms are highly collinear, the regressions fail to estimate the individual slopes precisely. The only interaction coefficients significant at the 5 percent level are those on the later rivals in the junior events on GOE, where the coefficient on the compatriot indicator times $W^+_p$ is positive and significant, and that on the compatriot indicator times the share of protected later rivals is negative and significant. It might be tempting to interpret the results as judges responding to the reputation of the rivals' compatriot judges beyond their presence on the panel, but we would like to read this evidence with caution, since both coefficients carry large shares and masses of negative implicit weights. We also carried out a similar unfruitful attempt trying to examine whether there is a nonlinear response to the colleagues' reputation, by adding $\mathrm{C}\times \mathrm{W}^2$ to the existing models. The results are likewise inconclusive because of the collinearity between $\mathrm{W}$ and $\mathrm{W}^2$. With these caveats, we read the results of this section as follows. Judges respond to the composition of the panel. The compatriot premium rises with the share of the skater's later rivals whose federation holds a seat on the panel, and, under a linearity assumption, with those judges' trained premia. These empirical findings support strategic nationalistic scoring. Positive estimates of $\beta^{+}_3$ in the Era A $\rightarrow$ Era B analysis show that the premium is, at least in part and in the Beijing Olympic cycle, a strategic response to the panel composition. 


\section{The Effect of Nationalistic Bias on Competition Outcomes} 
\label{sec:milan}

\subsection{Panel Composition and Competition Outcomes}

The previous section studies the impact of judge biases on synthetic judge scores. In this section, we ask whether such biases affect real competition outcomes. Since there is only one placement outcome for each performance, compared to 6-9 judge synthetic scores used in earlier sections, regressions run in this section are far less precisely estimated, even with our large dataset. 

Table~\ref{tab:tab1} report the regression results on all competitions and on a subsample of Championships and Olympics competitions whose panels are randomly drawn. Results are qualitatively similar across the columns, with having a compatriot judge increasing the final placement percentile of the skater. Specifically, in Championships and Olympics, having a compatriot judge in one of the two segments raises the placement percentile by 1.62, and having a compatriot judge in both segments raises the placement percentile by 2.49. A percentile point is $(n-1)/100$ places with $n$ being the field size. So these estimates are 1/4 to 2/5 of a place in an average event of 17.8 starters, and about 1/2 to 3/4 a place on the drawn competitions, whose fields are larger (30.3 starters on average).

\begin{table}[htbp]
  \centering
  \caption{Compatriot Premium and Competitor Penalty on Official Placement Percentiles}\label{tab:tab1}
  \begin{threeparttable}
  \makebox[\textwidth]{\begin{adjustbox}{max width=\textwidth}\begin{tabular}{lrrrrrr}
\toprule
 & \multicolumn{3}{c}{\textbf{All competitions}} & \multicolumn{3}{c}{\textbf{Championships and Olympics}} \\
\cmidrule(lr){2-4}\cmidrule(lr){5-7}
\textbf{Variable} & \textbf{(1) Segment 1} & \textbf{(2) Segment 2} & \textbf{(3) Event} & \textbf{(4) Segment 1} & \textbf{(5) Segment 2} & \textbf{(6) Event} \\
\midrule
Compatriot judge & +1.932*** & +0.961** &  & +2.725*** & -0.204 &  \\
 & (0.452) & (0.468) &  & (0.637) & (1.040) &  \\
Compatriot judge in one of the two segments &  &  & +2.115*** &  &  & +1.622* \\
 &  &  & (0.531) &  &  & (0.823) \\
Compatriot judge in both segments &  &  & +2.093*** &  &  & +2.494*** \\
 &  &  & (0.512) &  &  & (0.741) \\
Share of rivals protected, segment 1 & +0.968 &  & -0.130 & +1.697 &  & +0.062 \\
 & (1.032) &  & (0.988) & (1.537) &  & (1.471) \\
Share of rivals protected, segment 2 &  & -0.513 &  &  & +2.086 &  \\
 &  & (0.972) &  &  & (1.782) &  \\
Home ice & +1.136 & +0.754 & +0.332 & +2.613* & +1.882 & +0.197 \\
 & (0.723) & (0.673) & (0.752) & (1.303) & (1.434) & (1.134) \\
\midrule
Compatriot Technical Controller, Technical Specialist & $\checkmark$ & $\checkmark$ & $\checkmark$ & $\checkmark$ & $\checkmark$ & $\checkmark$ \\
$\tilde{Z}_{p}$ & $\checkmark$ & $\checkmark$ & $\checkmark$ & $\checkmark$ & $\checkmark$ & $\checkmark$ \\
Skater FE & $\checkmark$ & $\checkmark$ & $\checkmark$ & $\checkmark$ & $\checkmark$ & $\checkmark$ \\
Segment FE & $\checkmark$ & $\checkmark$ &  & $\checkmark$ & $\checkmark$ &  \\
Event FE &  &  & $\checkmark$ &  &  & $\checkmark$ \\
N & 11,971 & 10,768 & 10,767 & 3,633 & 2,511 & 2,511 \\
\# of Competitions (G) & 167 & 167 & 167 & 30 & 30 & 30 \\
Skaters & 2,470 & 2,375 & 2,374 & 1,131 & 788 & 788 \\
Mass of negative weights & 0.001--0.096 & 0.001--0.128 & 0.001--0.123 & 0.000--0.013 & 0.000--0.022 & 0.000--0.010 \\
Share of negative weights & 1\%--23\% & 2\%--25\% & 1\%--26\% & 0\%--7\% & 0\%--8\% & 0\%--5\% \\
\bottomrule
\end{tabular}
\end{adjustbox}}
    \begin{tablenotes}[flushleft]
      \footnotesize
      \item \textit{NOTE:} Two-way (competition + athlete) cluster-robust standard errors in
parentheses, $t$ with $G-1$ degrees of freedom, $G$ the column's competitions; *** $p<0.01$, ** $p<0.05$, * $p<0.10$. 
    \end{tablenotes}
  \end{threeparttable}
\end{table}

Whether the panel composition changes the podium outcomes is much harder to detect. Table~\ref{tab:tab2} studies the outcomes of finishing among the top-three and among the top-sixth using the championship and the Olympics competitions. None of the segment-level estimates are statistically significant. In the event standing, however, having a compatriot judge in one of the two segments raises the probability of a top-three finish by 3.46 percentage points and a top-six finish by 4.28 percentage points. The impact of rivals judges is small and statistically insignificant in almost all columns except for Column (3), where having all close rivals protected by a compatriot judge in segment 1 lowers the probability of the skater standing on the podium by about 7.5 percentage points. The share of rivals protected in Segment 2 is not used as a regressor in the event-level regressions as it is determined by the Segment 1 results, which are part of the event-level outcome. So adding the regressor would potentially cause a reverse-causality problem.

\begin{table}[htbp]
  \centering
  \caption{Table~\ref{tab:tab1} with the Top-three and Top-six Placements as the Outcome}\label{tab:tab2}
  \begin{threeparttable}
  \makebox[\textwidth]{\begin{adjustbox}{max width=\textwidth}\begin{tabular}{lrrrrrr}
\toprule
 & \multicolumn{3}{c}{\textbf{A top-three placement}} & \multicolumn{3}{c}{\textbf{A top-six placement}} \\
\cmidrule(lr){2-4}\cmidrule(lr){5-7}
\textbf{Variable} & \textbf{(1) Segment 1} & \textbf{(2) Segment 2} & \textbf{(3) Event} & \textbf{(4) Segment 1} & \textbf{(5) Segment 2} & \textbf{(6) Event} \\
\midrule
Compatriot judge & +1.814 & -1.147 &  & +1.640 & -0.150 &  \\
 & (1.327) & (1.627) &  & (1.694) & (1.881) &  \\
Compatriot judge in one of the two segments &  &  & +3.464* &  &  & +4.280** \\
 &  &  & (1.806) &  &  & (1.989) \\
Compatriot judge in both segments &  &  & +0.189 &  &  & +1.813 \\
 &  &  & (2.030) &  &  & (2.001) \\
Share of rivals protected, segment 1 & -3.988 &  & -7.492** & -6.393 &  & -7.118 \\
 & (3.403) &  & (3.541) & (4.129) &  & (5.045) \\
Share of rivals protected, segment 2 &  & -2.454 &  &  & +4.763 &  \\
 &  & (4.107) &  &  & (4.115) &  \\
Home ice & +2.099 & +5.896 & +2.695 & +1.228 & +3.250 & +0.244 \\
 & (2.591) & (4.767) & (3.912) & (2.507) & (3.770) & (2.848) \\
\midrule
Compatriot Technical Controller, Technical Specialist & $\checkmark$ & $\checkmark$ & $\checkmark$ & $\checkmark$ & $\checkmark$ & $\checkmark$ \\
$\tilde{Z}_{p}$ & $\checkmark$ & $\checkmark$ & $\checkmark$ & $\checkmark$ & $\checkmark$ & $\checkmark$ \\
Skater FE & $\checkmark$ & $\checkmark$ & $\checkmark$ & $\checkmark$ & $\checkmark$ & $\checkmark$ \\
Segment FE & $\checkmark$ & $\checkmark$ &  & $\checkmark$ & $\checkmark$ &  \\
Event FE &  &  & $\checkmark$ &  &  & $\checkmark$ \\
N & 3,633 & 2,511 & 2,511 & 3,633 & 2,511 & 2,511 \\
\# of Competitions (G) & 30 & 30 & 30 & 30 & 30 & 30 \\
Skaters & 1,131 & 788 & 788 & 1,131 & 788 & 788 \\
Mass of negative weights & 0.000--0.013 & 0.000--0.022 & 0.000--0.010 & 0.000--0.013 & 0.000--0.022 & 0.000--0.010 \\
Share of negative weights & 0\%--7\% & 0\%--8\% & 0\%--5\% & 0\%--7\% & 0\%--8\% & 0\%--5\% \\
\bottomrule
\end{tabular}
\end{adjustbox}}
    \begin{tablenotes}[flushleft]
      \footnotesize
      \item \textit{NOTE:} Two-way (competition + athlete) cluster-robust standard errors in
parentheses, $t$ with $G-1$ degrees of freedom, $G$ the column's competitions; *** $p<0.01$, ** $p<0.05$, * $p<0.10$. 
    \end{tablenotes}
  \end{threeparttable}
\end{table}

\subsection{A Counterfactual Analysis for the Milan Olympic Games}

We now return to the ice dance event at the 2026 Milan Olympic Games discussed in the introduction and carry out a counterfactual analysis. The goal of this analysis is not to question the official result nor the ISU's response to the controversy that followed, but, instead, to show that the compatriot premium, the competitor penalty, and the colleague-compatriot premium studied in this paper, together with the randomness of the panel draw, can change the result of a close competition. One might expect that the use of trimmed mean in official score calculation can protect skaters from a biased judge, but trimming in fact only removes a small part of a typical judge's biases. In Appendix~\ref{app:trimming}, we model a nine-judge panel in which every judge's mark carries independent normal noise and one judge adds a premium to her compatriot’s mark. The average pass-through rate is the share of the premium’s effect on the untrimmed mean that survives in the trimmed mean. At the median of our judge-specific estimates, the pass-through rate is about 97 percent. At the 95th percentile, it falls to 87.5 percent. For the French judge at Milan, the pass-through rates for her marks on the French and the American couples lie between 86 and 96 percent. Trimming therefore removes only a small part of a judge’s bias, and the panel’s composition can change outcomes.

At the Milan Olympic Games, the judge pool of an event typically had thirteen judges. A segment's panel is any nine of them, so a segment has $\binom{13}{9} = 715$ possible panels, and the event has $\binom{13}{9} \times \binom{9}{5} = 715 \times 126 =90,090$ possible pairs of panels, since the four judges not drawn for the first segment sit the second and its other five seats are drawn from the nine who sat the first. The panels that judged at Milan are one of these pairs. In the ice dance event, the French judge was randomly drawn twice for judging while the American judge was drawn only to mark the free dance. If the realized scores were pure signals of the performance quality and no judge biases, a redraw of the panels could change the result only through the judges' honest disagreement. To measure how much, our first counterfactual analysis treats each couple's realized score as the true performance quality. Every judge on every possible panel, whether she sat at Milan or is drawn in, gives a score that is equal to that quality plus an independent normal shock with the standard deviation set to the within-panel estimate at the event (3.00 points in the rhythm dance and 3.62 in the free dance). No compatriot premium, rival penalty or colleague-compatriot premium considerations enter this first counterfactual analysis. The analysis shows that, without the shocks, every panel would return the Milan scores, and the French couple would win on every draw. With the shocks, the French couple would lead the rhythm dance in 60.2 percent of the draws over the 715 possible panels and lead the free dance in 64.8 percent of the draws. The couple would also keep the title in 66.3 percent of the draws of admissible panel pairs. This exercise supports the ISU's statement after the Milan ice dance event that it is normal for different judges on a panel to give a range of scores\footnote{The ISU's statement, as reported by the Associated Press and the NBC news on Feb.\ 12, 2026, says that ``it is normal for there to be a range of scores given by different judge[s] in any panel'' and that ``a number of mechanisms are used to mitigate these variations''. U.S.\ Figure Skating did not appeal the result before the 24-hour window closed.}: with honest disagreement among the judges, which the random shocks in our model represent, the result of a close competition is naturally probabilistic. With judge biases, however, these probabilities \textit{do} depend on who sits on the panel, because the judges no longer mark the same way.

In the second counterfactual analysis, we illustrate how judge biases can work together with the random panel draw to shape competition results. Let $P$ denote the observed panel and $P^*$ a panel redraw. Let $\mathcal{P}$ denote the set of all admissible redraws. For a judge $j\in P^*$ and a performance $p$ of couple $i$, we set her counterfactual score, in percent of the scale, as 
\begin{align*}
   Y_{ijp}^*&=Y_{ijp}+\hat \beta_{\mathrm{O}}\left(\mathrm{OP}_{ijp}(P^*) -\mathrm{OP}_{ijp}(P)\right)  \text{ if } j \in P\\
   Y_{ijp}^*&=\nu_{ijp}+\hat \beta_{1,c(j)} \mathrm{C}_{ijp} +\hat \beta_2 \mathrm{R}_{ijp}+ \hat \beta_{\mathrm{O}} \mathrm{OP}_{ijp}(P^*)  \text{ if } j \notin P.
\end{align*}
That is, a judge who sat keeps the score she gave, with a colleague-compatriot premium adjustment with respect to the new panel composition, and a judge who did not sit marks at a baseline $\nu_{ijp}$ plus three strategic scoring terms estimated from Table~\ref{tab:t61}: $\hat \beta_{1,c(j)}$ is the estimated compatriot premium of the judge's federation, and $\hat \beta_{2}$ and $\hat \beta_{\mathrm{O}}$ are the segment's competitor penalty and colleague-compatriot premium estimates. The baseline $\nu_{ijp}$ is drawn from a normal distribution with the mean calculated from the average judge scores of the realized panel net of the three terms, excluding those from top-three's federations (i.e., France, the United States and Canada), and the variance $\sigma^2$ being either zero or from within-panel estimates. These counterfactual scores are then rescaled back to raw points for the analysis, and a couple's score under $P^*$ is the trimmed mean of its nine marks. Note that this trims the judges' totals, whereas the official score trims each element and component separately. In our data the two ways of trimming give close results. At the Milan Games, the two trimming methods give the same segment winner in all 16 segments of the individual and team events, and the same medalists, in order, in each of the four individual events.

\begin{table}[htbp]
  \centering
  \caption{Compatriot Premium, Competitor Penalty, and Colleague-compatriot Premium Used for the Counterfactual Analysis}\label{tab:t61}
  \begin{threeparttable}
  \makebox[\textwidth]{\begin{adjustbox}{max width=\textwidth}\begin{tabular}{lrrrrrrrr}
\toprule
 & \multicolumn{4}{c}{\textbf{Segment 1}} & \multicolumn{4}{c}{\textbf{Segment 2}} \\
\cmidrule(lr){2-5}\cmidrule(lr){6-9}
\textbf{Variable} & \textbf{(1) Men} & \textbf{(2) Women} & \textbf{(3) Pairs} & \textbf{(4) Ice dance} & \textbf{(5) Men} & \textbf{(6) Women} & \textbf{(7) Pairs} & \textbf{(8) Ice dance} \\
\midrule
Compatriot, CAN judges &  &  &  & +2.951*** &  &  &  & +3.665*** \\
 &  &  &  & (0.475) &  &  &  & (0.726) \\
Compatriot, FRA judges &  &  &  & +2.585*** &  &  &  & +1.411** \\
 &  &  &  & (0.379) &  &  &  & (0.514) \\
Compatriot, GEO judges &  &  & +2.313 &  &  &  &  &  \\
 &  &  & (1.383) &  &  &  &  &  \\
Compatriot, GER judges &  &  & +2.911*** &  &  &  & +1.372** &  \\
 &  &  & (0.817) &  &  &  & (0.611) &  \\
Compatriot, JPN judges &  & +1.623 & -0.119 &  &  & +0.923 & +0.629 &  \\
 &  & (0.991) & (0.933) &  &  & (0.836) & (1.588) &  \\
Compatriot, KAZ judges & +2.197** &  &  &  & +0.848 &  &  &  \\
 & (1.039) &  &  &  & (2.013) &  &  &  \\
Compatriot, USA judges &  & +1.701** &  & +3.390*** &  & +0.630 &  & +2.618*** \\
 &  & (0.740) &  & (0.508) &  & (0.830) &  & (0.454) \\
Rival & +0.068 & -0.035 & +0.061 & -0.277** & -0.271 & -0.161 & -0.286* & -0.110 \\
 & (0.138) & (0.158) & (0.162) & (0.113) & (0.162) & (0.171) & (0.142) & (0.158) \\
Other-panel compatriot & +0.829 & +0.459 & +0.718 & +0.197 & -0.311 & -0.581 & -1.201** & -0.142 \\
 & (0.543) & (0.423) & (0.681) & (0.262) & (0.668) & (0.542) & (0.587) & (0.410) \\
\midrule
Home ice, Compatriot Technical Controller and Specialist & $\checkmark$ & $\checkmark$ & $\checkmark$ & $\checkmark$ & $\checkmark$ & $\checkmark$ & $\checkmark$ & $\checkmark$ \\
$\tilde{Z}_{p}$ & $\checkmark$ & $\checkmark$ & $\checkmark$ & $\checkmark$ & $\checkmark$ & $\checkmark$ & $\checkmark$ & $\checkmark$ \\
Skater, segment and judge FE & $\checkmark$ & $\checkmark$ & $\checkmark$ & $\checkmark$ & $\checkmark$ & $\checkmark$ & $\checkmark$ & $\checkmark$ \\
N & 9,648 & 10,224 & 5,003 & 7,812 & 6,462 & 6,462 & 4,292 & 5,373 \\
\# of Competitions (G) & \multicolumn{4}{c}{30} & \multicolumn{4}{c}{30} \\
Mass of negative weights & 0.000--0.001 & 0.000--0.002 & 0.000--0.028 & 0.000--0.006 & 0.000--0.003 & 0.000--0.001 & 0.000--0.030 & 0.000--0.019 \\
Share of negative weights & 0\%--1\% & 0\%--1\% & 0\%--9\% & 0\%--4\% & 0\%--2\% & 0\%--2\% & 0\%--11\% & 0\%--6\% \\
\bottomrule
\end{tabular}
\end{adjustbox}}
    \begin{tablenotes}[flushleft]
      \footnotesize
      \item \textit{NOTE:}  Two-way (competition + athlete) cluster-robust standard errors in parentheses, t with G$-$1 degrees of freedom, G the column's competitions; *** $p<0.01$, ** $p<0.05$, * $p<0.10$.
    \end{tablenotes}
  \end{threeparttable}
\end{table}

Table~\ref{tab:t61} reports the coefficients the counterfactual analysis uses. The model is an extension of the cross-panel model studied in Section~\ref{subsec:cross-panel}, fitted for each segment and with the compatriot premium separately estimated for each federation that had a judge at the discipline's Milan competition, although the table only reports the coefficients for federations with both a judge and a skater on the podium. The sample used for Table~\ref{tab:t61} includes Championships and Olympics where the judge panel is randomly drawn. 

\begin{table}[htbp]
  \centering
  \caption{Share of the Possible Panels on Which the French Couple Leads the Segment}\label{tab:t62}
  \begin{threeparttable}
  \makebox[\textwidth]{\begin{adjustbox}{max width=\textwidth}\begin{tabular}{>{\raggedright\arraybackslash}p{0.25\linewidth}cccc}
\toprule
& \multicolumn{2}{c}{\textbf{No shock, $\sigma = 0$}} & \multicolumn{2}{c}{\textbf{$\sigma$ of the event's panels}} \\
\cmidrule(lr){2-3}\cmidrule(lr){4-5}
\textbf{Which of the two judges the draw seats} & \textbf{(1) Rhythm dance} & \textbf{(2) Free dance} & \textbf{(3) Rhythm dance} & \textbf{(4) Free dance} \\
\midrule
Neither judge seated & 0.0\% & 81.8\% & 24.1\% & 64.2\% \\
& [$-$0.96, $-$0.22] & [$-$0.31, +0.66] & [$-$1.97, +0.78] & [$-$1.42, +2.03] \\
Only the French judge & 80.0\% & 98.8\% & 64.8\% & 85.0\% \\
& [$-$0.20, +0.64] & [+0.11, +1.47] & [$-$1.05, +1.65] & [$-$0.68, +2.72] \\
Only the American judge & 0.0\% & 43.0\% & 11.6\% & 48.7\% \\
& [$-$1.57, $-$0.77] & [$-$0.64, +0.63] & [$-$2.60, +0.40] & [$-$1.73, +1.62] \\
Both seated & 11.2\% & 83.9\% & 41.8\% & 74.5\% \\
& [$-$0.80, +0.09] & [$-$0.23, +1.32] & [$-$1.67, +1.30] & [$-$1.02, +2.31] \\
Every panel & 23.6\% & 77.8\% & 38.8\% & 70.2\% \\
& [$-$1.36, +0.44] & [$-$0.42, +1.33] & [$-$2.02, +1.30] & [$-$1.25, +2.33] \\
Every panel, both judges struck & 3.9\% & 70.1\% & 26.0\% & 61.8\% \\
& [$-$1.11, $-$0.03] & [$-$0.53, +0.88] & [$-$2.15, +0.94] & [$-$1.66, +2.23] \\
\bottomrule
\end{tabular}
\end{adjustbox}}
    \begin{tablenotes}[flushleft]
      \footnotesize
      \item \textit{NOTE:} The percentage number is the share of the panels in the redraws on which the French couple leads the segment. The interval is the 5th and the 95th percentiles of the French couple winning margin over the best other couple, in points. The last row drops out both French and American judges from every panel in the redraws, which is not a case the current rules allow for.
    \end{tablenotes}
  \end{threeparttable}
\end{table}

Table~\ref{tab:t62} reports how often the French couple leads each segment among all the panel draws in the resampling exercise. Columns (1) and (2) have no shock, or $\sigma = 0$, so that each panel $P^* \in \mathcal{P}$ gives one score per couple. Columns (3) and (4) add the shock, and each panel $P^* \in \mathcal{P}$ is repeated drawn 1,000 times. In the table, the percentage number is the share of the panels for which the French couple leads the segment, so 100 means that they lead on every panel in the group and 0 that they lead on none. The bracket beneath it gives the 5th and the 95th percentiles of the French couple's winning margin, across all panel redraws. 

Seating the French judge turns out to raise the French couple's margin by about 0.8 points in the rhythm dance and 0.6 points in the free dance on average over the panels without the shock, and by about 0.9 and 0.7 points with it. Without the shock, the French couple leads the rhythm dance on 80.0\% of the panels that seat the French judge alone, on 11.2\% of those that seat both judges, and on none of those that seat neither judge or the American judge alone. With the shock, the shares are 64.8\%, 41.8\%, 24.1\% and 11.6\%, for the panels that seat the French judge alone, both judges, neither judge and the American judge alone. The free dance is less close, and the French couple leads with a higher share in the counterfactual analysis. The last row of Table~\ref{tab:t62} drops both judges from every panel on which they are drawn. The French couple then trails in the rhythm dance but leads the free dance in most of the panel redraws.

\begin{table}[htbp]
  \centering
  \caption{Counterfactual Analysis of All Four disciplines in Milan 2026 Olympic Games}\label{tab:t63}
  \begin{threeparttable}
  \makebox[\textwidth]{\begin{adjustbox}{max width=\textwidth}\begin{tabular}{lrcccccc}
\toprule
&  & \multicolumn{6}{c}{\textbf{Champion changes (\%)}} \\
\cmidrule(lr){3-8}
&  & \multicolumn{3}{c}{\textbf{Each discipline's coefficients}} & \multicolumn{3}{c}{\textbf{Men, women and pairs pooled}} \\
\cmidrule(lr){3-5}\cmidrule(lr){6-8}
&  & \textbf{(1) No shock,} & \textbf{(2) $\sigma$ of the} & \textbf{(3) $\sigma$ pooled over} & \textbf{(4) No shock,} & \textbf{(5) $\sigma$ of the} & \textbf{(6) $\sigma$ pooled over} \\
\textbf{Discipline} & \textbf{Panel pairs} & \textbf{$\sigma = 0$} & \textbf{event's panels} & \textbf{Table~\ref{tab:t61}'s rows} & \textbf{$\sigma = 0$} & \textbf{event's panels} & \textbf{Table~\ref{tab:t61}'s rows} \\
\midrule
Ice dance & 90,090 & 57.2 & 44.8 & 45.5 & -- & -- & -- \\
Men & 18,480 & 0.0 & 0.0 & 0.0 & 0.0 & 0.0 & 0.0 \\
Women & 90,090 & 19.7 & 25.5 & 26.8 & 24.2 & 28.2 & 29.2 \\
Pairs & 90,090 & 0.0 & 0.0 & 0.0 & 0.0 & 0.0 & 0.0 \\
\bottomrule
\end{tabular}
\end{adjustbox}}
    \begin{tablenotes}[flushleft]
      \footnotesize
      \item \textit{NOTE:} Every possible panel is scored judge by judge: a judge who sat keeps her real mark, moved by the change in her fitted response between the two panels; a judge the real panel did not have enters at the skater's mean residual mark from the real panel's judges who are not from one of the event's top three's federations, plus her fitted response, plus a random shock drawn from $N(0,\sigma^2)$ where $\sigma$ is a measure of within-panel score spread. The marks are then trimmed and the field re-ranked. Judge pool sizes is 13 for ice dance, 12 for men, 13 for women, and 13 for pairs.
    \end{tablenotes}
  \end{threeparttable}
\end{table}

Table~\ref{tab:t63} moves from the segment to the event and reports the counterfactual analysis results for all four disciplines. For counterfactual analysis of the event, one also needs to take into account the fact that the second segment's start order depends on the first, and the start order enters the regression both as a control and as an input for the competitor definition. Therefore, for each redrawn segment panel pairs, we re-form the second-segment start order from the counterfactual first-segment ranking, recompute the rival indicators, and adjust each skater's counterfactual second-segment scores with respect to these changes. 

The counterfactual analysis shows that, in ice dance, the top placement changes in 45-57\% of the 90,090 admissible pairs of panels, with the exact number depending on how the shock is handled. In the women's event the champion changes in 20\%-27\% of the redraws. Part of these changes would occur without any bias, because results of close competitions are probabilistic. In the first, pure-signal analysis, with the same shocks, honest disagreement alone changes the champion in 32-34\% of the ice dance redraws and in 14-19\% of the women’s. In the men's and pairs' events, no admissible panel redraw changes the top placement as the champions in these two events won by more than ten points. Table~\ref{tab:t63} also reports a robustness check that pools the estimation over the men's, women's, and pairs' events for more precise estimates, since these disciplines share the same judging license and are more likely to have the same judges. 

Overall, the draw of the panel matters for the result of close competitions. With judge biases, the draw decides which federations have a judge on each panel, and the compatriot premium, the competitor penalty, and the colleague-compatriot premium then move the result. We suggest keeping the federations of the leading skaters off the judging panel when feasible. This rule is already adopted for the technical panel at the Olympic Winter Games. In the individual events of the three Olympic Winter Games in our sample, no technical controller, technical specialist or referee came from a medalist's federation, in line with the ISU rule that no member of the technical panel may serve in a discipline if skaters of their own federation are expected to place in the top five of the Games. If keeping the federations of the leading skaters off the judging panel is not feasible, an alternative could be to seat the judges of the contending federations symmetrically, either all of them on a segment or none of them. The trimmed mean offers no alternative. As discussed above and in Appendix~\ref{app:trimming}, trimming removes only a small part of a typical judge's bias.

\section{Conclusion} \label{sec:conclusion}

In the data analysis above, we find that targeted scoring exists both in the compatriot premium and in the competitor penalty, which provide evidence for strategic nationalistic bias. The compatriot premium falls along a randomly drawn start order and rises when the compatriot's close rivals have their own judges on the panel. Judges also mark down close rivals who skate before their compatriot and close rivals who skates after while the compatriot skater is still in contention, and the penalty grows with the compatriot's chance of a good placement. We also find that the judge gives more compatriot premium when more of her compatriot's close rivals have a compatriot judge on the panel, and that the premium rises further when those judges are known to be biased in past competitions. In the counterfactual analysis of 2026 Milan Winter Olympic Games, we find that, in close events, the compatriot premium, the competitor penalty, and the colleague compatriot premium influence competition results through the random draw of the panel composition. The ISU's trimming rule does not offer much protection as it removes only a small part of a compatriot premium or a competitor penalty of a typical size. We recommend keeping the federations of the leading skaters off the judging panels where feasible, or seating their judges symmetrically, either all of them on a segment or none of them. 

A general remedy often proposed for judge bias is to let artificial intelligence judge. At the 2026 Artistic Gymnastics World Championships in Rotterdam, the International Gymnastics Federation used an AI-based judging support system to aid judge scoring. The system recognizes the elements a gymnast performs, on which the difficulty score is built. The federation's president has said that its long-term aim is for computers and AI to award all the scores\footnote{Morinari Watanabe, interview with The Athletic on 4 November 2026.}. Figure skating, in our view, is different from artistic gymnastics. Skaters perform in a large space at high speed, and much of judges' decisions rest on interpretation. Machines could probably take over some of the technical panel's calls, such as determining the rotation of a jump or the difficulty level of a spin, but the marks on grades of execution and program components should, in our view, remain the judges', even if unconscious bias rooted from taste or cultural differences cannot be avoided: the styles of skating, jumping and spinning, and in pairs of lifts and throws, as well as the music and the choreography, differ widely from one skater to the other, and this diversity is much of what defines the sport.

Over the ten seasons we study, the compatriot premium fell by 44-51\% (Section~\ref{sec:descriptive}), and its two sharp declines, in 2018/19 and 2021/22, coincide with the ISU's sanctions of judges for nationalistic bias and with its decision of ramping up statistical screening of judges' scores for national bias. We therefore suggest building on this approach: using statistical tools and judges' marking records to monitor their behavior and to find ways to reduce the intentional and strategic parts of the compatriot premium, rather than handing the judging to machines. The ISU should make sure to extend such monitoring efforts to the competitor penalty, which we show is strategic and targeting to increase the compatriot's chance of a good placement. Because we identify close rivals by the start order, which identifies a compatriot's competitors with noise, our estimates of the penalty are lower bounds. The true competitor penalty is, if anything, larger and therefore deserves careful investigations.

\clearpage
\theendnotes               

\clearpage
\bibliography{references}

\clearpage
\appendix

\pagenumbering{arabic} 
\setcounter{table}{0}
\renewcommand{\thetable}{A.\arabic{table}}
\setcounter{figure}{0}
\renewcommand{\thefigure}{A.\arabic{figure}}

\section{Official Protocols and Panel of Judges Documents}
\label{app:protocol}

This appendix works through the scoring rules described in Section 2.1 using the example of one performance.

\begin{figure}[H]
 \centering
 \includegraphics[width=0.9\linewidth]{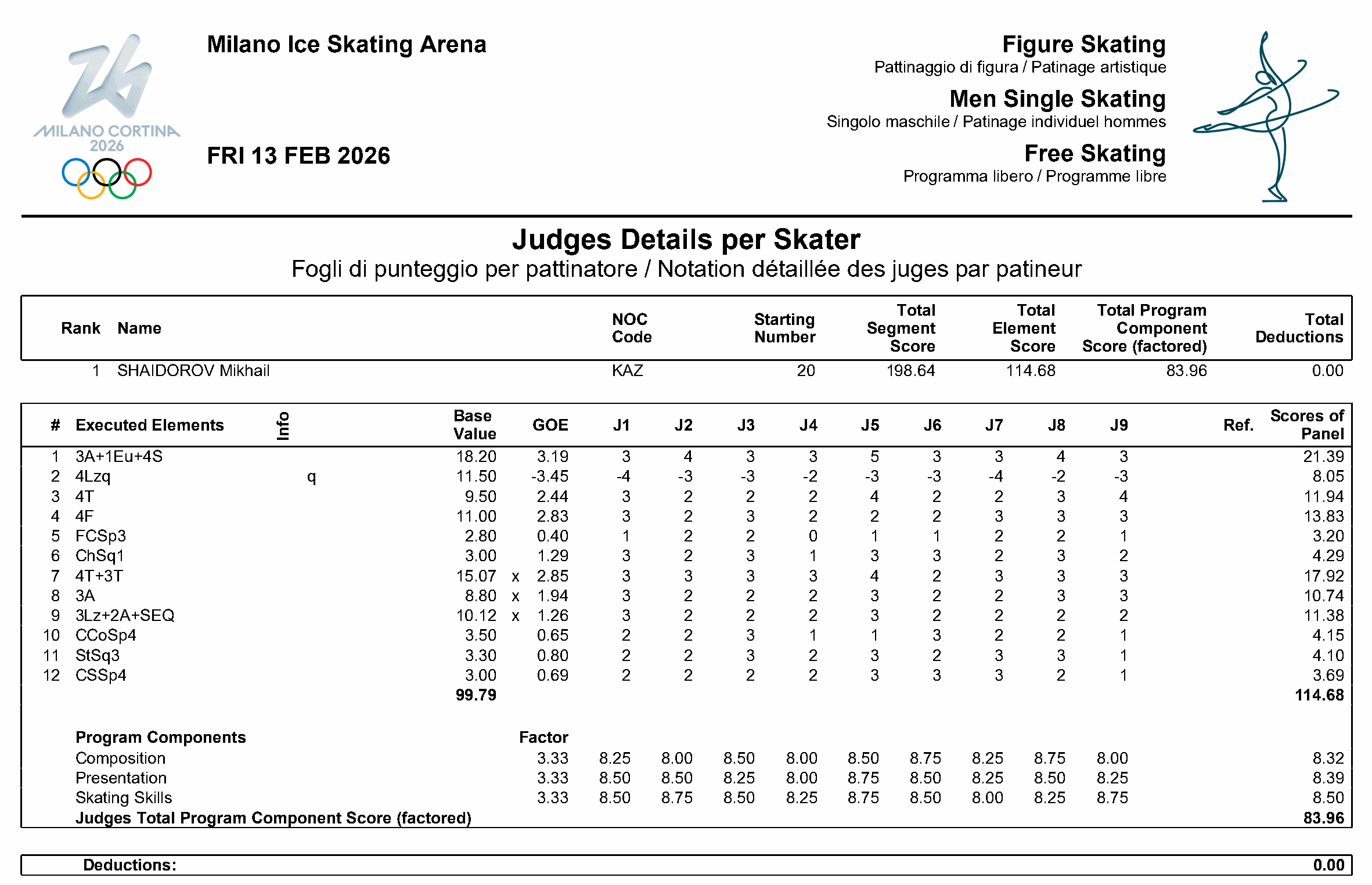}
 \caption{The official protocol for one performance}\label{fig:protocol}
 \begin{minipage}{0.95\linewidth}
  \footnotesize
  \textit{NOTE:} Judges Details per Skater: Men Single Skating, Free Skating, 2026 Olympic Winter Games (Milano Cortina).
 \end{minipage}
\end{figure}

Figure~\ref{fig:protocol} shows the official free skate protocol of Mikhail Shaidorov, the gold medalist in men's singles at the 2026 Olympic Winter Games in Milan. The header reports the TES (114.68), the PCS (83.96) and Deductions (0.00), which together give the TSS: $114.68+83.96-0.00=198.64$. A skater's final ranking in a category is determined by the sum of the two segment scores. Shaidorov was fifth after the short program (92.94) and won the free skate, taking the gold medal with 291.58.

The judges's execution marks are shown in the columns J1-J9 in the upper block of the protocol. For example, the grades for the quadruple flip (``4F'') are $\{3,2,3,2,2,2,3,3,3\}$. With one 3 and one 2 deleted, the trimmed mean of this element is $18/7=2.57$. The GOE column converts grades into points at 10\% of the base value per grade: $2.57\times1.10=2.83$. For a combination the scale is the BV of its most valuable jump, without any bonus. For ``3A+1Eu+4S'' that jump is the quadruple Salchow (9.70), so the GOE is $3.29\times0.97=3.19$. The Scores of Panel column adds the BV and the GOE, e.g.\ $11.00+2.83=13.83$ for the 4F. Summing over the twelve elements gives TES$=99.79+14.89=114.68$.

The program component marks are shown in the columns J1-J9 as well but in the lower block. Each component's trimmed mean is multiplied by the segment's factor (3.33 in the men's free skate) and the products are summed: $8.32\times3.33+8.39\times3.33+8.50\times3.33=27.71+27.94+28.31=83.96$. The PCS therefore consists of the judges' marks alone.

On the protocol the technical panel's calls appear under \textit{Executed Elements}, in the \textit{Info} column, and in the \textit{Base Value} column. In Figure~\ref{fig:protocol}, 4Lz is a quadruple Lutz with BV 11.50. The flag ``q'' on the 4Lz marks a jump landed a quarter short in revolution: it keeps its full base value of 11.50, but the judges are expected to reduce their GOE, and all nine grades are negative. If a 4Lz is under-rotated for over a quarter but less than a half of revolution, it will be marked ``<'' with BV discounted by 20\%. If it is called downgraded (``$<<$''), the 4Lz would be valued as a triple Lutz with BV 5.90. Level calls work the same way. The flag ``x'' marks jumps that earn the 10\% bonus for the last three jump elements executed in the second half of the program, e.g.\ $(9.50+4.20)\times1.10=15.07$ for ``4T+3T''. ``SEQ'' marks a jump sequence and ``ChSq1'' is the choreographic sequence, which carries a fixed base value of 3.00 and 0.50 points per GOE grade. The sum of base values is, therefore, set by the technical panel alone. Its calls affect the GOE both directly, as the basis for the multiplication, and indirectly, through their influences on judges' execution markings.

On the protocol displayed above, judges are not named. A separate Panel of Judges document names them, so that we can link each name to a federation. For example, for Shaidorov's free skate performance at the 2026 Olympic Winter Games, J8 was from Kazakhstan, Shaidorov's own federation. Judges' names are further linked to their federation. On GP, JGP and World Team Trophy Panel of Judges pages, federations are printed beside judges' names. On the pages of the World Championships and the Olympic and Youth Olympic Games, federations are not printed. We then link these judges to federations using pages that print the same judge's federation in the same season, then in any other season if she represented a single federation throughout, and otherwise the ISU's seasonal List of Officials, which also dates the eight judges who changed federation within our sample.

\section{The Impact of Trimming on Score} \label{app:trimming}

Suppose the panel includes nine judges, and judge $j$'s score follows the model $y_j = \lambda +a C_j +u_j$ where $u_j \sim i.i.d.\ N(0, \sigma^2)$ and $C_j = 1$ for the compatriot judge and 0 otherwise. The model assumes that the non-compatriot judges scores reflect an honest judgment on the performance quality $\lambda$ plus judge-specific idiosyncratic shocks $u_j$. The compatriot adds a premium $a$. The larger the premium relative to the shocks, the more likely the compatriot judge's score is the highest on the panel and gets trimmed off. Let judge 1 be the compatriot judge and normalize $\lambda=0$ and $\sigma=1$, so that $a$ is measured in units of $\sigma$. In a nine-judge panel, the expected trimmed average is
\begin{align*}
    \mathbb{E}[T(a)]\equiv (a - \mathbb{E}[\max(a+u_1, M')]-\mathbb{E}[\min(a+u_1, m')])/7,
\end{align*}
where $M'=\max(u_2,\cdots,u_9)$ and $m'=\min(u_2,\cdots,u_9)$; $\mathbb{E}[T(0)] = \lambda =0$ by symmetry. The average pass-through of the premium in the trimmed average (relative to an untrimmed average), is 
\begin{align*}
    \pi_a(a)=\frac{\mathbb{E}[T(a)]-\mathbb{E}[T(0)]}{a/9}=\frac{9}{7a}(a- \mathbb{E}[\max(a+u_1, M')]-\mathbb{E}[\min(a+u_1, m')]).
\end{align*}

Meanwhile, the marginal pass-through rate is the response of the expected trimmed average to marginal change of the premium, so
\begin{align*}
    \pi_m(a)&= \frac{d \mathbb{E}[9 T(a)]}{d a}=\frac{9}{7}\left(1-\mathbb P(a+u_1>M')-\mathbb P(a+u_1<m')\right)\\
    &=\frac{9}{7}\left(1-\int \phi(x-a)\Phi(x)^8dx -\int \phi(x-a)(1-\Phi(x))^8dx\right),
\end{align*}
where $\phi(.)$ and $\Phi(.)$ are the standard normal probability and cumulative distribution functions.

\begin{figure}[htbp]
  \centering
\includegraphics[width=0.9\linewidth,height=0.72\textheight,keepaspectratio]{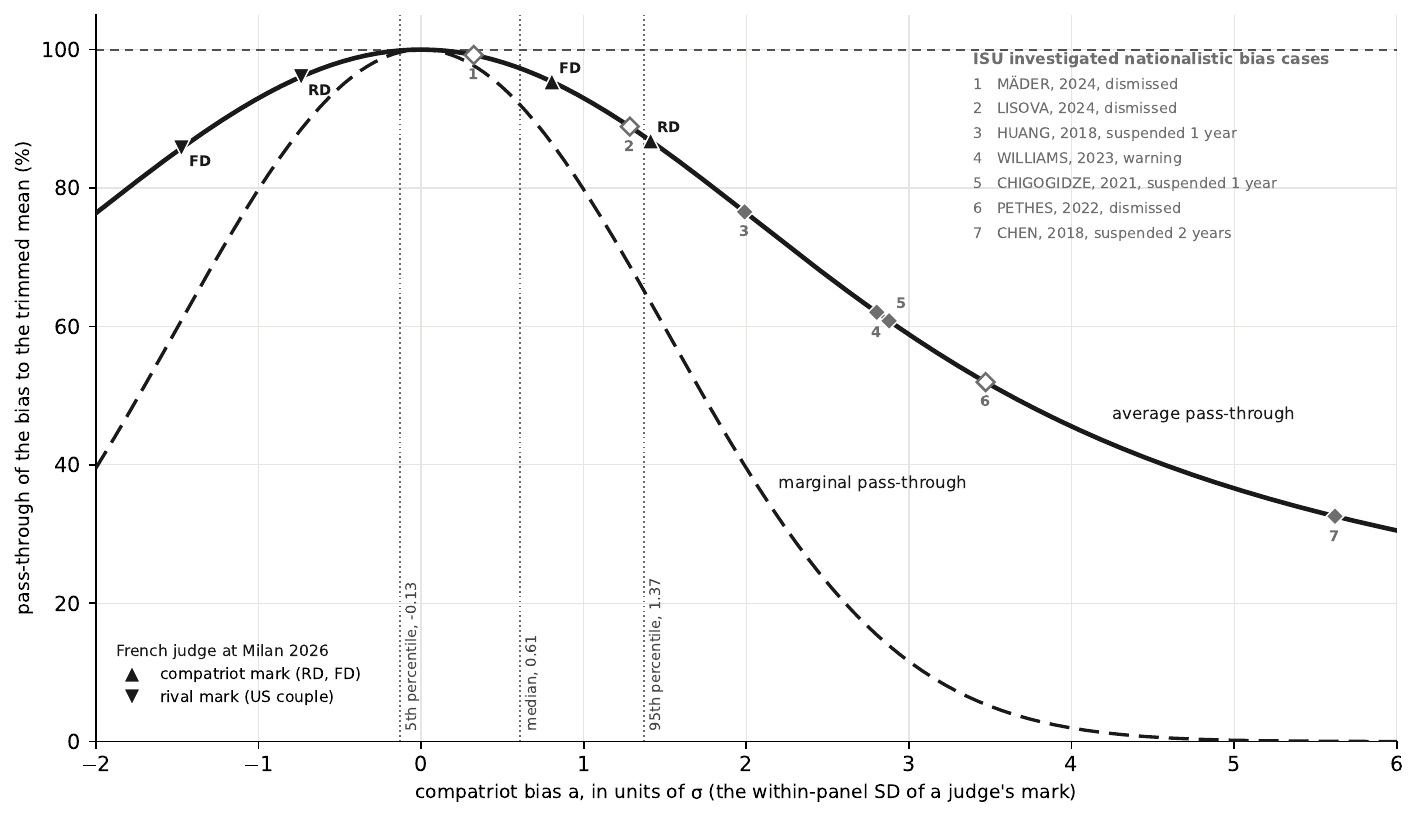}
  \caption{Pass-through of Compatriot Bias Under Trimming in a Nine-judge Panel}\label{fig:f5}
  \begin{minipage}{0.95\linewidth}
    \footnotesize
    \textit{NOTE:} The solid curve is the average pass-through rate and the dashed curve is the marginal pass-through rate. Both are functions of the premium $a$ in the unit of the within-panel standard deviation $\sigma$. The dotted vertical lines are 5th, 50th, and 95th percentiles of the judge-specific compatriot premium estimates. The French judge at the Milan 2026 ice dance event, in the rhythm dance (RD) and the free dance (FD), her mark on the French couple (pointing up) and on the American couple (pointing down), is her score minus the mean of the other judges' scores on the same performance minus gap on the non-compatriot skaters at the event (her own generosity at the event), normalized by the within-panel standard deviation. The seven ISU investigations marked on the graph: HUANG and CHEN were suspended for nationalistic bias at the 2018 Olympics; CHIGOGIDZE was suspended for nationalistic bias at the 2021 World Championships; WILLIAMS received a warning for the 2023 World Championships; PETHES, LISOVA and M\"ADER's cases were dismissed. 
  \end{minipage}
\end{figure}

Both functions are plotted in Figure~\ref{fig:f5} against $a/\sigma$, the premium in the unit of within-panel standard deviation. A pass-through close to one means that the premium moves the score as much as it would do without the trimming, so the trimming makes little difference. As a benchmark, Figure~\ref{fig:f5} marks the 5th, 50th, and 95th  percentiles of the judge-specific compatriot premium estimates of Figure~\ref{fig:f4}, re-estimated with outcome scores normalized by the within-panel standard deviation ($\sigma$) of its cell (age tier $\times$ discipline $\times$ program $\times$ season). Trimming removes little of a typical judge's compatriot premium: at the median, about 97\% of the premium reaches the trimmed score. Even at the 95th percentile, the average pass-through rate is still 87.5\%. Figure~\ref{fig:f5} also marks, in the same units, the compatriot bias of each ISU investigated cases as well as French judge's compatriot premium and rival penalty in each segment of the Milan ice dance event. The French judge's marks on the French couple and on the American couple pass through at 86\% or more. Even among the ISU-investigated cases, trimming removes a substantial share of the bias only at the extreme. 

\section{Supplementary Analysis}
\label{app:supp}

\subsection{Data Description}
\label{app:hetero}

Table~\ref{tab:tac1} estimates equation~\eqref{eq:baseline} on each season subsample and reports the compatriot premium season by season. Results of Table~\ref{tab:tac1} are reported in Figure~\ref{fig:f21a} in Section~\ref{sec:descriptive}. The seasons fall roughly into three periods: 2016/17 to 2017/18, 2018/19 to 2020/21, and 2021/22 to 2025/26. The premium is steady within each period and steps down in between: the TSS premium averages 2.43 percent in the first period, 1.92 in the second (COVID-impacted 2020/21 left out) and 1.44 in the third. 

\begin{table}[htbp]
  \centering
  \caption{Compatriot Premium by Season}\label{tab:tac1}
  \begin{threeparttable}
  \makebox[\textwidth]{\begin{adjustbox}{max width=\textwidth}\begin{tabular}{lrrr}
\toprule
\textbf{Season} & \textbf{(1) TSS pct} & \textbf{(2) GOE pct} & \textbf{(3) PCS pct} \\
\midrule
2016/17 & 2.465*** & 4.103*** & 2.724*** \\
& (0.127) & (0.241) & (0.123) \\
2017/18 & 2.390*** & 4.392*** & 2.609*** \\
& (0.124) & (0.249) & (0.125) \\
2018/19 & 1.887*** & 2.807*** & 1.969*** \\
& (0.115) & (0.174) & (0.130) \\
2019/20 & 1.962*** & 2.897*** & 2.060*** \\
& (0.125) & (0.205) & (0.115) \\
2020/21 & 2.304 & 3.576 & 2.077 \\
& (--) & (--) & (--) \\
2021/22 & 1.325*** & 2.112*** & 1.302*** \\
& (0.088) & (0.141) & (0.096) \\
2022/23 & 1.447*** & 1.906*** & 1.532*** \\
& (0.145) & (0.176) & (0.184) \\
2023/24 & 1.549*** & 2.211*** & 1.510*** \\
& (0.131) & (0.166) & (0.145) \\
2024/25 & 1.482*** & 2.048*** & 1.517*** \\
& (0.156) & (0.190) & (0.166) \\
2025/26 & 1.375*** & 2.049*** & 1.329*** \\
& (0.131) & (0.166) & (0.135) \\
Equal premium in every season, $F$(9, 171) & 11.04 (p < 0.001) & 13.52 (p < 0.001) & 14.87 (p < 0.001) \\
Performance FE & $\checkmark$ & $\checkmark$ & $\checkmark$ \\
Judge FE & $\checkmark$ & $\checkmark$ & $\checkmark$ \\
\bottomrule
\end{tabular}
\end{adjustbox}}
    \begin{tablenotes}[flushleft]
      \footnotesize
      \item \textit{NOTE:} Two-way (competition + athlete) cluster-robust standard errors in parentheses, $t$ with $G-1$ degrees of freedom, $G$ the column's competitions; *** $p<0.01$, ** $p<0.05$, * $p<0.10$.
    \end{tablenotes}
  \end{threeparttable}
\end{table}

Table~\ref{tab:tac2} estimates equation~\eqref{eq:baseline} on each discipline by segment subsample and reports the compatriot premium of each cell. Results of Table~\ref{tab:tac2} are reported in Figure~\ref{fig:f21b} in Section~\ref{sec:descriptive}. The discipline differences are large. Ice dance carries a TSS premium about twice the size of men's and women's, and the point estimates follow the ordering of men $<$ women $<$ pairs $<$ ice dance within each program on both TSS and GOE. On PCS, the three disciplines with jumps lie between 1.4-1.8 percent, while ice dance is above 2.5.

\begin{table}[htbp]
  \centering
  \caption{Compatriot Premium by Discipline $\times$ Segment}\label{tab:tac2}
  \begin{threeparttable}
  \makebox[\textwidth]{\begin{adjustbox}{max width=\textwidth}\begin{tabular}{lrrr}
\toprule
\textbf{Cell} & \textbf{(1) TSS pct} & \textbf{(2) GOE pct} & \textbf{(3) PCS pct} \\
\midrule
Men SP & +1.187*** & +2.040*** & +1.376*** \\
& (0.093) & (0.160) & (0.121) \\
Men FS & +1.407*** & +2.077*** & +1.761*** \\
& (0.094) & (0.150) & (0.121) \\
Women SP & +1.525*** & +2.322*** & +1.678*** \\
& (0.089) & (0.141) & (0.114) \\
Women FS & +1.461*** & +2.191*** & +1.624*** \\
& (0.089) & (0.134) & (0.114) \\
Pairs SP & +1.811*** & +2.758*** & +1.712*** \\
& (0.140) & (0.215) & (0.153) \\
Pairs FS & +1.724*** & +2.608*** & +1.697*** \\
& (0.118) & (0.194) & (0.133) \\
IceDance RD & +2.714*** & +4.168*** & +2.606*** \\
& (0.097) & (0.200) & (0.108) \\
IceDance FD & +2.665*** & +4.231*** & +2.491*** \\
& (0.107) & (0.216) & (0.108) \\
N (TSS), per cell (total) & 14,071--38,475 (206,983) &  &  \\
Equal premium in the eight cells, $F$(7, 171) & 31.68 (p < 0.001) & 19.75 (p < 0.001) & 14.35 (p < 0.001) \\
\bottomrule
\end{tabular}
\end{adjustbox}}
    \begin{tablenotes}[flushleft]
      \footnotesize
      \item \textit{NOTE:} Two-way (competition + athlete) cluster-robust standard errors in parentheses, $t$ with $G-1$ degrees of freedom, $G$ the column's competitions; *** $p<0.01$, ** $p<0.05$, * $p<0.10$.
    \end{tablenotes}
  \end{threeparttable}
\end{table}

Table~\ref{tab:tac3} estimates the compatriot premium separately for each program component score before and after the number of components changed to three. In both periods,  the premium is smallest on skating skills and largest on composition, except for pairs in 2016/17-2021/22, where it is slightly larger on interpretation. Across disciplines, ice dance has the largest premium in every component under both component systems. Table~\ref{tab:tac4} groups the premium in singles' GOE scores by element type. Step and choreographic sequences carry the largest premium in all four categories, from 2.354 (men's short program) to 3.021 (men's free skate).

\begin{table}[htbp]
  \centering
  \caption{Compatriot Premium of Different Components}\label{tab:tac3}
  \begin{threeparttable}
  \makebox[\textwidth]{\begin{adjustbox}{max width=\textwidth}\begin{tabular}{lrrrr}
\toprule
\textbf{Component} & \textbf{(1) Men} & \textbf{(2) Women} & \textbf{(3) Pairs} & \textbf{(4) Ice dance} \\
\midrule
\multicolumn{5}{l}{\textit{2016/17--2021/22: five components}} \\
Skating skills & +1.293*** & +1.483*** & +1.572*** & +2.550*** \\
& (0.138) & (0.120) & (0.139) & (0.131) \\
Transitions & +1.885*** & +2.127*** & +2.160*** & +3.044*** \\
& (0.157) & (0.129) & (0.168) & (0.145) \\
Performance & +1.517*** & +1.666*** & +2.038*** & +2.942*** \\
& (0.150) & (0.143) & (0.152) & (0.144) \\
Composition & +2.078*** & +2.234*** & +2.428*** & +3.175*** \\
& (0.155) & (0.128) & (0.157) & (0.135) \\
Interpretation & +1.872*** & +2.058*** & +2.466*** & +3.024*** \\
& (0.169) & (0.143) & (0.172) & (0.156) \\
N & 33,275 & 39,287 & 15,976 & 24,456 \\
\midrule
\multicolumn{5}{l}{\textit{2022/23--2025/26: three components}} \\
Composition & +1.793*** & +1.789*** & +1.556*** & +2.333*** \\
& (0.162) & (0.169) & (0.208) & (0.123) \\
Presentation & +1.291*** & +1.267*** & +1.243*** & +1.949*** \\
& (0.153) & (0.164) & (0.192) & (0.116) \\
Skating skills & +1.017*** & +0.883*** & +0.826*** & +1.826*** \\
& (0.149) & (0.133) & (0.199) & (0.113) \\
N & 27,545 & 33,571 & 13,036 & 19,837 \\
\midrule
Performance FE & $\checkmark$ & $\checkmark$ & $\checkmark$ & $\checkmark$ \\
Judge FE & $\checkmark$ & $\checkmark$ & $\checkmark$ & $\checkmark$ \\
\bottomrule
\end{tabular}
\end{adjustbox}}
    \begin{tablenotes}[flushleft]
      \footnotesize
      \item \textit{NOTE:} Two-way (competition + athlete) cluster-robust standard errors in parentheses, $t$ with $G-1$ degrees of freedom, $G$ the column's competitions; *** $p<0.01$, ** $p<0.05$, * $p<0.10$. Joint $F$ tests reject the equality of premiums in the four disciplines: in 2016/17-2021/22, $F(3, 95)$ = 20.95, 13.95, 22.29, 14.13 and 12.50 for skating skills, transitions, performance, composition and interpretation ($p<0.001$ each); in 2022/23-2025/26, $F(3, 75)$ = 5.79 ($p=0.001$), 7.51 and 14.13 ($p<0.001$) for composition, presentation and skating skills. Joint $F$ tests also reject the equality of premiums in the components: $F(4, G-1)$ = 29.02, 28.63, 18.53 and 23.48 in 2016/17-2021/22, and $F(2, G-1)$ = 29.82, 36.10, 14.60 and 18.21 in 2022/23-2025/26, for men, women, pairs and ice dance ($p<0.001$ each).
    \end{tablenotes}
  \end{threeparttable}
\end{table}

\begin{table}[htbp]
  \centering
  \caption{Compatriot Premium of Different Element Types in Singles}\label{tab:tac4}
  \begin{threeparttable}
  \makebox[\textwidth]{\begin{adjustbox}{max width=\textwidth}\begin{tabular}{lrrrr}
\toprule
\textbf{Elements} & \textbf{(1) Men SP} & \textbf{(2) Men FS} & \textbf{(3) Women SP} & \textbf{(4) Women FS} \\
\midrule
Solo jumps & +2.054*** & +1.757*** & +2.377*** & +1.788*** \\
& (0.196) & (0.153) & (0.184) & (0.147) \\
Jump combinations and sequences & +2.063*** & +2.269*** & +2.082*** & +2.383*** \\
& (0.229) & (0.192) & (0.190) & (0.166) \\
Spins & +1.880*** & +1.863*** & +2.230*** & +2.031*** \\
& (0.202) & (0.219) & (0.187) & (0.180) \\
Step and choreographic sequences & +2.354*** & +3.021*** & +2.595*** & +2.925*** \\
& (0.231) & (0.226) & (0.197) & (0.211) \\
N, the cell (the fewest of its groups) & 32,252 (32,081) & 28,568 (28,379) & 38,475 (38,165) & 34,383 (34,239) \\
\midrule
Performance FE & $\checkmark$ & $\checkmark$ & $\checkmark$ & $\checkmark$ \\
Judge FE & $\checkmark$ & $\checkmark$ & $\checkmark$ & $\checkmark$ \\
\bottomrule
\end{tabular}
\end{adjustbox}}
    \begin{tablenotes}[flushleft]
      \footnotesize
      \item \textit{NOTE:} Two-way (competition + athlete) cluster-robust standard errors in parentheses, $t$ with $G-1$ degrees of freedom, $G$ the column's competitions; *** $p<0.01$, ** $p<0.05$, * $p<0.10$. Joint $F$ tests reject the equality of premiums in the four columns for solo jumps ($F(3, 171) = 5.47$, $p = 0.001$) and for step and choreographic sequences ($F(3, 171) = 2.97$, $p = 0.033$), but not for jump combinations and sequences ($F(3, 171) = 1.68$, $p = 0.174$) or spins ($F(3, 171) = 1.02$, $p = 0.385$). Joint $F$ tests reject equality in the four groups in the free skating, for men ($F(3, 170) = 18.07$) and women ($F(3, 171) = 15.36$; $p<0.001$ both), but not at the 5 percent level in the short program, for men ($F(3, 171) = 1.68$, $p = 0.174$) or women ($F(3, 171) = 2.16$, $p = 0.095$).
    \end{tablenotes}
  \end{threeparttable}
\end{table}

\subsection{Performance  Controls}
\label{app:choreographer}

Table~\ref{tab:tac5-1} and~\ref{tab:tac5-2} report the slope estimates of the performance controls used in columns (4)-(9) of Table~\ref{tab:t32}. As Table~\ref{tab:tac5-1} shows, the career indicators affect more on the PCS than on the GOE. A prior Worlds placement raises the PCS by 3.2 percent of the scale and the GOE by 2.0, and a prior World Juniors placement raises the PCS. The first season in junior events lowers both the GOE and the PCS by similar amounts. Relative to popular music, film, stage and classical programs receive a PCS that is 0.2 to 0.3 percent higher. In the championships and Olympics sample, most career slopes lose significance, except the prior-placement effects on the PCS, and stage music. The start-order slope is large and significant everywhere, and it roughly doubles in championships.

\begin{table}[htbp]
  \centering
  \caption{Slope Estimates of the Performance Quality Controls}\label{tab:tac5-1}
  \begin{threeparttable}
  \makebox[\textwidth]{\begin{adjustbox}{max width=\textwidth}\begin{tabular}{lrrrrrr}
\toprule
 & \multicolumn{3}{c}{\textbf{All rows}} & \multicolumn{3}{c}{\textbf{Championships and Olympics}} \\
\cmidrule(lr){2-4}\cmidrule(lr){5-7}
\textbf{Variable} & \textbf{(4) TSS pct} & \textbf{(5) GOE pct} & \textbf{(6) PCS pct} & \textbf{(7) TSS pct} & \textbf{(8) GOE pct} & \textbf{(9) PCS pct} \\
\midrule
Prior Worlds placement & +2.757*** & +1.993** & +3.213*** & +0.949 & -0.691 & +1.478** \\
 & (0.729) & (0.790) & (0.530) & (0.852) & (1.080) & (0.566) \\
Prior Worlds best place (0 if none) & -0.062* & -0.004 & -0.076*** & +0.009 & +0.102** & -0.008 \\
 & (0.032) & (0.036) & (0.023) & (0.035) & (0.047) & (0.022) \\
Prior World Juniors placement & +1.346*** & +0.395 & +1.919*** & +0.365 & -0.970 & +1.087** \\
 & (0.509) & (0.587) & (0.376) & (0.700) & (0.982) & (0.407) \\
Prior World Juniors best place (0 if none) & -0.017 & +0.006 & -0.037** & +0.033 & +0.032 & -0.018 \\
 & (0.021) & (0.026) & (0.015) & (0.035) & (0.050) & (0.020) \\
First season in junior events & -1.778*** & -1.466*** & -1.589*** & -0.278 & -0.808 & -0.301 \\
 & (0.301) & (0.372) & (0.230) & (0.688) & (0.868) & (0.518) \\
First season in senior events & -0.451 & +0.197 & -0.540 & -0.036 & +0.654 & -0.174 \\
 & (0.533) & (0.646) & (0.388) & (0.682) & (0.954) & (0.411) \\
Music: film & +0.284* & +0.235 & +0.222** & +0.240 & +0.390 & +0.159 \\
 & (0.160) & (0.216) & (0.095) & (0.308) & (0.391) & (0.163) \\
Music: stage & +0.344 & +0.380 & +0.306** & +0.734* & +1.181* & +0.587** \\
 & (0.236) & (0.331) & (0.152) & (0.429) & (0.584) & (0.242) \\
Music: classical & +0.319 & +0.280 & +0.246* & +0.166 & +0.403 & +0.201 \\
 & (0.208) & (0.299) & (0.126) & (0.327) & (0.475) & (0.165) \\
Music: opera & -0.044 & +0.148 & +0.145 & +0.019 & -0.047 & +0.133 \\
 & (0.404) & (0.567) & (0.229) & (0.788) & (1.309) & (0.359) \\
Music: contemporary instrumental & +0.144 & +0.391 & +0.131 & +0.415 & +0.475 & +0.242 \\
 & (0.207) & (0.304) & (0.123) & (0.318) & (0.463) & (0.160) \\
Music: others & -0.016 & +0.086 & -0.016 & +0.669 & +1.719* & +0.512 \\
 & (0.347) & (0.545) & (0.239) & (0.653) & (0.958) & (0.431) \\
Music: mixed & +0.151 & +0.030 & +0.138 & +0.254 & -0.082 & +0.241* \\
 & (0.168) & (0.230) & (0.108) & (0.235) & (0.306) & (0.141) \\
$\mathrm{Start}_{p}$ & +3.863*** & +1.450*** & +5.562*** & +8.353*** & +4.545*** & +10.809*** \\
 & (0.312) & (0.363) & (0.263) & (0.582) & (0.751) & (0.386) \\
\midrule
The 30 choreographer indicators (shown apart) & $\checkmark$ & $\checkmark$ & $\checkmark$ & $\checkmark$ & $\checkmark$ & $\checkmark$ \\
Compatriot, Rival, Other-panel compatriot, Home ice, the technical panel & $\checkmark$ & $\checkmark$ & $\checkmark$ & $\checkmark$ & $\checkmark$ & $\checkmark$ \\
Compatriot $\times$ home ice & $\checkmark$ & $\checkmark$ & $\checkmark$ & $\checkmark$ & $\checkmark$ & $\checkmark$ \\
Other-segment compatriot &  &  &  & $\checkmark$ & $\checkmark$ & $\checkmark$ \\
Skater FE & $\checkmark$ & $\checkmark$ & $\checkmark$ & $\checkmark$ & $\checkmark$ & $\checkmark$ \\
Segment FE & $\checkmark$ & $\checkmark$ & $\checkmark$ & $\checkmark$ & $\checkmark$ & $\checkmark$ \\
Judge FE & $\checkmark$ & $\checkmark$ & $\checkmark$ & $\checkmark$ & $\checkmark$ & $\checkmark$ \\
N & 206,983 & 206,983 & 206,983 & 55,276 & 55,276 & 55,276 \\
\# of Competitions (G) & 172 & 172 & 172 & 30 & 30 & 30 \\
\bottomrule
\end{tabular}
\end{adjustbox}}
    \begin{tablenotes}[flushleft]
      \footnotesize
      \item \textit{NOTE:} Prior Worlds (World Juniors) placement means that the biography lists a placement at the championships in a season before; the best place is 0 where there is none. For music type dummies, the latent type is popular music, the most frequent type. Slopes of the top 10 choreographers in each Era are reported in Table~\ref{tab:tac5-2}. Two-way (competition + athlete) cluster-robust standard errors in parentheses, $t$ with $G-1$ degrees of freedom, $G$ the column's competitions; *** $p<0.01$, ** $p<0.05$, * $p<0.10$.
    \end{tablenotes}
  \end{threeparttable}
\end{table}

Table~\ref{tab:tac5-2} reports the slopes of the choreographer indicators in columns (4)-(9) of Table~\ref{tab:t32}. For each Era, we take the ten choreographers credited with the most programs in the Era, and create choreographer indicator, which equals one if the performance is instructed by one of the ten choreographers in the era. With skater fixed effects, each slope compares a skater's programs by that choreographer with her programs by choreographers outside the Era's top ten. In the full sample, none of the Era A choreographers has a significant slope on TSS pct. In Era B, programs by Benoit Richaud and Mark Pillay score 1.7 and 2.9 percent of the scale more on TSS pct, both significant on all three outcomes, and programs by Daniil Gleikhengauz score 3.1 percent more, significant at the 10 percent level. In Era C, programs by Benoit Richaud and Adam Solya score 1.7 and 2.2 percent more. Most of the other indicators are insignificant, and the few negative ones are marginally significant. In the Championships and Olympics sample, the estimates are larger and less precise with only 30 competitions, and in Era A they take both signs, from $-3.716$ for David Wilson to $+5.356$ for Julie Marcotte on TSS pct. Still, Benoit Richaud's programs score 2.3 percent more in Era B and Era C in this sample as well, the only choreographer with a positive and significant slope on TSS pct in both samples and both Eras. As we discuss in Section~\ref{subsec:cross-panel}, these coefficients are not necessarily causal, since a skater may choose a better-known choreographer in seasons when she is in better form.

\begin{table}[htbp]
  \centering
  \caption{Table~\ref{tab:t32} (continued) Slope Estimates of Top Choreographers Dummies}\label{tab:tac5-2}
  \begin{threeparttable}
  \makebox[\textwidth]{\begin{adjustbox}{max width=\textwidth}\begin{tabular}{lrrrrrr}
\toprule
 & \multicolumn{3}{c}{\textbf{All rows}} & \multicolumn{3}{c}{\textbf{Championships and Olympics}} \\
\cmidrule(lr){2-4}\cmidrule(lr){5-7}
\textbf{Variable} & \textbf{(4) TSS pct} & \textbf{(5) GOE pct} & \textbf{(6) PCS pct} & \textbf{(7) TSS pct} & \textbf{(8) GOE pct} & \textbf{(9) PCS pct} \\
\midrule
Era A \#1 choreographer: \textbf{David Wilson} & -0.587 & -0.431 & -0.000 & -3.716*** & -1.970 & -0.833 \\
 & (0.965) & (1.413) & (0.659) & (1.196) & (2.839) & (0.671) \\
Era A \#2 choreographer: \textbf{Benoit Richaud} & -0.219 & +0.481 & -0.099 & +1.307 & +0.551 & +1.129 \\
 & (1.171) & (1.155) & (0.740) & (1.391) & (2.431) & (0.800) \\
Era A \#3 choreographer: \textbf{Nikolai Morozov} & +0.423 & -0.314 & -0.407 & +2.437* & -0.026 & +0.626 \\
 & (1.022) & (1.281) & (0.619) & (1.325) & (1.862) & (0.726) \\
Era A \#4 choreographer: \textbf{Massimo Scali} & +0.287 & +1.391 & +0.122 & -3.016* & -3.125 & -1.569** \\
 & (1.341) & (1.794) & (1.099) & (1.480) & (1.982) & (0.752) \\
Era A \#5 choreographer: \textbf{Mark Pillay} & +1.572 & +0.613 & +0.650 & +1.462 & -0.623 & +1.112 \\
 & (1.487) & (1.781) & (0.883) & (2.359) & (2.790) & (1.260) \\
Era A \#6 choreographer: \textbf{Julie Marcotte} & +1.368 & +2.965 & -0.242 & +5.356*** & +8.710*** & +1.202 \\
 & (1.174) & (1.880) & (0.793) & (0.828) & (1.701) & (0.781) \\
Era A \#7 choreographer: \textbf{Shae-Lynn Bourne} & +1.459 & +0.582 & +1.288** & +0.964 & -1.328 & +1.282 \\
 & (1.103) & (1.429) & (0.652) & (3.210) & (4.814) & (1.323) \\
Era A \#8 choreographer: \textbf{Corrado Giordani} & -2.347 & -1.857 & -1.236 & -1.463 & -0.243 & -0.753 \\
 & (1.488) & (1.877) & (1.312) & (1.248) & (1.834) & (1.087) \\
Era A \#9 choreographer: \textbf{Marie-France Dubreuil} & +1.311 & +2.545 & +0.981 & +2.346 & +4.187** & +1.449 \\
 & (1.343) & (1.693) & (1.267) & (1.529) & (1.784) & (1.193) \\
Era A \#10 choreographer: \textbf{Valentin Molotov} & -2.222 & -1.486 & -1.324 & -0.641 & +3.910 & +0.428 \\
 & (2.069) & (1.870) & (1.887) & (2.451) & (2.582) & (0.983) \\
Era B \#1 choreographer: \textbf{Benoit Richaud} & +1.739** & +1.840** & +1.589*** & +2.304** & +2.102 & +1.218** \\
 & (0.743) & (0.920) & (0.553) & (0.962) & (1.384) & (0.592) \\
Era B \#2 choreographer: \textbf{Nikolai Morozov} & +1.043 & +1.434 & +0.507 & +2.189* & +2.610 & +0.956 \\
 & (0.973) & (1.281) & (0.797) & (1.211) & (1.956) & (1.022) \\
Era B \#3 choreographer: \textbf{Romain Haguenauer} & +0.514 & +0.785 & +0.519 & +2.059** & +2.676* & +1.820** \\
 & (0.692) & (0.975) & (0.628) & (0.935) & (1.356) & (0.726) \\
Era B \#4 choreographer: \textbf{Marie-France Dubreuil} & +0.604 & +1.770 & +0.541 & +1.496* & +2.779** & +0.717 \\
 & (0.879) & (1.151) & (0.736) & (0.855) & (1.255) & (0.707) \\
Era B \#5 choreographer: \textbf{Shae-Lynn Bourne} & +0.956 & +0.195 & +0.776 & -0.504 & -1.438 & +0.073 \\
 & (0.887) & (1.280) & (0.514) & (1.219) & (2.332) & (0.558) \\
Era B \#6 choreographer: \textbf{David Wilson} & +0.911 & -0.423 & +1.158** & -0.448 & -2.654* & +0.194 \\
 & (1.042) & (1.286) & (0.545) & (1.179) & (1.423) & (0.635) \\
Era B \#7 choreographer: \textbf{Pasquale Camerlengo} & -0.111 & -0.427 & +0.161 & -1.165 & -0.064 & -1.175 \\
 & (0.879) & (1.053) & (0.741) & (1.222) & (1.429) & (0.886) \\
Era B \#8 choreographer: \textbf{Daniil Gleikhengauz} & +3.137* & +1.565 & +1.757** & +2.691 & +1.815 & +1.113 \\
 & (1.856) & (1.892) & (0.760) & (1.985) & (2.474) & (0.703) \\
Era B \#9 choreographer: \textbf{Corrado Giordani} & +0.237 & +0.279 & +0.025 & +0.179 & +0.735 & -0.343 \\
 & (1.118) & (1.589) & (1.000) & (1.134) & (1.834) & (0.919) \\
Era B \#10 choreographer: \textbf{Mark Pillay} & +2.923*** & +3.341** & +2.173*** & +2.566* & +1.433 & +2.027** \\
 & (0.936) & (1.668) & (0.639) & (1.490) & (1.993) & (0.771) \\
Era C \#1 choreographer: \textbf{Benoit Richaud} & +1.717** & +0.832 & +1.652*** & +2.266** & +0.198 & +1.537** \\
 & (0.696) & (0.731) & (0.499) & (0.974) & (1.079) & (0.629) \\
Era C \#2 choreographer: \textbf{Adam Solya} & +2.212** & +1.968* & +1.617** & +2.148* & +2.185 & +1.389* \\
 & (1.021) & (1.117) & (0.770) & (1.052) & (1.695) & (0.754) \\
Era C \#3 choreographer: \textbf{Yeaji Shin} & -0.504 & -1.909* & -0.980* & -0.192 & -0.615 & -1.063 \\
 & (0.716) & (1.037) & (0.520) & (1.262) & (1.320) & (1.087) \\
Era C \#4 choreographer: \textbf{David Wilson} & -1.380 & -1.649 & -0.290 & -1.316 & -1.475 & -0.227 \\
 & (0.932) & (1.107) & (0.559) & (1.363) & (1.702) & (0.784) \\
Era C \#5 choreographer: \textbf{Marie-France Dubreuil} & -0.734 & -1.240 & -1.038 & -1.181 & -2.790* & -1.282 \\
 & (1.203) & (1.249) & (0.962) & (1.197) & (1.397) & (0.877) \\
Era C \#6 choreographer: \textbf{Shae-Lynn Bourne} & -0.922 & -0.714 & -0.836 & +1.329 & +0.476 & +0.002 \\
 & (1.092) & (1.031) & (0.619) & (1.156) & (0.968) & (0.439) \\
Era C \#7 choreographer: \textbf{Romain Haguenauer} & +1.030 & +0.312 & +0.919 & +1.318 & -0.073 & +1.113 \\
 & (0.828) & (1.190) & (0.733) & (1.203) & (1.772) & (0.917) \\
Era C \#8 choreographer: \textbf{Massimo Scali} & -0.830 & -1.274 & -0.669 & -0.547 & -0.229 & +0.088 \\
 & (0.685) & (0.883) & (0.513) & (0.970) & (1.129) & (0.599) \\
Era C \#9 choreographer: \textbf{Samuel Chouinard} & +0.562 & +2.002* & +0.454 & +0.532 & +2.221 & +0.554 \\
 & (1.069) & (1.206) & (0.832) & (1.114) & (1.445) & (0.961) \\
Era C \#10 choreographer: \textbf{Mark Pillay} & +1.279 & -0.062 & +0.624 & +3.566 & +2.151 & +1.720 \\
 & (1.440) & (1.591) & (0.969) & (2.142) & (2.109) & (1.201) \\
\midrule
The other components of $\tilde{Z}_{p}$ (shown apart) & $\checkmark$ & $\checkmark$ & $\checkmark$ & $\checkmark$ & $\checkmark$ & $\checkmark$ \\
Compatriot, Rival, Other-panel compatriot, Home ice, the technical panel & $\checkmark$ & $\checkmark$ & $\checkmark$ & $\checkmark$ & $\checkmark$ & $\checkmark$ \\
Compatriot $\times$ home ice & $\checkmark$ & $\checkmark$ & $\checkmark$ & $\checkmark$ & $\checkmark$ & $\checkmark$ \\
Other-segment compatriot &  &  &  & $\checkmark$ & $\checkmark$ & $\checkmark$ \\
Skater FE & $\checkmark$ & $\checkmark$ & $\checkmark$ & $\checkmark$ & $\checkmark$ & $\checkmark$ \\
Segment FE & $\checkmark$ & $\checkmark$ & $\checkmark$ & $\checkmark$ & $\checkmark$ & $\checkmark$ \\
Judge FE & $\checkmark$ & $\checkmark$ & $\checkmark$ & $\checkmark$ & $\checkmark$ & $\checkmark$ \\
N & 206,983 & 206,983 & 206,983 & 55,276 & 55,276 & 55,276 \\
\# of Competitions (G) & 172 & 172 & 172 & 30 & 30 & 30 \\
\bottomrule
\end{tabular}
\end{adjustbox}}
    \begin{tablenotes}[flushleft]
      \footnotesize
      \item \textit{NOTE:} Continuation of Table~\ref{tab:t32}, reporting slopes of the top 10 choreographers in each Era. Two-way (competition + athlete) cluster-robust standard errors in parentheses, $t$ with $G-1$ degrees of freedom, $G$ the column's competitions; *** $p<0.01$, ** $p<0.05$, * $p<0.10$.
    \end{tablenotes}
  \end{threeparttable}
\end{table}

\subsection{Alternative Definitions on Being in Contention and on Close Rivals}
\label{app:altdef}

Tables~\ref{tab:tac6} and~\ref{tab:tac7} reproduce columns (4)-(6) of Table~\ref{tab:t41} in Section~\ref{subsec:uncertainty} under alternative definitions of being in contention and of a close rival. In the baseline, the judge's own skater is in contention if she ranks in the top 3 among the skaters who have already skated in the segment, and in the second segment of an individual event, she also placed 1-8 after the first segment; a rival is defined as a skater from the judge's federation within three start order of the marked skater. Table~\ref{tab:tac6} uses the top 3 in both segments without the placement requirement (columns (1)-(3)), and the top 4 with places 1-10 after the first segment (columns (4)-(6)); since the first definition coincides with the baseline in the first segment, its columns (1) and (2) are the same as columns (4) and (5) of Table~\ref{tab:t41}. Table~\ref{tab:tac7} narrows the rival window to two start numbers (columns (1)-(3)) and widens it to four (columns (4)-(6)). Accordingly, the side of the rival, the in contention split and the compatriots' start position are adjusted as well. From these tables, we see the same pattern as we found in Section~\ref{subsec:uncertainty}. 

In the first segment, there is no significant penalty on a rival under any alternative definitions. In the second segment, a rival starting earlier is marked down by 0.21-0.25 percent of the scale (against $-0.232$ in Table~\ref{tab:t41}), and a rival starting later is marked down while the compatriot is still in contention (all significant at the 1\% level). Once the compatriot is out of contention, the penalty on a rival starting later is close to zero (between $-0.031$ and $+0.010$). The in-contention penalty is smaller under both alternative contention definitions, which are broader than the baseline and therefore include compatriots with weaker chances of a top placement. The compatriot premium (1.64-1.86 percent of the scale) and its decline along the start order are also unchanged. The slope on Compatriot $\times$ start position is significantly negative at the 1\% level in every column.

In the second segment, all three rival slopes along the compatriots' start position remain negative under every alternative definition. The slopes for a rival starting earlier and for a rival starting later while the compatriot is in contention are significant at the 1\% level in every column but one (5\%), and the in-contention slope is the steepest, as in the baseline. The out-of-contention slope, significant in the baseline, is marginally significant under the alternative rival windows. In the seeded first segments, the slopes are mostly smaller and less precisely estimated. In the open-draw first segments, a few rival slopes are significant, but  all at the 10\% level only.

Overall, the alternative definitions leave the main conclusions of Section~\ref{subsec:uncertainty} unchanged.

\begin{table}[htbp]
  \centering
  \caption{Table~\ref{tab:t41} Robustness Check: Alternative In Contention Definitions}\label{tab:tac6}
  \begin{threeparttable}
  \makebox[\textwidth]{\begin{adjustbox}{max width=\textwidth}\begin{tabular}{lrrrrrr}
\toprule
 & \multicolumn{3}{c}{\textbf{In contention: top 3 in both segments}} & \multicolumn{3}{c}{\textbf{In contention: top 4, Segment 2 places 1--10}} \\
\cmidrule(lr){2-4}\cmidrule(lr){5-7}
\textbf{Variable} & \textbf{(1) Open draw} & \textbf{(2) Seeded order} & \textbf{(3) Segment 2} & \textbf{(4) Open draw} & \textbf{(5) Seeded order} & \textbf{(6) Segment 2} \\
\midrule
Compatriot & +1.641*** & +1.863*** & +1.714*** & +1.643*** & +1.863*** & +1.711*** \\
 & (0.099) & (0.083) & (0.065) & (0.099) & (0.083) & (0.065) \\
Rival starting earlier & -0.044 & -0.027 & -0.223*** & -0.043 & -0.024 & -0.225*** \\
 & (0.046) & (0.038) & (0.029) & (0.047) & (0.037) & (0.029) \\
Rival starting later, in contention & -0.047 & +0.022 & -0.110*** & -0.045 & +0.013 & -0.166*** \\
 & (0.069) & (0.051) & (0.034) & (0.057) & (0.047) & (0.041) \\
Rival starting later, out of contention & -0.018 & -0.026 & -0.031 & -0.014 & -0.013 & +0.000 \\
 & (0.069) & (0.053) & (0.045) & (0.075) & (0.070) & (0.036) \\
Compatriot $\times$ start position & -0.697*** & -0.536*** & -0.697*** & -0.704*** & -0.525*** & -0.699*** \\
 & (0.192) & (0.185) & (0.128) & (0.190) & (0.184) & (0.129) \\
Rival starting earlier $\times$ compatriots' start position & -0.176 & -0.238 & -0.466*** & -0.174 & -0.242 & -0.485*** \\
 & (0.149) & (0.168) & (0.130) & (0.149) & (0.171) & (0.130) \\
Rival starting later, in contention $\times$ compatriots' start position & -0.434 & -0.419** & -0.717*** & -0.470* & -0.362** & -0.566*** \\
 & (0.269) & (0.205) & (0.148) & (0.254) & (0.181) & (0.179) \\
Rival starting later, out of contention $\times$ compatriots' start position & +0.176 & -0.642* & -0.666** & +0.385 & -1.603*** & -0.616*** \\
 & (0.240) & (0.340) & (0.261) & (0.281) & (0.435) & (0.173) \\
\midrule
$\mathrm{C}_{jp}$, $\mathrm{R}^{+}_{jp}$, $\mathrm{R}^{-,in}_{jp}$ and $\mathrm{R}^{-,out}_{jp}$ $\cdot$ $H_{p}$ & $\checkmark$ & $\checkmark$ & $\checkmark$ & $\checkmark$ & $\checkmark$ & $\checkmark$ \\
Performance FE & $\checkmark$ & $\checkmark$ & $\checkmark$ & $\checkmark$ & $\checkmark$ & $\checkmark$ \\
Judge FE & $\checkmark$ & $\checkmark$ & $\checkmark$ & $\checkmark$ & $\checkmark$ & $\checkmark$ \\
N & 46,208 & 62,918 & 97,857 & 46,208 & 62,918 & 97,857 \\
\# of Competitions (G) & 70 & 102 & 172 & 70 & 102 & 172 \\
Mass of negative weights, main regressors & 0.000 & 0.000 & 0.000 & 0.000 & 0.000 & 0.000 \\
Mass of negative weights, Compatriot $\times$ start position & 0.010 & 0.063 & 0.045 & 0.010 & 0.063 & 0.045 \\
Mass of negative weights, the three Rival $\times$ compatriots' start position & 0.012--0.034 & 0.050--0.066 & 0.045--0.064 & 0.012--0.030 & 0.049--0.087 & 0.036--0.064 \\
Share of negative weights, main regressors & 0\% & 0\% & 0\% & 0\% & 0\% & 0\% \\
Share of negative weights, Compatriot $\times$ start position & 10\% & 21\% & 21\% & 10\% & 22\% & 21\% \\
Share of negative weights, the three Rival $\times$ compatriots' start position & 12\%--18\% & 22\%--24\% & 21\%--22\% & 12\%--18\% & 21\%--24\% & 20\%--23\% \\
\bottomrule
\end{tabular}
\end{adjustbox}}
    \begin{tablenotes}[flushleft]
      \footnotesize
      \item \textit{NOTE:} Two-way (competition + athlete) cluster-robust standard errors in parentheses, $t$ with $G-1$ degrees of freedom, $G$ the column's competitions; *** $p<0.01$, ** $p<0.05$, * $p<0.10$. 
    \end{tablenotes}
  \end{threeparttable}
\end{table}

\begin{table}[htbp]
  \centering
  \caption{Table~\ref{tab:t41} Robustness Check: Alternative Close Rival Definitions}\label{tab:tac7}
  \begin{threeparttable}
  \makebox[\textwidth]{\begin{adjustbox}{max width=\textwidth}\begin{tabular}{lrrrrrr}
\toprule
 & \multicolumn{3}{c}{\textbf{Rival within $\pm$2 start numbers}} & \multicolumn{3}{c}{\textbf{Rival within $\pm$4 start numbers}} \\
\cmidrule(lr){2-4}\cmidrule(lr){5-7}
\textbf{Variable} & \textbf{(1) Open draw} & \textbf{(2) Seeded order} & \textbf{(3) Segment 2} & \textbf{(4) Open draw} & \textbf{(5) Seeded order} & \textbf{(6) Segment 2} \\
\midrule
Compatriot & +1.655*** & +1.860*** & +1.729*** & +1.648*** & +1.864*** & +1.704*** \\
 & (0.099) & (0.081) & (0.063) & (0.100) & (0.085) & (0.065) \\
Rival starting earlier & -0.058 & -0.023 & -0.247*** & -0.046 & -0.020 & -0.213*** \\
 & (0.052) & (0.037) & (0.030) & (0.044) & (0.039) & (0.028) \\
Rival starting later, in contention & +0.019 & +0.047 & -0.248*** & -0.021 & -0.002 & -0.234*** \\
 & (0.086) & (0.055) & (0.049) & (0.069) & (0.054) & (0.047) \\
Rival starting later, out of contention & -0.025 & -0.081 & +0.009 & +0.002 & +0.032 & +0.010 \\
 & (0.085) & (0.063) & (0.038) & (0.062) & (0.050) & (0.030) \\
Compatriot $\times$ start position & -0.710*** & -0.525*** & -0.687*** & -0.722*** & -0.519*** & -0.660*** \\
 & (0.189) & (0.187) & (0.127) & (0.194) & (0.185) & (0.128) \\
Rival starting earlier $\times$ compatriots' start position & -0.308* & -0.099 & -0.451*** & -0.269* & -0.207 & -0.283** \\
 & (0.174) & (0.180) & (0.142) & (0.137) & (0.159) & (0.126) \\
Rival starting later, in contention $\times$ compatriots' start position & -0.499* & -0.476** & -0.760*** & -0.452* & -0.324 & -0.802*** \\
 & (0.292) & (0.215) & (0.214) & (0.242) & (0.204) & (0.211) \\
Rival starting later, out of contention $\times$ compatriots' start position & +0.223 & -0.863* & -0.341* & -0.070 & -0.474* & -0.345** \\
 & (0.291) & (0.483) & (0.175) & (0.220) & (0.258) & (0.139) \\
\midrule
$\mathrm{C}_{jp}$, $\mathrm{R}^{+}_{jp}$, $\mathrm{R}^{-,in}_{jp}$ and $\mathrm{R}^{-,out}_{jp}$ $\cdot$ $H_{p}$ & $\checkmark$ & $\checkmark$ & $\checkmark$ & $\checkmark$ & $\checkmark$ & $\checkmark$ \\
Performance FE & $\checkmark$ & $\checkmark$ & $\checkmark$ & $\checkmark$ & $\checkmark$ & $\checkmark$ \\
Judge FE & $\checkmark$ & $\checkmark$ & $\checkmark$ & $\checkmark$ & $\checkmark$ & $\checkmark$ \\
N & 46,208 & 62,918 & 97,857 & 46,208 & 62,918 & 97,857 \\
\# of Competitions (G) & 70 & 102 & 172 & 70 & 102 & 172 \\
Mass of negative weights, main regressors & 0.000 & 0.000 & 0.000 & 0.000 & 0.000 & 0.000 \\
Mass of negative weights, Compatriot $\times$ start position & 0.010 & 0.062 & 0.045 & 0.010 & 0.063 & 0.045 \\
Mass of negative weights, the three Rival $\times$ compatriots' start position & 0.012--0.041 & 0.058--0.097 & 0.037--0.058 & 0.013--0.033 & 0.044--0.056 & 0.029--0.046 \\
Share of negative weights, main regressors & 0\% & 0\% & 0\% & 0\% & 0\% & 0\% \\
Share of negative weights, Compatriot $\times$ start position & 10\% & 21\% & 21\% & 10\% & 22\% & 21\% \\
Share of negative weights, the three Rival $\times$ compatriots' start position & 12\%--19\% & 21\%--23\% & 19\%--22\% & 12\%--18\% & 20\%--24\% & 18\%--22\% \\
\bottomrule
\end{tabular}
\end{adjustbox}}
    \begin{tablenotes}[flushleft]
      \footnotesize
      \item \textit{NOTE:} Two-way (competition + athlete) cluster-robust standard errors in parentheses, $t$ with $G-1$ degrees of freedom, $G$ the column's competitions; *** $p<0.01$, ** $p<0.05$, * $p<0.10$.
    \end{tablenotes}
  \end{threeparttable}
\end{table}

Tables~\ref{tab:tac8} and~\ref{tab:tac9} reproduce columns (1)-(3) of Table~\ref{tab:t42} in Section~\ref{sec:escalation}, the second segment in Era B, under the same alternative definitions of being in contention and of a close rival as in Tables~\ref{tab:tac6} and~\ref{tab:tac7}. The results are similar to what we found in Section~\ref{sec:escalation}. Compatriot $\times$ $\mathrm{W}_{\mathrm{later}}$ remains positive and significant in eleven of the twelve columns; on TSS pct it is between $0.176$ and $0.215$, against $0.192$ in Table~\ref{tab:t42}. Compatriot $\times$ $\mathrm{W}_{\mathrm{earlier}}$, in and out of contention, remains insignificant, except on PCS pct out of contention in column (3) of both tables. The penalty on a rival starting earlier remains negatively significant in all columns. The penalty on a rival starting later while the judge's compatriot skater is still in contention is robust to the rival window, but, as in Table~\ref{tab:tac6}, it is smaller and less precisely estimated under the alternative in contention definitions.

\begin{table}[htbp]
  \centering
  \caption{Table~\ref{tab:t42} Robustness Check: Alternative In Contention Definitions}\label{tab:tac8}
  \begin{threeparttable}
  \makebox[\textwidth]{\begin{adjustbox}{max width=\textwidth}\begin{tabular}{lrrrrrr}
\toprule
 & \multicolumn{3}{c}{\textbf{In contention: top 3 so far}} & \multicolumn{3}{c}{\textbf{In contention: top 4 so far, places 1--10 after Segment 1}} \\
\cmidrule(lr){2-4}\cmidrule(lr){5-7}
\textbf{Variable} & \textbf{(1) TSS pct} & \textbf{(2) GOE pct} & \textbf{(3) PCS pct} & \textbf{(4) TSS pct} & \textbf{(5) GOE pct} & \textbf{(6) PCS pct} \\
\midrule
Compatriot & +1.753*** & +2.654*** & +1.782*** & +1.754*** & +2.655*** & +1.785*** \\
 & (0.090) & (0.145) & (0.098) & (0.090) & (0.144) & (0.098) \\
Rival starting earlier & -0.205*** & -0.338*** & -0.183** & -0.203*** & -0.333*** & -0.180** \\
 & (0.061) & (0.104) & (0.073) & (0.059) & (0.101) & (0.072) \\
Rival starting later, in contention & -0.094 & -0.203** & -0.104 & -0.058 & -0.167 & -0.051 \\
 & (0.061) & (0.097) & (0.072) & (0.075) & (0.116) & (0.092) \\
Rival starting later, out of contention & +0.020 & -0.111 & +0.121 & -0.074 & -0.196** & -0.041 \\
 & (0.063) & (0.124) & (0.099) & (0.055) & (0.089) & (0.089) \\
Compatriot $\times$ $\mathrm{W}_{\mathrm{earlier}}$, in contention & +0.052 & +0.110 & +0.043 & +0.065 & +0.108 & +0.064 \\
 & (0.063) & (0.099) & (0.074) & (0.066) & (0.109) & (0.077) \\
Compatriot $\times$ $\mathrm{W}_{\mathrm{earlier}}$, out of contention & +0.094 & -0.190 & +0.307** & +0.031 & -0.049 & +0.110 \\
 & (0.102) & (0.185) & (0.135) & (0.074) & (0.121) & (0.096) \\
Compatriot $\times$ $\mathrm{W}_{\mathrm{later}}$ & +0.200*** & +0.290*** & +0.192** & +0.190*** & +0.260** & +0.191** \\
 & (0.065) & (0.105) & (0.081) & (0.064) & (0.105) & (0.082) \\
Rival starting earlier $\times$ $\mathrm{P}_{p}$ & +0.059 & +0.088 & +0.061 & +0.061 & +0.089 & +0.066 \\
 & (0.037) & (0.059) & (0.048) & (0.038) & (0.060) & (0.049) \\
Rival starting later, in contention $\times$ $\mathrm{P}_{p}$ & +0.046 & +0.036 & +0.070 & +0.064 & +0.072 & +0.084 \\
 & (0.056) & (0.096) & (0.067) & (0.074) & (0.134) & (0.084) \\
Rival starting later, out of contention $\times$ $\mathrm{P}_{p}$ & +0.039 & -0.015 & +0.101 & +0.016 & +0.024 & +0.015 \\
 & (0.084) & (0.127) & (0.119) & (0.058) & (0.095) & (0.077) \\
\midrule
$\mathrm{C}_{jp}$, $\mathrm{R}^{+}_{jp}$, $\mathrm{R}^{-,in}_{jp}$ and $\mathrm{R}^{-,out}_{jp}$ $\cdot$ $\tilde{H}_{p}$ & $\checkmark$ & $\checkmark$ & $\checkmark$ & $\checkmark$ & $\checkmark$ & $\checkmark$ \\
Performance FE & $\checkmark$ & $\checkmark$ & $\checkmark$ & $\checkmark$ & $\checkmark$ & $\checkmark$ \\
Judge FE & $\checkmark$ & $\checkmark$ & $\checkmark$ & $\checkmark$ & $\checkmark$ & $\checkmark$ \\
N & 31,312 & 31,312 & 31,312 & 31,312 & 31,312 & 31,312 \\
\# of Competitions (G) & 57 & 57 & 57 & 57 & 57 & 57 \\
Mass of negative weights, $\mathrm{C}_{jp}$, $\mathrm{R}^{+}_{jp}$, $\mathrm{R}^{-,in}_{jp}$, $\mathrm{R}^{-,out}_{jp}$ & \multicolumn{3}{c}{0.000} & \multicolumn{3}{c}{0.000} \\
Mass of negative weights, $\mathrm{C}_{jp} \times \mathrm{W}_{\mathrm{later}}$ & \multicolumn{3}{c}{0.034} & \multicolumn{3}{c}{0.035} \\
Mass of negative weights, all other interactions & \multicolumn{3}{c}{0.026--0.047} & \multicolumn{3}{c}{0.029--0.066} \\
Share of negative weights, $\mathrm{C}_{jp}$, $\mathrm{R}^{+}_{jp}$, $\mathrm{R}^{-,in}_{jp}$, $\mathrm{R}^{-,out}_{jp}$ & \multicolumn{3}{c}{0\%} & \multicolumn{3}{c}{0\%} \\
Share of negative weights, $\mathrm{C}_{jp} \times \mathrm{W}_{\mathrm{later}}$ & \multicolumn{3}{c}{20\%} & \multicolumn{3}{c}{20\%} \\
Share of negative weights, all other interactions & \multicolumn{3}{c}{15\%--19\%} & \multicolumn{3}{c}{15\%--24\%} \\
\bottomrule
\end{tabular}
\end{adjustbox}}
    \begin{tablenotes}[flushleft]
      \footnotesize
      \item \textit{NOTE:} Era A: 2016/17-2017/18. Era B: 2018/19-2021/22. Two-way cluster-robust standard errors in parentheses. Each interaction is centered at its mean over the rows of its indicator. $t$ with $G-1$ degrees of freedom, $G$ the column's competitions; *** $p<0.01$, ** $p<0.05$, * $p<0.10$. 
    \end{tablenotes}
  \end{threeparttable}
\end{table}

\begin{table}[htbp]
  \centering
  \caption{Table~\ref{tab:t42} Robustness Check: Alternative Close Rival Definitions}\label{tab:tac9}
  \begin{threeparttable}
  \makebox[\textwidth]{\begin{adjustbox}{max width=\textwidth}\begin{tabular}{lrrrrrr}
\toprule
 & \multicolumn{3}{c}{\textbf{Rivals within $\pm$2 start numbers}} & \multicolumn{3}{c}{\textbf{Rivals within $\pm$4 start numbers}} \\
\cmidrule(lr){2-4}\cmidrule(lr){5-7}
\textbf{Variable} & \textbf{(1) TSS pct} & \textbf{(2) GOE pct} & \textbf{(3) PCS pct} & \textbf{(4) TSS pct} & \textbf{(5) GOE pct} & \textbf{(6) PCS pct} \\
\midrule
Compatriot & +1.773*** & +2.692*** & +1.801*** & +1.745*** & +2.654*** & +1.772*** \\
 & (0.085) & (0.136) & (0.093) & (0.091) & (0.151) & (0.097) \\
Rival starting earlier & -0.191*** & -0.287*** & -0.197*** & -0.207*** & -0.322*** & -0.188** \\
 & (0.059) & (0.104) & (0.071) & (0.059) & (0.104) & (0.071) \\
Rival starting later, in contention & -0.155* & -0.319*** & -0.130 & -0.111 & -0.231* & -0.100 \\
 & (0.083) & (0.119) & (0.107) & (0.077) & (0.129) & (0.097) \\
Rival starting later, out of contention & -0.029 & -0.142 & +0.002 & +0.007 & -0.026 & +0.001 \\
 & (0.068) & (0.109) & (0.092) & (0.047) & (0.084) & (0.063) \\
Compatriot $\times$ $\mathrm{W}_{\mathrm{earlier}}$, in contention & +0.060 & +0.171 & +0.017 & +0.055 & +0.163 & +0.003 \\
 & (0.060) & (0.111) & (0.065) & (0.080) & (0.149) & (0.091) \\
Compatriot $\times$ $\mathrm{W}_{\mathrm{earlier}}$, out of contention & +0.056 & -0.013 & +0.147* & +0.034 & -0.001 & +0.091 \\
 & (0.060) & (0.098) & (0.075) & (0.071) & (0.108) & (0.092) \\
Compatriot $\times$ $\mathrm{W}_{\mathrm{later}}$ & +0.215*** & +0.343*** & +0.190** & +0.176** & +0.265** & +0.161* \\
 & (0.055) & (0.089) & (0.074) & (0.068) & (0.115) & (0.084) \\
Rival starting earlier $\times$ $\mathrm{P}_{p}$ & +0.031 & +0.026 & +0.049 & +0.078* & +0.100* & +0.095* \\
 & (0.045) & (0.073) & (0.055) & (0.040) & (0.059) & (0.054) \\
Rival starting later, in contention $\times$ $\mathrm{P}_{p}$ & +0.020 & +0.058 & +0.011 & +0.040 & +0.086 & +0.043 \\
 & (0.098) & (0.169) & (0.109) & (0.080) & (0.143) & (0.094) \\
Rival starting later, out of contention $\times$ $\mathrm{P}_{p}$ & +0.117** & +0.170** & +0.135** & -0.004 & -0.022 & +0.006 \\
 & (0.051) & (0.081) & (0.068) & (0.040) & (0.066) & (0.051) \\
\midrule
$\mathrm{C}_{jp}$, $\mathrm{R}^{+}_{jp}$, $\mathrm{R}^{-,in}_{jp}$ and $\mathrm{R}^{-,out}_{jp}$ $\cdot$ $\tilde{H}_{p}$ & $\checkmark$ & $\checkmark$ & $\checkmark$ & $\checkmark$ & $\checkmark$ & $\checkmark$ \\
Performance FE & $\checkmark$ & $\checkmark$ & $\checkmark$ & $\checkmark$ & $\checkmark$ & $\checkmark$ \\
Judge FE & $\checkmark$ & $\checkmark$ & $\checkmark$ & $\checkmark$ & $\checkmark$ & $\checkmark$ \\
N & 31,312 & 31,312 & 31,312 & 31,312 & 31,312 & 31,312 \\
\# of Competitions (G) & 57 & 57 & 57 & 57 & 57 & 57 \\
Mass of negative weights, $\mathrm{C}_{jp}$, $\mathrm{R}^{+}_{jp}$, $\mathrm{R}^{-,in}_{jp}$, $\mathrm{R}^{-,out}_{jp}$ & \multicolumn{3}{c}{0.000} & \multicolumn{3}{c}{0.000} \\
Mass of negative weights, $\mathrm{C}_{jp} \times \mathrm{W}_{\mathrm{later}}$ & \multicolumn{3}{c}{0.020} & \multicolumn{3}{c}{0.040} \\
Mass of negative weights, all other interactions & \multicolumn{3}{c}{0.031--0.075} & \multicolumn{3}{c}{0.025--0.068} \\
Share of negative weights, $\mathrm{C}_{jp}$, $\mathrm{R}^{+}_{jp}$, $\mathrm{R}^{-,in}_{jp}$, $\mathrm{R}^{-,out}_{jp}$ & \multicolumn{3}{c}{0\%} & \multicolumn{3}{c}{0\%} \\
Share of negative weights, $\mathrm{C}_{jp} \times \mathrm{W}_{\mathrm{later}}$ & \multicolumn{3}{c}{15\%} & \multicolumn{3}{c}{21\%} \\
Share of negative weights, all other interactions & \multicolumn{3}{c}{15\%--25\%} & \multicolumn{3}{c}{15\%--28\%} \\
\bottomrule
\end{tabular}
\end{adjustbox}}
    \begin{tablenotes}[flushleft]
      \footnotesize
      \item \textit{NOTE:} Era A: 2016/17-2017/18. Era B: 2018/19-2021/22. Two-way cluster-robust standard errors in parentheses. Each interaction is centered at its mean over the rows of its indicator. $t$ with $G-1$ degrees of freedom, $G$ the column's competitions; *** $p<0.01$, ** $p<0.05$, * $p<0.10$. 
    \end{tablenotes}
  \end{threeparttable}
\end{table}

\subsection{Heterogeneity Analysis on Escalation Effect}
\label{app:hetescal}

Tables~\ref{tab:tac10} and~\ref{tab:tac11} estimate columns (1)-(3) of Tables~\ref{tab:t42} and~\ref{tab:t42b} separately for senior and junior events. As is discussed in Section~\ref{sec:escalation}, we see that in Era A $\rightarrow$ Era B, most of the escalation comes from the junior events. $\mathrm{C}\times W^{+}_p$ is positive and significant on TSS ($0.253$) and PCS ($0.392$), and $\mathrm{C}\times W^{-,in}_p$ on TSS ($0.296$) and GOE ($0.758$). In the senior events, the only escalation term significant at the 5 percent level is $\mathrm{C}\times W^{+}_p$ on GOE ($0.302$). The competitor penalty responds to $P_p$ only in the senior events and only for later rivals of a compatriot already out of contention ($0.178$, $0.266$ and $0.183$ on TSS, GOE and PCS). The three $\beta_4$ terms, however, are not jointly significant. In Era B $\rightarrow$ Era C, no interaction is significant at the 5 percent level in either tier. In both tables, the mass of negative weights is at most $0.04$ for $\mathrm{C}\times W^{+}_p$ and at most $0.09$ for the other interactions.

\begin{table}[htbp]
  \centering
  \caption{Table~\ref{tab:t42} Heterogeneity: Senior and Junior Events, Era A $\rightarrow$ Era B}\label{tab:tac10}
  \begin{threeparttable}
  \makebox[\textwidth]{\begin{adjustbox}{max width=\textwidth}\begin{tabular}{lrrrrrr}
\toprule
 & \multicolumn{3}{c}{\textbf{Senior events}} & \multicolumn{3}{c}{\textbf{Junior events}} \\
\cmidrule(lr){2-4}\cmidrule(lr){5-7}
\textbf{Variable} & \textbf{(1) TSS pct} & \textbf{(2) GOE pct} & \textbf{(3) PCS pct} & \textbf{(4) TSS pct} & \textbf{(5) GOE pct} & \textbf{(6) PCS pct} \\
\midrule
Compatriot & +1.926*** & +2.757*** & +1.887*** & +1.578*** & +2.581*** & +1.697*** \\
 & (0.125) & (0.204) & (0.124) & (0.118) & (0.196) & (0.154) \\
Rival starting earlier & -0.207** & -0.303* & -0.189** & -0.211*** & -0.369*** & -0.191 \\
 & (0.089) & (0.166) & (0.085) & (0.075) & (0.115) & (0.114) \\
Rival starting later, in contention & -0.090 & -0.132 & -0.123 & -0.185 & -0.462** & -0.107 \\
 & (0.095) & (0.150) & (0.111) & (0.143) & (0.219) & (0.189) \\
Rival starting later, out of contention & +0.071 & -0.074 & +0.152* & -0.120** & -0.200* & -0.139 \\
 & (0.081) & (0.147) & (0.088) & (0.057) & (0.112) & (0.096) \\
Compatriot $\times$ $\mathrm{W}_{\mathrm{earlier}}$, in contention & -0.042 & -0.109 & -0.005 & +0.296** & +0.758*** & +0.131 \\
 & (0.081) & (0.139) & (0.097) & (0.112) & (0.198) & (0.135) \\
Compatriot $\times$ $\mathrm{W}_{\mathrm{earlier}}$, out of contention & +0.070 & -0.027 & +0.183 & +0.064 & +0.056 & +0.111 \\
 & (0.094) & (0.126) & (0.120) & (0.091) & (0.167) & (0.140) \\
Compatriot $\times$ $\mathrm{W}_{\mathrm{later}}$ & +0.107 & +0.302** & -0.019 & +0.253** & +0.183 & +0.392*** \\
 & (0.086) & (0.141) & (0.098) & (0.092) & (0.156) & (0.131) \\
Rival starting earlier $\times$ $\mathrm{P}_{p}$ & +0.065 & +0.113 & +0.041 & +0.066 & +0.091 & +0.085 \\
 & (0.052) & (0.080) & (0.064) & (0.052) & (0.092) & (0.068) \\
Rival starting later, in contention $\times$ $\mathrm{P}_{p}$ & +0.088 & +0.194 & +0.053 & -0.024 & -0.095 & +0.056 \\
 & (0.127) & (0.225) & (0.124) & (0.105) & (0.165) & (0.167) \\
Rival starting later, out of contention $\times$ $\mathrm{P}_{p}$ & +0.178** & +0.266* & +0.183** & -0.050 & -0.090 & -0.056 \\
 & (0.080) & (0.140) & (0.080) & (0.050) & (0.083) & (0.077) \\
\midrule
$\mathrm{C}_{jp}$, $\mathrm{R}^{+}_{jp}$, $\mathrm{R}^{-,in}_{jp}$ and $\mathrm{R}^{-,out}_{jp}$ $\cdot$ $\tilde{H}_{p}$ & $\checkmark$ & $\checkmark$ & $\checkmark$ & $\checkmark$ & $\checkmark$ & $\checkmark$ \\
Performance FE & $\checkmark$ & $\checkmark$ & $\checkmark$ & $\checkmark$ & $\checkmark$ & $\checkmark$ \\
Judge FE & $\checkmark$ & $\checkmark$ & $\checkmark$ & $\checkmark$ & $\checkmark$ & $\checkmark$ \\
N & 14,501 & 14,501 & 14,501 & 16,811 & 16,811 & 16,811 \\
\# of Competitions (G) & 32 & 32 & 32 & 27 & 27 & 27 \\
Mass of negative weights, $\mathrm{C}_{jp}$, $\mathrm{R}^{+}_{jp}$, $\mathrm{R}^{-,in}_{jp}$, $\mathrm{R}^{-,out}_{jp}$ & \multicolumn{3}{c}{0.000} & \multicolumn{3}{c}{0.000} \\
Mass of negative weights, $\mathrm{C}_{jp} \times \mathrm{W}_{\mathrm{later}}$ & \multicolumn{3}{c}{0.022} & \multicolumn{3}{c}{0.040} \\
Mass of negative weights, all other interactions & \multicolumn{3}{c}{0.029--0.069} & \multicolumn{3}{c}{0.028--0.077} \\
Share of negative weights, $\mathrm{C}_{jp}$, $\mathrm{R}^{+}_{jp}$, $\mathrm{R}^{-,in}_{jp}$, $\mathrm{R}^{-,out}_{jp}$ & \multicolumn{3}{c}{0\%} & \multicolumn{3}{c}{0\%} \\
Share of negative weights, $\mathrm{C}_{jp} \times \mathrm{W}_{\mathrm{later}}$ & \multicolumn{3}{c}{17\%} & \multicolumn{3}{c}{20\%} \\
Share of negative weights, all other interactions & \multicolumn{3}{c}{18\%--25\%} & \multicolumn{3}{c}{14\%--28\%} \\
\bottomrule
\end{tabular}
\end{adjustbox}}
    \begin{tablenotes}[flushleft]
      \scriptsize
      \item \textit{NOTE:} Columns (1) to (3) of Table~\ref{tab:t42} (Segment 2 of Era B) estimated separately on the senior and the junior events. Era A: 2016/17-2017/18. Era B: 2018/19-2021/22. Each interaction is centered at its mean over the rows of its indicator in the column. Two-way (competition + athlete) cluster-robust standard errors in parentheses, with a term for the estimation error of the first-step trained premia added to the variance; $t$ with $G-1$ degrees of freedom, $G$ the column's competitions; *** $p<0.01$, ** $p<0.05$, * $p<0.10$. 
    \end{tablenotes}
  \end{threeparttable}
\end{table}

\begin{table}[htbp]
  \centering
  \caption{Table~\ref{tab:t42b} Heterogeneity: Senior and Junior Events, Era B $\rightarrow$ Era C}\label{tab:tac11}
  \begin{threeparttable}
  \makebox[\textwidth]{\begin{adjustbox}{max width=\textwidth}\begin{tabular}{lrrrrrr}
\toprule
 & \multicolumn{3}{c}{\textbf{Senior events}} & \multicolumn{3}{c}{\textbf{Junior events}} \\
\cmidrule(lr){2-4}\cmidrule(lr){5-7}
\textbf{Variable} & \textbf{(1) TSS pct} & \textbf{(2) GOE pct} & \textbf{(3) PCS pct} & \textbf{(4) TSS pct} & \textbf{(5) GOE pct} & \textbf{(6) PCS pct} \\
\midrule
Compatriot & +1.703*** & +2.155*** & +1.759*** & +1.056*** & +1.431*** & +1.252*** \\
 & (0.116) & (0.160) & (0.140) & (0.099) & (0.127) & (0.126) \\
Rival starting earlier & -0.182*** & -0.313*** & -0.122 & -0.238*** & -0.330*** & -0.255*** \\
 & (0.052) & (0.081) & (0.076) & (0.058) & (0.074) & (0.089) \\
Rival starting later, in contention & -0.039 & -0.192 & +0.066 & -0.303** & -0.671*** & -0.175 \\
 & (0.085) & (0.128) & (0.105) & (0.124) & (0.177) & (0.161) \\
Rival starting later, out of contention & +0.040 & +0.089 & +0.012 & +0.058 & -0.160** & +0.225** \\
 & (0.075) & (0.121) & (0.101) & (0.066) & (0.073) & (0.098) \\
Compatriot $\times$ $\mathrm{W}_{\mathrm{earlier}}$, in contention & -0.030 & +0.019 & -0.076 & +0.092 & +0.228 & +0.002 \\
 & (0.121) & (0.198) & (0.136) & (0.156) & (0.205) & (0.220) \\
Compatriot $\times$ $\mathrm{W}_{\mathrm{earlier}}$, out of contention & +0.008 & -0.100 & +0.072 & +0.207* & +0.244 & +0.307* \\
 & (0.097) & (0.149) & (0.146) & (0.111) & (0.155) & (0.176) \\
Compatriot $\times$ $\mathrm{W}_{\mathrm{later}}$ & -0.045 & +0.043 & -0.116 & +0.098 & +0.028 & +0.174 \\
 & (0.078) & (0.133) & (0.080) & (0.086) & (0.150) & (0.110) \\
Rival starting earlier $\times$ $\mathrm{P}_{p}$ & -0.123* & -0.152 & -0.145 & +0.074 & +0.088 & +0.098 \\
 & (0.072) & (0.093) & (0.088) & (0.055) & (0.082) & (0.087) \\
Rival starting later, in contention $\times$ $\mathrm{P}_{p}$ & -0.197* & -0.280* & -0.178 & +0.281* & +0.155 & +0.439* \\
 & (0.103) & (0.157) & (0.121) & (0.151) & (0.175) & (0.223) \\
Rival starting later, out of contention $\times$ $\mathrm{P}_{p}$ & -0.031 & +0.014 & -0.083 & -0.107 & -0.161 & -0.107 \\
 & (0.076) & (0.101) & (0.093) & (0.068) & (0.101) & (0.104) \\
\midrule
$\mathrm{C}_{jp}$, $\mathrm{R}^{+}_{jp}$, $\mathrm{R}^{-,in}_{jp}$ and $\mathrm{R}^{-,out}_{jp}$ $\cdot$ $\tilde{H}_{p}$ & $\checkmark$ & $\checkmark$ & $\checkmark$ & $\checkmark$ & $\checkmark$ & $\checkmark$ \\
Performance FE & $\checkmark$ & $\checkmark$ & $\checkmark$ & $\checkmark$ & $\checkmark$ & $\checkmark$ \\
Judge FE & $\checkmark$ & $\checkmark$ & $\checkmark$ & $\checkmark$ & $\checkmark$ & $\checkmark$ \\
N & 20,293 & 20,293 & 20,293 & 24,386 & 24,386 & 24,386 \\
\# of Competitions (G) & 44 & 44 & 44 & 36 & 36 & 36 \\
Mass of negative weights, $\mathrm{C}_{jp}$, $\mathrm{R}^{+}_{jp}$, $\mathrm{R}^{-,in}_{jp}$, $\mathrm{R}^{-,out}_{jp}$ & \multicolumn{3}{c}{0.000} & \multicolumn{3}{c}{0.000} \\
Mass of negative weights, $\mathrm{C}_{jp} \times \mathrm{W}_{\mathrm{later}}$ & \multicolumn{3}{c}{0.016} & \multicolumn{3}{c}{0.035} \\
Mass of negative weights, all other interactions & \multicolumn{3}{c}{0.028--0.057} & \multicolumn{3}{c}{0.031--0.087} \\
Share of negative weights, $\mathrm{C}_{jp}$, $\mathrm{R}^{+}_{jp}$, $\mathrm{R}^{-,in}_{jp}$, $\mathrm{R}^{-,out}_{jp}$ & \multicolumn{3}{c}{0\%} & \multicolumn{3}{c}{0\%} \\
Share of negative weights, $\mathrm{C}_{jp} \times \mathrm{W}_{\mathrm{later}}$ & \multicolumn{3}{c}{14\%} & \multicolumn{3}{c}{21\%} \\
Share of negative weights, all other interactions & \multicolumn{3}{c}{17\%--25\%} & \multicolumn{3}{c}{12\%--30\%} \\
\bottomrule
\end{tabular}
\end{adjustbox}}
    \begin{tablenotes}[flushleft]
      \scriptsize
      \item \textit{NOTE:} Columns (1) to (3) of Table~\ref{tab:t42b} (Segment 2 of Era C) estimated separately on the senior and the junior events. Era B: 2018/19-2021/22. Era C: 2022/23-2025/26. Each interaction is centered at its mean over the rows of its indicator in the column. Two-way (competition + athlete) cluster-robust standard errors in parentheses, with a term for the estimation error of the first-step trained premia added to the variance; $t$ with $G-1$ degrees of freedom, $G$ the column's competitions; *** $p<0.01$, ** $p<0.05$, * $p<0.10$. 
    \end{tablenotes}
  \end{threeparttable}
\end{table}

Table~\ref{tab:tac12} adds the unweighted terms of Table~\ref{tab:t42c} ($\mathrm{C}\times$share and $\mathrm{R}\times$compatriot judge) to Table~\ref{tab:tac10}. Table~\ref{tab:tac13} then drops the rival terms. Holding the share fixed, $\mathrm{C}\times W$ is identified only from differences in the trained premia of the protecting judges. As is discussed in Section~\ref{sec:escalation}, because the two measures are close to collinear, the standard errors of the $\mathrm{C}\times W$ terms are four to five times those in Table~\ref{tab:tac10}. Over the same change, the mass of negative weights on $\mathrm{C}\times W^{+}_p$ rises from $0.02$-$0.04$ to $0.39$-$0.55$. The only interactions significant at the 5 percent level are on junior GOE, where $\mathrm{C}\times W^{+}_p$ is positive and $\mathrm{C}\times$share of later rivals is negative. But both carry 30-47\% negative weights, so we do not interpret them. Dropping the rival terms barely moves the compatriot-side estimates, so this imprecision mainly comes from the collinearity of the weighted and unweighted measures.

\begin{table}[htbp]
  \centering
  \caption{Table~\ref{tab:tac10} With the Unweighted Shares and the Compatriot-judge Indicator Added}\label{tab:tac12}
  \begin{threeparttable}
  \makebox[\textwidth]{\begin{adjustbox}{max width=\textwidth}\begin{tabular}{lrrrrrr}
\toprule
 & \multicolumn{3}{c}{\textbf{Senior events}} & \multicolumn{3}{c}{\textbf{Junior events}} \\
\cmidrule(lr){2-4}\cmidrule(lr){5-7}
\textbf{Variable} & \textbf{(1) TSS pct} & \textbf{(2) GOE pct} & \textbf{(3) PCS pct} & \textbf{(4) TSS pct} & \textbf{(5) GOE pct} & \textbf{(6) PCS pct} \\
\midrule
Compatriot & +1.928*** & +2.755*** & +1.891*** & +1.577*** & +2.578*** & +1.698*** \\
 & (0.124) & (0.204) & (0.124) & (0.117) & (0.194) & (0.154) \\
Rival starting earlier & -0.208** & -0.305* & -0.189** & -0.212*** & -0.373*** & -0.190 \\
 & (0.089) & (0.166) & (0.085) & (0.075) & (0.114) & (0.114) \\
Rival starting later, in contention & -0.090 & -0.129 & -0.125 & -0.186 & -0.467** & -0.106 \\
 & (0.094) & (0.150) & (0.111) & (0.144) & (0.219) & (0.189) \\
Rival starting later, out of contention & +0.071 & -0.077 & +0.155* & -0.119* & -0.199* & -0.138 \\
 & (0.081) & (0.146) & (0.088) & (0.058) & (0.112) & (0.096) \\
Compatriot $\times$ $\mathrm{W}_{\mathrm{earlier}}$, in contention & -0.246 & -0.785 & -0.023 & +0.807 & +1.552 & +0.469 \\
 & (0.377) & (0.686) & (0.434) & (0.531) & (1.202) & (0.418) \\
Compatriot $\times$ $\mathrm{W}_{\mathrm{earlier}}$, out of contention & +0.075 & +0.188 & +0.062 & -0.373 & -0.526 & -0.279 \\
 & (0.332) & (0.565) & (0.326) & (0.577) & (1.208) & (0.647) \\
Compatriot $\times$ $\mathrm{W}_{\mathrm{later}}$ & +0.392 & +0.595 & +0.335 & +0.533 & +1.253** & +0.160 \\
 & (0.374) & (0.587) & (0.499) & (0.412) & (0.582) & (0.529) \\
Compatriot $\times$ share earlier, in contention & +0.556 & +1.840 & +0.055 & -1.478 & -2.301 & -0.972 \\
 & (1.045) & (1.807) & (1.242) & (1.432) & (3.297) & (1.163) \\
Compatriot $\times$ share earlier, out of contention & -0.014 & -0.593 & +0.339 & +1.127 & +1.485 & +1.019 \\
 & (0.822) & (1.445) & (0.781) & (1.527) & (3.184) & (1.650) \\
Compatriot $\times$ share later & -0.802 & -0.829 & -0.993 & -0.762 & -2.947** & +0.657 \\
 & (0.947) & (1.517) & (1.250) & (1.008) & (1.384) & (1.306) \\
Rival starting earlier $\times$ $\mathrm{P}_{p}$ & -0.042 & +0.230 & -0.222 & +0.059 & +0.053 & +0.036 \\
 & (0.129) & (0.238) & (0.154) & (0.237) & (0.409) & (0.267) \\
Rival starting later, in contention $\times$ $\mathrm{P}_{p}$ & +0.068 & +0.101 & +0.064 & -0.145 & +0.305 & -0.384 \\
 & (0.236) & (0.454) & (0.252) & (0.378) & (0.655) & (0.488) \\
Rival starting later, out of contention $\times$ $\mathrm{P}_{p}$ & +0.227 & +0.384 & +0.127 & +0.020 & -0.126 & +0.159 \\
 & (0.212) & (0.400) & (0.233) & (0.237) & (0.559) & (0.274) \\
Rival starting earlier $\times$ compatriot judge & +0.336 & -0.370 & +0.830* & +0.021 & +0.100 & +0.139 \\
 & (0.330) & (0.616) & (0.408) & (0.597) & (1.025) & (0.704) \\
Rival starting later, in contention $\times$ compatriot judge & +0.072 & +0.347 & -0.050 & +0.385 & -1.275 & +1.409 \\
 & (0.629) & (1.167) & (0.723) & (1.019) & (1.982) & (1.270) \\
Rival starting later, out of contention $\times$ compatriot judge & -0.147 & -0.351 & +0.165 & -0.190 & +0.094 & -0.579 \\
 & (0.619) & (1.110) & (0.697) & (0.657) & (1.452) & (0.800) \\
\midrule
$\mathrm{C}_{jp}$, $\mathrm{R}^{+}_{jp}$, $\mathrm{R}^{-,in}_{jp}$ and $\mathrm{R}^{-,out}_{jp}$ $\cdot$ $\tilde{H}_{p}$ & $\checkmark$ & $\checkmark$ & $\checkmark$ & $\checkmark$ & $\checkmark$ & $\checkmark$ \\
Performance FE & $\checkmark$ & $\checkmark$ & $\checkmark$ & $\checkmark$ & $\checkmark$ & $\checkmark$ \\
Judge FE & $\checkmark$ & $\checkmark$ & $\checkmark$ & $\checkmark$ & $\checkmark$ & $\checkmark$ \\
N & 14,501 & 14,501 & 14,501 & 16,811 & 16,811 & 16,811 \\
\# of Competitions (G) & 32 & 32 & 32 & 27 & 27 & 27 \\
Mass of negative weights, $\mathrm{C}_{jp}$, $\mathrm{R}^{+}_{jp}$, $\mathrm{R}^{-,in}_{jp}$, $\mathrm{R}^{-,out}_{jp}$ & \multicolumn{3}{c}{0.000} & \multicolumn{3}{c}{0.000} \\
Mass of negative weights, $\mathrm{C}_{jp} \times \mathrm{W}_{\mathrm{later}}$ & \multicolumn{3}{c}{0.394} & \multicolumn{3}{c}{0.545} \\
Mass of negative weights, $\mathrm{C}_{jp} \times$ share later & \multicolumn{3}{c}{0.397} & \multicolumn{3}{c}{0.617} \\
Mass of negative weights, all other interactions & \multicolumn{3}{c}{0.137--0.781} & \multicolumn{3}{c}{0.384--1.033} \\
Share of negative weights, $\mathrm{C}_{jp}$, $\mathrm{R}^{+}_{jp}$, $\mathrm{R}^{-,in}_{jp}$, $\mathrm{R}^{-,out}_{jp}$ & \multicolumn{3}{c}{0\%} & \multicolumn{3}{c}{0\%} \\
Share of negative weights, $\mathrm{C}_{jp} \times \mathrm{W}_{\mathrm{later}}$ & \multicolumn{3}{c}{39\%} & \multicolumn{3}{c}{47\%} \\
Share of negative weights, $\mathrm{C}_{jp} \times$ share later & \multicolumn{3}{c}{33\%} & \multicolumn{3}{c}{30\%} \\
Share of negative weights, all other interactions & \multicolumn{3}{c}{30\%--49\%} & \multicolumn{3}{c}{22\%--59\%} \\
\bottomrule
\end{tabular}
\end{adjustbox}}
    \begin{tablenotes}[flushleft]
      \scriptsize
      \item \textit{NOTE:} The columns of Table~\ref{tab:tac10} with the terms of Table~\ref{tab:t42c} added. Era A: 2016/17-2017/18. Era B: 2018/19-2021/22. Each interaction is centered at its mean over the rows of its indicator in the column. Two-way (competition + athlete) cluster-robust standard errors in parentheses, with a term for the estimation error of the first-step trained premia added to the variance; $t$ with $G-1$ degrees of freedom, $G$ the column's competitions; *** $p<0.01$, ** $p<0.05$, * $p<0.10$. 
    \end{tablenotes}
  \end{threeparttable}
\end{table}

\begin{table}[htbp]
  \centering
  \caption{Table~\ref{tab:tac12} Without the Rival Terms}\label{tab:tac13}
  \begin{threeparttable}
  \makebox[\textwidth]{\begin{adjustbox}{max width=\textwidth}\begin{tabular}{lrrrrrr}
\toprule
 & \multicolumn{3}{c}{\textbf{Senior events}} & \multicolumn{3}{c}{\textbf{Junior events}} \\
\cmidrule(lr){2-4}\cmidrule(lr){5-7}
\textbf{Variable} & \textbf{(1) TSS pct} & \textbf{(2) GOE pct} & \textbf{(3) PCS pct} & \textbf{(4) TSS pct} & \textbf{(5) GOE pct} & \textbf{(6) PCS pct} \\
\midrule
Compatriot & +1.957*** & +2.828*** & +1.904*** & +1.643*** & +2.698*** & +1.756*** \\
 & (0.119) & (0.183) & (0.116) & (0.111) & (0.180) & (0.137) \\
Compatriot $\times$ $\mathrm{W}_{\mathrm{earlier}}$, in contention & -0.108 & -0.577 & +0.116 & +0.787 & +1.491 & +0.482 \\
 & (0.379) & (0.667) & (0.445) & (0.540) & (1.217) & (0.422) \\
Compatriot $\times$ $\mathrm{W}_{\mathrm{earlier}}$, out of contention & +0.077 & +0.166 & +0.085 & -0.261 & -0.324 & -0.161 \\
 & (0.317) & (0.547) & (0.315) & (0.543) & (1.162) & (0.617) \\
Compatriot $\times$ $\mathrm{W}_{\mathrm{later}}$ & +0.365 & +0.608 & +0.280 & +0.606 & +1.383** & +0.219 \\
 & (0.377) & (0.595) & (0.494) & (0.410) & (0.597) & (0.516) \\
Compatriot $\times$ share earlier, in contention & +0.152 & +1.239 & -0.347 & -1.487 & -2.140 & -1.132 \\
 & (1.066) & (1.781) & (1.278) & (1.463) & (3.333) & (1.184) \\
Compatriot $\times$ share earlier, out of contention & -0.065 & -0.572 & +0.213 & +0.888 & +0.985 & +0.802 \\
 & (0.782) & (1.393) & (0.750) & (1.429) & (3.059) & (1.572) \\
Compatriot $\times$ share later & -0.708 & -0.855 & -0.805 & -0.866 & -3.065** & +0.528 \\
 & (0.947) & (1.527) & (1.224) & (1.012) & (1.412) & (1.294) \\
\midrule
$\mathrm{C}_{jp}$ $\cdot$ $\tilde{H}_{p}$ & $\checkmark$ & $\checkmark$ & $\checkmark$ & $\checkmark$ & $\checkmark$ & $\checkmark$ \\
Performance FE & $\checkmark$ & $\checkmark$ & $\checkmark$ & $\checkmark$ & $\checkmark$ & $\checkmark$ \\
Judge FE & $\checkmark$ & $\checkmark$ & $\checkmark$ & $\checkmark$ & $\checkmark$ & $\checkmark$ \\
N & 14,501 & 14,501 & 14,501 & 16,811 & 16,811 & 16,811 \\
\# of Competitions (G) & 32 & 32 & 32 & 27 & 27 & 27 \\
Mass of negative weights, $\mathrm{C}_{jp}$ & \multicolumn{3}{c}{0.000} & \multicolumn{3}{c}{0.000} \\
Mass of negative weights, $\mathrm{C}_{jp} \times \mathrm{W}_{\mathrm{later}}$ & \multicolumn{3}{c}{0.394} & \multicolumn{3}{c}{0.547} \\
Mass of negative weights, $\mathrm{C}_{jp} \times$ share later & \multicolumn{3}{c}{0.397} & \multicolumn{3}{c}{0.618} \\
Mass of negative weights, all other interactions & \multicolumn{3}{c}{0.471--0.784} & \multicolumn{3}{c}{0.697--1.034} \\
Share of negative weights, $\mathrm{C}_{jp}$ & \multicolumn{3}{c}{0\%} & \multicolumn{3}{c}{0\%} \\
Share of negative weights, $\mathrm{C}_{jp} \times \mathrm{W}_{\mathrm{later}}$ & \multicolumn{3}{c}{39\%} & \multicolumn{3}{c}{46\%} \\
Share of negative weights, $\mathrm{C}_{jp} \times$ share later & \multicolumn{3}{c}{33\%} & \multicolumn{3}{c}{30\%} \\
Share of negative weights, all other interactions & \multicolumn{3}{c}{34\%--49\%} & \multicolumn{3}{c}{31\%--56\%} \\
\bottomrule
\end{tabular}
\end{adjustbox}}
    \begin{tablenotes}[flushleft]
      \scriptsize
      \item \textit{NOTE:} Table~\ref{tab:tac12} without the three rival indicators and their interactions. Era A: 2016/17-2017/18. Era B: 2018/19-2021/22. Each interaction is centered at its mean over the rows of its indicator in the column. Two-way (competition + athlete) cluster-robust standard errors in parentheses, with a term for the estimation error of the first-step trained premia added to the variance; $t$ with $G-1$ degrees of freedom, $G$ the column's competitions; *** $p<0.01$, ** $p<0.05$, * $p<0.10$. 
    \end{tablenotes}
  \end{threeparttable}
\end{table}

\section{Robustness Checks With Alternative Cluster-Robust Inference Methods} 
\label{app:fedcluster}

In this section, we report the results of Sections~\ref{sec:scorebias} and~\ref{sec:strategic} with two alternatives to our default standard errors, which cluster by competition and by athlete \citep{cameron2011robust}. The first clusters by competition and by athlete federation instead. The second keeps the clustering by competition and by athlete but replaces the standard variance estimator with the two-way cluster jackknife version discussed in \citet{mackinnon2026jackknife}. Clustering by athlete federation follows \cite{zitzewitz2014does}, who clusters by competitor country and by competition, and it allows the residuals of different athletes from the same federation to be correlated, for example through the federation's judges or its national selection. We do not use it as the default because the athlete federation clusters are few in effect and very unequal in size. In our data, the ten largest federations account for 56 percent of the observations, while 52 of the 78 federations have under 1 percent each. With a few dominant clusters and many small ones, the cluster-robust variance estimator relies mainly on the largest federations, and its $t$ statistics can be far from their nominal distribution, so that inference may not be reliable \citep{mackinnon2023cluster}. The jackknife re-estimates each regression without one competition, one athlete, or one competition-athlete pair at a time. \citet{mackinnon2026jackknife} argues that standard two-way cluster-robust tests can perform poorly in finite samples, while the proposed jackknife alternative performs well in simulations. For tables reported in this section, both alternatives keep the correction for the first-step estimation error and add the same first-step variance term as our default.

Tables~\ref{tab:tad1}-\ref{tab:tad4} reproduce Tables~\ref{tab:t41}-\ref{tab:t42c} and report the slopes and the interactions with both alternative standard errors. For Tables~\ref{tab:t31}-~\ref{tab:t32}, the alternative inference results are reported in the table footnotes, recording every estimate whose significance level changes.

Table~\ref{tab:tad1} reproduces Table~\ref{tab:t41}. Under clustering by athlete federation, the second segment estimates in column (6) keep their significance levels. On GOE in the pooled columns, Compatriot $\times$ start position moves from the 1 to the 10 percent level. Rival starting later, out of contention $\times$ compatriots' start position moves from the 10 to the 5 percent level on PCS and is no longer significant in column (5), and Rival starting later, in contention $\times$ compatriots' start position in column (5) moves from the 5 to the 1 percent level. The jackknife standard errors are 1.14 to 1.36 times ours. Under the jackknife, Compatriot $\times$ start position and Rival starting earlier $\times$ compatriots' start position remain significant at the 5 percent level or better wherever they are significant in Table 6, and so does Rival starting later, in contention $\times$ compatriots' start position on TSS pct and PCS pct and in the second segment. 

\begin{table}[H]
\centering
\footnotesize
\caption{Table~\ref{tab:t41} with Alternative Cluster-Robust Standard Errors} \label{tab:tad1}
  \begin{threeparttable}
  \makebox[\textwidth]{\begin{adjustbox}{max width=\textwidth}\begin{tabular}{lrrrrrr}
\toprule
 & \multicolumn{3}{c}{\textbf{All rows}} & \multicolumn{3}{c}{\textbf{TSS pct, by how the start order is set}} \\
\cmidrule(lr){2-4}\cmidrule(lr){5-7}
\textbf{Variable} & \textbf{(1) TSS pct} & \textbf{(2) GOE pct} & \textbf{(3) PCS pct} & \textbf{(4) Segment 1, open draw} & \textbf{(5) Segment 1, seeded order} & \textbf{(6) Segment 2} \\
\midrule
Compatriot $\times$ start position & -0.711*** & -0.454*** & -1.162*** & -0.697*** & -0.536*** & -0.702*** \\
 & (0.104)*** & (0.254)* & (0.135)*** & (0.166)*** & (0.185)*** & (0.156)*** \\
 & [0.114]*** & [0.197]** & [0.149]*** & [0.237]*** & [0.241]** & [0.152]*** \\
Rival starting earlier $\times$ compatriots' start position & -0.346*** & -0.525*** & -0.379*** & -0.176 & -0.238 & -0.468*** \\
 & (0.100)*** & (0.167)*** & (0.121)*** & (0.154) & (0.167) & (0.159)*** \\
 & [0.108]*** & [0.170]*** & [0.148]** & [0.184] & [0.196] & [0.148]*** \\
Rival starting later, in contention $\times$ compatriots' start position & -0.553*** & -0.494** & -0.757*** & -0.434 & -0.419** & -0.734*** \\
 & (0.135)*** & (0.242)** & (0.165)*** & (0.430) & (0.150)*** & (0.180)*** \\
 & [0.153]*** & [0.258]* & [0.192]*** & [0.367] & [0.255] & [0.253]*** \\
Rival starting later, out of contention $\times$ compatriots' start position & -0.215* & -0.148 & -0.324* & +0.176 & -0.642* & -0.358** \\
 & (0.116)* & (0.204) & (0.145)** & (0.227) & (0.443) & (0.166)** \\
 & [0.150] & [0.241] & [0.214] & [0.316] & [0.441] & [0.187]* \\
\midrule
$\mathrm{C}_{jp}$, $\mathrm{R}^{+}_{jp}$, $\mathrm{R}^{-,in}_{jp}$ and $\mathrm{R}^{-,out}_{jp}$ $\cdot$ $H_{p}$ & $\checkmark$ & $\checkmark$ & $\checkmark$ & $\checkmark$ & $\checkmark$ & $\checkmark$ \\
Performance FE & $\checkmark$ & $\checkmark$ & $\checkmark$ & $\checkmark$ & $\checkmark$ & $\checkmark$ \\
Judge FE & $\checkmark$ & $\checkmark$ & $\checkmark$ & $\checkmark$ & $\checkmark$ & $\checkmark$ \\
N & 206,983 & 206,983 & 206,983 & 46,208 & 62,918 & 97,857 \\
\# of Competitions (G) & 172 & 172 & 172 & 70 & 102 & 172 \\
\# of Skaters (H) & 2,470 & 2,470 & 2,470 & 2,109 & 1,338 & 2,368 \\
\# of Federations (F) & 78 & 78 & 78 & 78 & 63 & 77 \\
\bottomrule
\end{tabular}
\end{adjustbox}}
    \begin{tablenotes}[flushleft]
      \footnotesize
        \item \textit{NOTE:} Each estimate is the slope in Table~\ref{tab:t41}, with the stars of that table's standard error. In parentheses: the two-way (competition + athlete federation) cluster-robust standard errors in parentheses, $t$ with $\min(G, F)-1$ degrees of freedom, $G$ the column’s competitions and $F$ its athlete federations. In brackets: the two-way cluster jackknife standard error by competition and athlete, $t$ with $\min(G,H)-1$ degrees of freedom, $H$ the number of athletes. Each standard error carries the stars of its own p-value; *** $p<0.01$, ** $p<0.05$, * $p<0.10$.
    \end{tablenotes}
  \end{threeparttable}
\end{table}

Table~\ref{tab:tad2} reproduces Table~\ref{tab:t42}. Compatriot $\times$ $\mathrm{W}_{\mathrm{later}}$ remains significant at the 5 percent level on TSS and GOE under both alternatives and at the 10 percent level on PCS, while in column (6) it keeps the 5 percent level under clustering by athlete federation and moves to the 10 percent level under the jackknife, whose standard errors are 1.12 to 1.69 times the default. Rival starting later, in contention $\times P_p$ in column (5) is no longer significant under either alternative, and in column (6) it moves from the 5 to the 10 percent level under the jackknife. Under the jackknife, Compatriot $\times W$ for earlier rivals, out of contention in column (4) is also no longer significant.

\begin{table}[H]
\centering
\footnotesize
\caption{Table~\ref{tab:t42} with Alternative Cluster-Robust Standard Errors} \label{tab:tad2}
  \begin{threeparttable}
  \makebox[\textwidth]{\begin{adjustbox}{max width=\textwidth}\begin{tabular}{lrrrrrr}
\toprule
 & \multicolumn{3}{c}{\textbf{Segment 2}} & \multicolumn{3}{c}{\textbf{TSS pct}} \\
\cmidrule(lr){2-4}\cmidrule(lr){5-7}
\textbf{Variable} & \textbf{(1) TSS pct} & \textbf{(2) GOE pct} & \textbf{(3) PCS pct} & \textbf{(4) Segment 1, open draw} & \textbf{(5) Segment 1, seeded order} & \textbf{(6) All rows} \\
\midrule
Compatriot $\times$ $\mathrm{W}_{\mathrm{earlier}}$, in contention & +0.052 & +0.153 & +0.009 & -0.116 & +0.000 & +0.002 \\
 & (0.064) & (0.115) & (0.060) & (0.170) & (0.071) & (0.056) \\
 & [0.094] & [0.173] & [0.101] & [0.187] & [0.095] & [0.067] \\
Compatriot $\times$ $\mathrm{W}_{\mathrm{earlier}}$, out of contention & +0.045 & -0.039 & +0.129 & -0.376* & -0.092 & -0.047 \\
 & (0.062) & (0.044) & (0.117) & (0.182)* & (0.166) & (0.073) \\
 & [0.082] & [0.141] & [0.103] & [0.233] & [0.149] & [0.070] \\
Compatriot $\times$ $\mathrm{W}_{\mathrm{later}}$ & +0.192*** & +0.262** & +0.193** & -0.020 & +0.036 & +0.114** \\
 & (0.071)*** & (0.096)*** & (0.103)* & (0.150) & (0.081) & (0.048)** \\
 & [0.076]** & [0.118]** & [0.102]* & [0.154] & [0.091] & [0.060]* \\
Rival starting earlier $\times$ $\mathrm{P}_{p}$ & +0.059 & +0.085 & +0.063 & -0.011 & -0.001 & +0.020 \\
 & (0.039) & (0.059) & (0.055) & (0.086) & (0.064) & (0.038) \\
 & [0.050] & [0.076] & [0.066] & [0.085] & [0.065] & [0.035] \\
Rival starting later, in contention $\times$ $\mathrm{P}_{p}$ & +0.056 & +0.073 & +0.077 & -0.039 & +0.141* & +0.113** \\
 & (0.079) & (0.147) & (0.092) & (0.131) & (0.096) & (0.045)** \\
 & [0.118] & [0.203] & [0.132] & [0.160] & [0.105] & [0.066]* \\
Rival starting later, out of contention $\times$ $\mathrm{P}_{p}$ & +0.035 & +0.051 & +0.032 & +0.014 & -0.169 & -0.022 \\
 & (0.051) & (0.100) & (0.055) & (0.108) & (0.100) & (0.039) \\
 & [0.065] & [0.114] & [0.077] & [0.167] & [0.185] & [0.051] \\
\midrule
$\mathrm{C}_{jp}$, $\mathrm{R}^{+}_{jp}$, $\mathrm{R}^{-,in}_{jp}$ and $\mathrm{R}^{-,out}_{jp}$ $\cdot$ $\tilde{H}_{p}$ & $\checkmark$ & $\checkmark$ & $\checkmark$ & $\checkmark$ & $\checkmark$ & $\checkmark$ \\
Performance FE & $\checkmark$ & $\checkmark$ & $\checkmark$ & $\checkmark$ & $\checkmark$ & $\checkmark$ \\
Judge FE & $\checkmark$ & $\checkmark$ & $\checkmark$ & $\checkmark$ & $\checkmark$ & $\checkmark$ \\
N & 31,312 & 31,312 & 31,312 & 13,814 & 20,862 & 65,988 \\
\# of Competitions (G) & 57 & 57 & 57 & 21 & 36 & 57 \\
\# of Skaters (H) & 1,084 & 1,084 & 1,084 & 809 & 669 & 1,147 \\
\# of Federations (F) & 71 & 71 & 71 & 69 & 56 & 71 \\
\bottomrule
\end{tabular}
\end{adjustbox}}
    \begin{tablenotes}[flushleft]
      \footnotesize
        \item \textit{NOTE:} Era A: 2016/17-2017/18. Era B: 2018/19-2021/22. Each estimate is the slope in Table~\ref{tab:t42}, with the stars of that table's standard error. In parentheses: the two-way (competition + athlete federation) cluster-robust standard errors in parentheses, $t$ with $\min(G, F)-1$ degrees of freedom, $G$ the column’s competitions and $F$ its athlete federations. In brackets: the two-way cluster jackknife standard error by competition and athlete, $t$ with $\min(G,H)-1$ degrees of freedom, $H$ the number of athletes. Each standard error carries the stars of its own p-value; *** $p<0.01$, ** $p<0.05$, * $p<0.10$.
    \end{tablenotes}
  \end{threeparttable}
\end{table}

Table~\ref{tab:tad3} reproduces Table~\ref{tab:t42b}. Rival starting earlier $\times P_p$ remains significant at the 5 percent level or better in columns (5) and (6) under both alternatives. Under clustering by athlete federation, Compatriot $\times W$ for earlier rivals in contention in column (6) becomes significant at the 10 percent level, and Rival starting later, out of contention  $\times P_p$ in column (4) is no longer significant. The jackknife standard errors are 1.15 to 1.66 times the default, and under the jackknife, Rival starting later, out of contention $\times P_p$ is no longer significant on TSS pct and PCS pct in the second segment or in column (4).

\begin{table}[H]
\centering
\footnotesize
\caption{Table~\ref{tab:t42b} with Alternative Cluster-Robust Standard Errors} \label{tab:tad3}
  \begin{threeparttable}
  \makebox[\textwidth]{\begin{adjustbox}{max width=\textwidth}\begin{tabular}{lrrrrrr}
\toprule
 & \multicolumn{3}{c}{\textbf{Segment 2}} & \multicolumn{3}{c}{\textbf{TSS pct}} \\
\cmidrule(lr){2-4}\cmidrule(lr){5-7}
\textbf{Variable} & \textbf{(1) TSS pct} & \textbf{(2) GOE pct} & \textbf{(3) PCS pct} & \textbf{(4) Segment 1, open draw} & \textbf{(5) Segment 1, seeded order} & \textbf{(6) All rows} \\
\midrule
Compatriot $\times$ $\mathrm{W}_{\mathrm{earlier}}$, in contention & +0.006 & +0.079 & -0.038 & +0.150 & +0.065 & +0.065 \\
 & (0.076) & (0.140) & (0.104) & (0.220) & (0.100) & (0.036)* \\
 & [0.122] & [0.197] & [0.146] & [0.333] & [0.128] & [0.075] \\
Compatriot $\times$ $\mathrm{W}_{\mathrm{earlier}}$, out of contention & +0.060 & +0.026 & +0.119 & -0.155 & -0.206 & -0.023 \\
 & (0.058) & (0.138) & (0.075) & (0.194) & (0.219) & (0.076) \\
 & [0.087] & [0.135] & [0.127] & [0.296] & [0.212] & [0.073] \\
Compatriot $\times$ $\mathrm{W}_{\mathrm{later}}$ & -0.004 & +0.042 & -0.039 & +0.252 & +0.097 & +0.053 \\
 & (0.107) & (0.151) & (0.106) & (0.164) & (0.083) & (0.076) \\
 & [0.071] & [0.112] & [0.092] & [0.217] & [0.125] & [0.069] \\
Rival starting earlier $\times$ $\mathrm{P}_{p}$ & -0.027 & -0.036 & -0.025 & -0.092 & -0.179*** & -0.084** \\
 & (0.040) & (0.059) & (0.059) & (0.083) & (0.064)*** & (0.031)*** \\
 & [0.054] & [0.077] & [0.073] & [0.108] & [0.084]** & [0.040]** \\
Rival starting later, in contention $\times$ $\mathrm{P}_{p}$ & -0.016 & -0.119 & +0.062 & +0.022 & -0.064 & -0.040 \\
 & (0.082) & (0.141) & (0.113) & (0.128) & (0.078) & (0.060) \\
 & [0.121] & [0.148] & [0.174] & [0.214] & [0.090] & [0.070] \\
Rival starting later, out of contention $\times$ $\mathrm{P}_{p}$ & -0.108** & -0.110 & -0.146** & +0.220* & +0.040 & +0.004 \\
 & (0.049)** & (0.073) & (0.066)** & (0.138) & (0.126) & (0.048) \\
 & [0.069] & [0.099] & [0.094] & [0.156] & [0.128] & [0.055] \\
\midrule
$\mathrm{C}_{jp}$, $\mathrm{R}^{+}_{jp}$, $\mathrm{R}^{-,in}_{jp}$ and $\mathrm{R}^{-,out}_{jp}$ $\cdot$ $\tilde{H}_{p}$ & $\checkmark$ & $\checkmark$ & $\checkmark$ & $\checkmark$ & $\checkmark$ & $\checkmark$ \\
Performance FE & $\checkmark$ & $\checkmark$ & $\checkmark$ & $\checkmark$ & $\checkmark$ & $\checkmark$ \\
Judge FE & $\checkmark$ & $\checkmark$ & $\checkmark$ & $\checkmark$ & $\checkmark$ & $\checkmark$ \\
N & 44,679 & 44,679 & 44,679 & 20,290 & 29,020 & 93,989 \\
\# of Competitions (G) & 76 & 76 & 76 & 28 & 48 & 76 \\
\# of Skaters (H) & 1,237 & 1,237 & 1,237 & 972 & 718 & 1,279 \\
\# of Federations (F) & 76 & 76 & 76 & 75 & 60 & 78 \\
\bottomrule
\end{tabular}
\end{adjustbox}}
    \begin{tablenotes}[flushleft]
      \footnotesize
        \item \textit{NOTE:} Era B: 2018/19-2021/22. Era C: 2022/23-2025/26.  Each estimate is the slope in Table~\ref{tab:t42b}, with the stars of that table's standard error. In parentheses: the two-way (competition + athlete federation) cluster-robust standard errors in parentheses, $t$ with $\min(G, F)-1$ degrees of freedom, $G$ the column’s competitions and $F$ its athlete federations. In brackets: the two-way cluster jackknife standard error by competition and athlete, $t$ with $\min(G,H)-1$ degrees of freedom, $H$ the number of athletes. Each standard error carries the stars of its own p-value; *** $p<0.01$, ** $p<0.05$, * $p<0.10$.
    \end{tablenotes}
  \end{threeparttable}
\end{table}

Table~\ref{tab:tad4} reproduces Table~\ref{tab:t42c}. Under clustering by athlete federation, Compatriot $\times$ share later remains significant on all three outcomes in the second segment, moving from the 1 to the 5 percent level on TSS and from the 5 to the 10 percent level on PCS. Rival starting later, in contention $\times$ compatriot judge in column (5) moves from the 1 to the 5 percent level. The jackknife standard errors are 1.16 to 1.54 times the default. Under the jackknife, Compatriot $\times$ share later also remains significant on all three outcomes in the second segment,
at the 5 percent level on TSS pct and at the 10 percent level on GOE pct and PCS pct. Rival starting later, in contention $\times$ compatriot judge moves to the 10 percent level in column (5) and to the 5 percent level in column (6), while Rival starting earlier $\times$  compatriot judge is no longer significant on TSS pct and PCS pct. Compatriot $\times$ share earlier, out of contention in column (4) moves from the 5 to the 10 percent level.

\begin{table}[H]
\centering
\footnotesize
\caption{Table~\ref{tab:t42c} with Alternative Cluster-Robust Standard Errors} \label{tab:tad4}
  \begin{threeparttable}
  \makebox[\textwidth]{\begin{adjustbox}{max width=\textwidth}\begin{tabular}{lrrrrrr}
\toprule
 & \multicolumn{3}{c}{\textbf{Segment 2}} & \multicolumn{3}{c}{\textbf{TSS pct}} \\
\cmidrule(lr){2-4}\cmidrule(lr){5-7}
\textbf{Variable} & \textbf{(1) TSS pct} & \textbf{(2) GOE pct} & \textbf{(3) PCS pct} & \textbf{(4) Segment 1, open draw} & \textbf{(5) Segment 1, seeded order} & \textbf{(6) All rows} \\
\midrule
Compatriot $\times$ share earlier, in contention & +0.147 & +0.433 & +0.039 & -0.369 & +0.035 & -0.002 \\
 & (0.135) & (0.236)* & (0.155) & (0.365) & (0.182) & (0.137) \\
 & [0.230] & [0.420] & [0.287] & [0.462] & [0.258] & [0.174] \\
Compatriot $\times$ share earlier, out of contention & +0.129 & -0.114 & +0.359 & -1.129** & -0.172 & -0.136 \\
 & (0.161) & (0.140) & (0.299) & (0.469)** & (0.518) & (0.199) \\
 & [0.214] & [0.369] & [0.268] & [0.588]* & [0.402] & [0.185] \\
Compatriot $\times$ share later & +0.423*** & +0.525** & +0.467** & -0.139 & +0.100 & +0.238* \\
 & (0.183)** & (0.251)** & (0.240)* & (0.397) & (0.196) & (0.119)* \\
 & [0.184]** & [0.294]* & [0.248]* & [0.412] & [0.217] & [0.144] \\
Rival starting earlier $\times$ compatriot judge & +0.182* & +0.189 & +0.243* & -0.013 & -0.033 & +0.056 \\
 & (0.093)* & (0.135) & (0.135)* & (0.228) & (0.153) & (0.086) \\
 & [0.126] & [0.191] & [0.173] & [0.225] & [0.173] & [0.088] \\
Rival starting later, in contention $\times$ compatriot judge & +0.206 & +0.105 & +0.370 & +0.029 & +0.535*** & +0.367*** \\
 & (0.255) & (0.334) & (0.354) & (0.277) & (0.207)** & (0.128)*** \\
 & [0.330] & [0.542] & [0.363] & [0.372] & [0.272]* & [0.179]** \\
Rival starting later, out of contention $\times$ compatriot judge & +0.075 & +0.108 & +0.074 & -0.005 & -0.411 & -0.073 \\
 & (0.140) & (0.251) & (0.165) & (0.257) & (0.255) & (0.098) \\
 & [0.175] & [0.297] & [0.210] & [0.351] & [0.421] & [0.121] \\
\midrule
$\mathrm{C}_{jp}$, $\mathrm{R}^{+}_{jp}$, $\mathrm{R}^{-,in}_{jp}$ and $\mathrm{R}^{-,out}_{jp}$ $\cdot$ $\tilde{H}_{p}$ & $\checkmark$ & $\checkmark$ & $\checkmark$ & $\checkmark$ & $\checkmark$ & $\checkmark$ \\
Performance FE & $\checkmark$ & $\checkmark$ & $\checkmark$ & $\checkmark$ & $\checkmark$ & $\checkmark$ \\
Judge FE & $\checkmark$ & $\checkmark$ & $\checkmark$ & $\checkmark$ & $\checkmark$ & $\checkmark$ \\
N & 31,312 & 31,312 & 31,312 & 13,814 & 20,862 & 65,988 \\
\# of Competitions (G) & 57 & 57 & 57 & 21 & 36 & 57 \\
\# of Skaters (H) & 1,084 & 1,084 & 1,084 & 809 & 669 & 1,147 \\
\# of Federations (F) & 71 & 71 & 71 & 69 & 56 & 71 \\
\bottomrule
\end{tabular}
\end{adjustbox}}
    \begin{tablenotes}[flushleft]
      \footnotesize
        \item \textit{NOTE:} Era A: 2016/17-2017/18. Era B: 2018/19-2021/22. Each estimate is the slope in Table~\ref{tab:t42c}, with the stars of that table's standard error. In parentheses: the two-way (competition + athlete federation) cluster-robust standard errors in parentheses, $t$ with $\min(G, F)-1$ degrees of freedom, $G$ the column’s competitions and $F$ its athlete federations. In brackets: the two-way cluster jackknife standard error by competition and athlete, $t$ with $\min(G,H)-1$ degrees of freedom, $H$ the number of athletes. Each standard error carries the stars of its own p-value; *** $p<0.01$, ** $p<0.05$, * $p<0.10$.
    \end{tablenotes}
  \end{threeparttable}
\end{table}

\end{document}